\documentclass[reprint, superscriptaddress]{revtex4-1}
\usepackage{amsmath, amssymb, amsfonts}
\usepackage{bm}
\usepackage{xr-hyper}
\usepackage{xr}
\usepackage{hyperref}
\usepackage{ulem}
\usepackage{xcolor}
\usepackage{float}
\usepackage{graphicx}
\usepackage{dcolumn}

\hypersetup{
	colorlinks = false,
	urlcolor = {1 1 1}
}

\graphicspath{{}}

\begin{document}

\title{Relations between large-scale brain connectivity and effects of regional stimulation depend on collective dynamical state}

\author{Lia Papadopoulos} 
\affiliation{Department of Physics \& Astronomy, University of Pennsylvania, Philadelphia, PA 19104, USA}
\author{Christopher W. Lynn} 
\affiliation{Department of Physics \& Astronomy, University of Pennsylvania, Philadelphia, PA 19104, USA}
\author{Demian Battaglia} 
\affiliation{Institute for Systems Neuroscience, University Aix-Marseille, Boulevard Jean Moulin 27, 13005 Marseille, France}
\author{Danielle S. Bassett}
\email{dsb@seas.upenn.edu}
\affiliation{Department of Physics \& Astronomy, University of Pennsylvania, Philadelphia, PA 19104, USA}
\affiliation{Department of Bioengineering, University of Pennsylvania, Philadelphia, PA 19104, USA}
\affiliation{Department of Electrical \& Systems Engineering, University of Pennsylvania, Philadelphia, PA 19104, USA}
\affiliation{Department of Neurology, University of Pennsylvania, Philadelphia, PA 19104, USA}
\affiliation{Department of Psychiatry, University of Pennsylvania, Philadelphia, PA 19104, USA}
\affiliation{Santa Fe Institute, Santa Fe, NM 87501, USA}

\begin{abstract}

At the macroscale, the brain operates as a network of interconnected neuronal populations, which display coordinated rhythmic dynamics that support interareal communication. Understanding how stimulation of different brain areas impacts such activity is important for gaining basic insights into brain function and for further developing therapeautic neurmodulation. However, the complexity of brain structure and dynamics hinders predictions regarding the downstream effects of focal stimulation. More specifically, little is known about how the collective oscillatory regime of brain network activity -- in concert with network stucture -- affects the outcomes of perturbations. Here, we combine human connectome data and biophysical modeling to begin filling these gaps. By tuning parameters that control collective system dynamics, we identify distinct states of simulated brain activity, and investigate how the distributed effects of stimulation manifest at different dynamical working points. When baseline oscillations are weak, the stimulated area exhibits enhanced power and frequency, and due to network interactions, nearby regions develop phase-locked activity in the excited frequency band. Notably, beyond these linear effects, we further find that focal stimulation causes more distributed modifications to interareal coherence in a band containing regions’ baseline oscillation frequencies. Imporantly, depending on the dynamical state of the system, these broadband effects can be better predicted by functional rather than structural connectivity, emphasizing a complex interplay between anatomical organization, dynamics, and response to perturbation. In contrast, when the network operates in a regime of strong regional oscillations, stimulation causes only slight shifts in power and frequency, and structural connectivity becomes most predictive of stimulation-induced changes in network activity. In sum, this work builds upon and extends previous computational studies investigating the impacts of stimulation, and underscores the fact that both the stimulation site, and, crucially, the regime of brain network dynamics, can influence the network-wide responses to local perturbations.

\end{abstract}

\maketitle

\section{Introduction \label{s:intro}}

The brain is a multiscale system composed of many dynamical units that interact to produce a vast array of functions. At a large scale, macroscopic regions -- each containing tens of thousands of neurons -- are linked by a physical web of white matter tracts that facilitate the propagation of activity between distributed network elements. At the level of large neuronal ensembles or brain areas, collective activity is often rhythmic in nature \cite{Buzsaki2006:RhythmsOfTheBrain}, and these rhythms can become temporally coordinated between distant regions, giving rise to so-called functional interactions \cite{Varela2001:TheBrainweb}. Importantly, oscillations have been implicated in a number of cognitive processes \cite{Ward2003:Synchronous,Wang2010:Neurophysiological,Salinas2001:Correlated,Pascal2009:NeuronalGamma,Thut2012:TheFunctional,Kopell2010:AreDifferentRhythms,Cannon2014:Neurosystems}, and coherent activity is hypothesized to play an important role in interareal communication and information transfer among distributed brain areas \cite{Pascal2009:NeuronalGamma,Fries2015:Rhythms,Salinas2001:Correlated}. Nonetheless, despite progress in mapping and characterizing the brain’s anatomical pathways and measuring neural oscillations, a number of questions remain as to how individual components in a brain network shape and modulate system-wide dynamics.

Among these questions, understanding how large-scale, oscillatory brain dynamics respond to localized perturbations is of critical importance \cite{Shafi2012:Exploration,Polania2018:Studying,To2018:Changing,Luft2014:Best,Thut2012:TheFunctional}. Because the brain is far from a closed or static system, such activity changes could be induced by sensory inputs \cite{Brosch2002:Stimulus,Henrie2005:LFPPowerSpectra}, different tasks \cite{Jokisch2007:Modulation,Feige1993:Oscillatory}, or other internal or regulatory processes \cite{Hirata2010:Neocortex,Kastner1999:Increased,BatistaBrito2018:Modulation,Gazzaley2012:TopDown}. In addition to naturally-induced changes, stimulation techniques such as transcranial magnetic stimulation \cite{Hallett2007:TranscranialMagnetic}, direct current stimulation \cite{Filmer2014:Applications}, and alternating current stimulation \cite{Vosskuhl2018:NonInvasive} can also be employed to invoke modulations of dynamics in a specific brain area. By combining these techniques with imaging methods like EEG and MEG \cite{Thut2009:NewInsights,Witkowski2016:Mapping,Antal2004:Oscillatory,Neuling2015:Friends,Siebner2009:Consensus,Bortoletto2015:TheContribution}, it is possible to examine how the act of exciting a particular network component modifies rhythmic neural activity. Furthermore, in addition to its utility for basic science, neurostimulation has emerged as a promising approach for treating a number of neurological and psychiatric conditions \cite{Johnson2013:Neuromodulation,Schulz2013:Noninvasive,Fisher2014:Electrical}. 

Yet, while prior work has mainly focused on characterizing the proximal effects of sensory input or exogenous neurostimulation, a growing body of literature indicates that local excitations of neural activity can have widespread downstream consequences \cite{Luft2014:Best,To2018:Changing,Shafi2012:Exploration,Polania2018:Studying}. The realization that stimulation can have network-wide effects necessitates further investigations into the operating principles underlying such phenomena \cite{Muldoon2016:StimulationBased,Gollo2017:MappingHowLocal,Spiegler2016:SelectiveActivation,Kunze2016:Transcranial,Witt2013:Controlling,Kirst2016:Dynamic,Stiso2019:WhiteMatter,Khambhati2019:FunctionalControl}. Furthermore, a crucial but seemingly understudied point is that the effects of stimulating a particular brain area can depend not only on the nature or location of the perturbation, but also on the intrinsic dynamical state of the system at baseline \cite{Bergmann2018:BrainState,Thut2017:Guiding,Silvanto2008:StateDependency}. In particular, recent efforts have investigated the state-dependent effects of stimulation via precise experiments \cite{Neuling2013:Orchestrating,Ruhnau2016:EyesWide} -- focusing largely on alpha-band activity in single cortical areas -- and via modeling \cite{Alagapan2016:Modulation,Lefebvre2017:Stochastic,Li2017:Unified}. These studies have uncovered robust relationships between the endogenous state of rhythmic activity and the capacity of external stimulation to modulate cortical oscillations in a given brain area. However, a pivotal next step is to extend the notion of state-dependence to the case of whole-brain networks, which acknowledge the fact that regions do not operate in isolation. Rather, in the case of large-scale brain networks, the macroscopic dynamical regime of the system arises from an interplay between units’ local activity and long-range anatomical coupling \cite{Breakspear2017:DynamicModels}, leading to the emergence of collective oscillatory modes \cite{Atasoy2016:HumanBrain,Kirst2016:Dynamic}. Although it is reasonable to hypothesize that the global state of brain network activity should play a role in determining how a focal stimulation will manifest and influence distributed functional interactions, these ideas have yet to be systematically examined.

Thus, there is now a need to concurrently investigate and merge two outstanding questions: \textit{(1)} how regional stimulation spreads to induce distributed effects on rhythmic brain network dynamics, and \textit{(2)} how the global dynamical regime of the system impacts these effects. Here, we begin to answer these questions by constructing a biophysically-motivated model of large-scale brain activity, in which individual brain areas are modeled as Wilson-Cowan neural masses \cite{Wilson1972:ExcitatoryandInhibitory} coupled according to empirically-derived anatomical connectivity \cite{Breakspear2017:DynamicModels}. We first demonstrate that, in the absence of stimulation, the interareal coupling strength and the baseline excitation of the network transition the system between qualitatively distinct collective dynamical states. By providing additional excitation to a single brain area, we then systematically examine the consequences of such local stimulation on network activity. The primary and crucial contribution of this study is an exploration of how the effects of focal perturbations can depend not only on which area is stimulated, but also on the baseline dynamical regime of the non-linear model. Hence, this work builds upon and extends previous whole-brain modeling efforts that have examined the effects of regional perturbations \cite{Gollo2017:MappingHowLocal,Spiegler2016:SelectiveActivation,Muldoon2016:StimulationBased} with other work examining the state-dependent effects of stimulation in single cortical areas, but not large-scale networks \cite{Alagapan2016:Modulation,Lefebvre2017:Stochastic}.

We find that in states of low baseline excitation, stimulation can significantly enhance the frequency and power of regional activity, whereas in states of high baseline drive, local dynamics are less sensitive to perturbations. Importantly, these results show qualitative similarities and agreement with past work examining the focal effects of stimulation \cite{Alagapan2016:Modulation,Lefebvre2017:Stochastic}. We further find that, due to network interactions, regional perturbations can propagate and interact with brain areas' ongoing rhythms. Depending on the system working point, changes in interareal phase-locking can be induced at the excited frequency of the stimulated region, and also in a broader band comprising brain areas' spontaneous oscillations, which may be well-separated from the excited frequency. Moreover, changing the dynamical regime of the system also modulates the strength of associations between network-wide responses to perturbations and structural or functional network connectivity. Hence, changing the collective oscillatory state of the system -- which need not be entirely determined by the anatomical network -- qualitatively changes the distributed effects of focal perturbations, and alters the relations between those effects and measures of either structural or dynamical organization. In sum, this study integrates but also expands past computational studies, and highlights that the effects of localized activity changes can depend on both the location of the perturbed region and on the collective state of brain network dynamics.

\section{Materials and Methods \label{s:methods}}

\subsection{Acquisition of empirical human structural brain data}

Human anatomical brain networks were reconstructed by applying deterministic tractography algorithms to diffusion-weighted MRI. In this study, we used a single, group-representative composite network assembled from 30 subject-level networks \cite{Betzel2018:SpecificityAndRobustness,Betzel2017:TheModular,Betzel2018:Diversity}. The mean age of participants was 26.2 years, the standard deviation was 5.7 years, and 14 of the subjects were female. In order to map anatomical networks, diffusion spectrum and T1-weighted anatomical images were acquired for each individual. For the DSI scans, 257 directions were sampled using a Q5 half-shell acquisition scheme with a maximum $b$-value of 5000 and an isotropic voxel size of 2.4 mm. We used an axial acquisition with repetition time TR = 5 seconds, echo time TE = 138 ms, 52 slices, and field of view of [231, 231, 125]mm. The T1 sequences used a voxel size of [0.9, 0.9, 1.0]mm, repetition time TR = 1.85 seconds, echo time TE = 4ms, and field of view of [240, 180, 160]mm. All participants gave informed consent in writing and all protocols were authorized by the Institutional Review Board of the University of Pennsylvania. These same data have been used in several prior studies \cite{Betzel2018:SpecificityAndRobustness,Betzel2017:TheModular,Betzel2018:Diversity,Medaglia2018:Functional,Betzel2016:Optimally}.

\begin{figure*}
	\centering
	\includegraphics[width=0.9\textwidth]{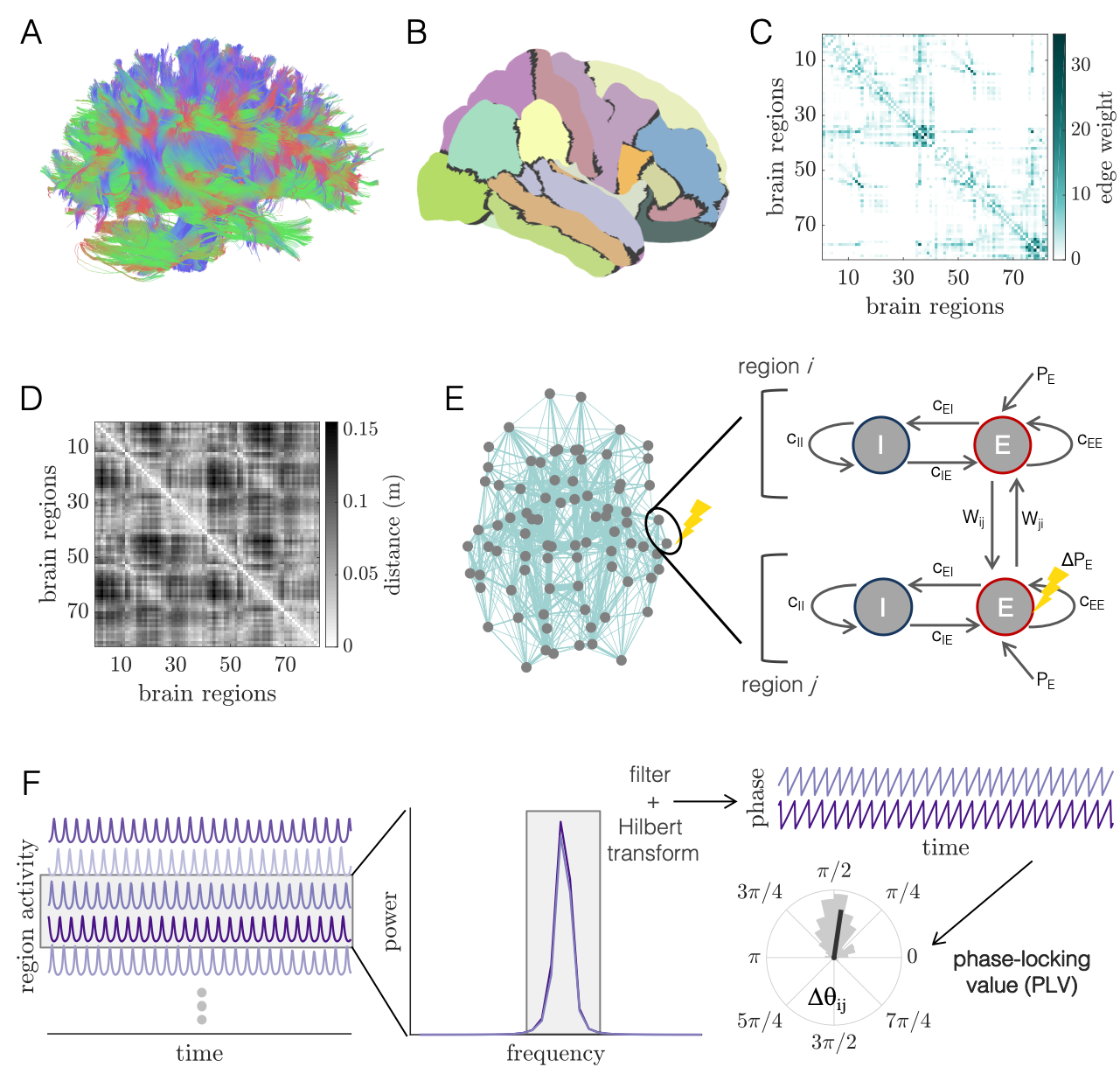}
	\caption{\textbf{Whole-brain imaging data, computational model of large-scale brain dynamics, and schematic of analysis}. \textit{(A)} An example of white matter streamlines that are reconstructed from diffusion spectrum imaging and tractography of a human brain. \textit{(B)} Noninvasive magnetic resonance imaging scans of human brain anatomy are used to segment the cortex and subcortex into 82 regions. \textit{(C)} Adjacency matrix for a group-averaged structural brain network, in which individual brain areas are represented as network nodes, and white matter streamline counts between region pairs are represented as weighted network edges. \textit{(D)} Matrix of Euclidean distances between the centers of mass of all pairs of brain regions. \textit{(E)} Left: Structural brain network representation in anatomical space; location of gray circles correspond to region centers of mass, and teal lines show the strongest 20\% of interareal connections, with line thickness proportional to connection strength. The two encircled nodes correspond to an unperturbed region $i$ and an excited region $j$ in the large-scale brain network, where the perturbed region is indicated by the yellow lightning bolt. Right: Schematic of the computational model of large-scale brain dynamics. The activity of a given brain region $i$ is modeled as a Wilson-Cowan neural mass unit, composed of interacting sub-populations of excitatory $E$ and inhibitory $I$ neurons. Neural masses are then coupled through their excitatory pools according to the anatomical brain network structure. A perturbation to region $j$ (pictorially represented with the lightning bolt) is modeled as an increase in the excitatory input from $P_{E}$ to $P_{E} + \Delta P_{E}$. \textit{(F)} The computational model generates oscillatory time-series of neural population activity for each brain region. These time-series can then be analyzed in Fourier space to determine relevant frequency bands for further analysis. After filtering time-series within a frequency band of interest, functional interactions between brain region pairs are determined by extracting phase variables from each region's filtered activity via the Hilbert transform, and then computing the phase-locking value to assess the consistency of phase relations over time and trials.}
	\label{f:methods}
\end{figure*}

DSI Studio (www.dsi-studio.labsolver.org) was used to reconstruct DSI data using $q$-space diffeomorphic reconstruction (QSDR) \cite{Yeh2011:Estimation}, which reconstructs diffusion-weighted images in native space and computes the quantitative anisotropy (QA) of each voxel. Using the statistical parametric mapping nonlinear registration algorithm \cite{Frackowiak2004:Experimental}, the image is then warped to a template QA volume in Montreal Neurological Institute (MNI) space. Finally, spin-density functions were reconstructed with a mean diffusion distance of 1.25 mm with three fiber orientations per voxel. A modified FACT algorithm  \cite{Mori2002:FiberTracking} was then used to perform deterministic fiber tracking with an angular cutoff of $55^{\mathrm{o}}$, step size of 1.0 mm, minimum length of 10 mm, spin density function smoothing of 0.00, maximum length of 400 mm, and a QA threshold determined by DWI signal in the colony-stimulating factor \cite{Betzel2018:SpecificityAndRobustness,Betzel2017:TheModular,Betzel2018:Diversity,Medaglia2018:Functional,Betzel2016:Optimally,Gu2015:Controllability}. The algorithm terminated when 1,000,000 streamlines were reconstructed for each individual \cite{Betzel2018:SpecificityAndRobustness,Betzel2017:TheModular,Betzel2018:Diversity,Medaglia2018:Functional,Betzel2016:Optimally,Gu2015:Controllability} (Fig.~\ref{f:methods}A).

T1 anatomical scans were segmented using FreeSurfer \cite{Fischl2012:FreeSurfer} and parcellated using the Connectome Mapping Toolkit (http://www.connectomics.org) according to an $N = 82$ area atlas \cite{Cammoun2012:Mapping} of 68 cortical and 14 subcortical areas (Fig.~\ref{f:methods}B; Table S1). The parcellation was registered to the $b=0$ volume of each subject's DSI data, and region labels were mapped from native space to MNI coordinates using a $b=0$-to-MNI voxel mapping \cite{Betzel2018:SpecificityAndRobustness,Betzel2017:TheModular,Betzel2018:Diversity,Medaglia2018:Functional,Betzel2016:Optimally,Gu2015:Controllability}. While we use a relatively coarse-grained atlas, it aligns with atlas sizes used in other computational modeling studies (e.g., \cite{Muldoon2016:StimulationBased,Abeysuriya2018:ABiophysicalModel,Cabral2011:RoleOfLocal,Tewarie2018:Relationships,Cabral2014:Exploring}), and was chosen to reduce the computational costs of numerical simulations. However, we do mention limitations involved with this choice in the Discussion. 

\subsection{Network representation of anatomical brain data}

To include the structure of interareal connections in the model of large-scale brain activity, we represented the human structural brain data as a network. This representation was achieved by first mapping each of the $N = 82$ brain regions to a unique node in a structural brain network $\mathbf{A}$ (see Table S1 for the mapping between node numbering and brain region labels). The edge weight $A_{ij}$ between two brain areas $i$ and $j$ was then given by the strength of the anatomical connectivity between the two regions. Following \cite{Betzel2018:Diversity,Betzel2018:SpecificityAndRobustness}, we defined the edge weights in the anatomical brain networks to be the total number of streamlines $s_{ij}$ that either started in area $i$ and terminated in area $j$ or that started in area $j$ and terminated in area $i$, normalized by the geometric mean of the volumes of regions $i$ and $j$.

Because the brain is a spatially-embedded system \cite{Stiso2018:Spatial}, each region $i$ also has a location $\mathbf{r}_{i} = (x_{i},y_{i},z_{i})$ in real space. In the network representation, we defined the location of each node to be the center of mass of the corresponding region, allowing us to calculate matrix elements $D_{ij}$ representing the Euclidean distance between nodes $i$ and $j$. These interareal distances are then used to approximate time delays for signal transmission in the computational model of neural activity (see Sec.~\ref{s:model}) \cite{Ritter2013:TheVirtual,Deco2009:KeyRoleofCoupling,Muldoon2016:StimulationBased,Abeysuriya2018:ABiophysicalModel,Cabral2011:RoleOfLocal}.

In this study, we report results using a group-representative structural brain network derived by combining individual brain networks across multiple subjects. We used a previously-established consensus method for constructing the group representative network that preserves both the average binary connection density of the individual brain networks, as well as the approximate edge-length distribution of intra- and inter-hemispheric connections \cite{Betzel2018:SpecificityAndRobustness}. More details on this pooling procedure can be found in \cite{Betzel2017:TheModular}. A group-representative interareal distance matrix was constructed by averaging the pairwise Euclidean distance matrices across subjects. In what follows, we assume that $\mathbf{A}$ (or $A_{ij}$) refers to the group-representative structural brain network, and that $\mathbf{D}$ (or $D_{ij}$) refers to the group-averaged interareal Euclidean distance matrix. We show the group-representative anatomical connectivity matrix in Fig.~\ref{f:methods}C, and we show the group-averaged Euclidean distance matrix in Fig.~\ref{f:methods}D.

\subsection{Biophysical model of large-scale brain dynamics}
\label{s:model}

To model large-scale brain dynamics, we use a biophysically-motivated approach in which simulated activity is generated by a network of interacting neural masses \cite{Breakspear2017:DynamicModels}. In particular, the activity of each brain area is modeled as a Wilson-Cowan (WC) neural mass \cite{Wilson1972:ExcitatoryandInhibitory} and individual units are coupled according to the edges in the anatomical network, which are derived from empirical measurements of white matter connectivity \cite{Muldoon2016:StimulationBased,Abeysuriya2018:ABiophysicalModel,Deco2009:KeyRoleofCoupling,Hlinka2012:UsingComputationalModels,Roberts2019:Metastable}. We then utilize this computational approach to conduct a basic examination of how localized (regional) changes in neural activity affect dynamics across the brain. We offer a schematic of the whole-brain computational model in Fig.~\ref{f:methods}E. On the left, we show the structural brain network in anatomical space.
We focus in particular on the two interconnected regions $i$ and $j$ encircled in black, of which the lower one ($j$) receives additional excitation (as denoted by the yellow lightning bolt). On the right, we show the basic setup of the coupled WC system for these two units. In the WC model, the activity of a particular brain region is defined by a coupled system of excitatory ($E$) and inhibitory ($I$) neuronal populations, and the dynamical variables are the mean firing rates of the $E$ and $I$ pools comprising a given brain region. The time-evolution of the average firing rates are thus in general governed by both intrinsic properties of the sub-populations in a single region, as well as delayed, long-range input from other areas as dictated by the pattern of anatomical connectivity. In line with several previous studies \cite{Muldoon2016:StimulationBased,Abeysuriya2018:ABiophysicalModel,Deco2009:KeyRoleofCoupling,Honey2009:Predicting,Gollo2015:DwellingQuietly,Roberts2019:Metastable,Hlinka2012:UsingComputationalModels,Glomb2017:RestingState}, we consider long-range connections to couple only the excitatory subpopulations of distinct brain areas.

The dynamics of the $j^{th}$ brain area are governed by the following set of coupled differential equations:

\begin{equation}
\begin{aligned}
\tau_{E}\frac{dE_{j}(t)}{dt} = & -E_{j}(t) + [1-E_{j}(t)]\mathcal{S}_{E}[c_{EE}E_{j}(t) - c_{IE}I_{j}(t) \\
											& +C\sum_{i}W_{ij}E_{i}(t-\tau_{ij}) + P_{E,j}] + \sigma_{E}\xi(t)											
\label{eq:WC_network_E}	
\end{aligned}		
\end{equation}
\noindent and					
\begin{equation}
\begin{aligned}
\tau_{I}\frac{dI_{j}(t)}{dt} = & -I_{j}(t) + [1 - I_{j}(t)]\mathcal{S}_{I}[c_{EI}E_{j}(t)-c_{II}I_{j}(t) \\
										& +P_{I}]+\sigma_{I}\xi(t).
\label{eq:WC_network_I}
\end{aligned}
\end{equation}

\noindent The variables $E_{j}(t)$ and $I_{j}(t)$ correspond to the mean firing rates of the excitatory and inhibitory subpopulations of region $j$, and $\tau_{E}$ and $\tau_{I}$ are the excitatory and inhibitory time constants, respectively. The non-linear activation functions $\mathcal{S}_{E}$ and $\mathcal{S}_{I}$ of the excitatory and inhibitory pools are given by sigmoidals of the form

\begin{equation}
\mathcal{S}_{E}(x) = \frac{1}{1+e^{-a_{E}(x-\mu_{E})}}  
\label{eq:activations_functionsE}
\end{equation}
and
\begin{equation}
\mathcal{S}_{I}(x) = \frac{1}{1+e^{-a_{I}(x-\mu_{I})}}.
\label{eq:activations_functionsI}
\end{equation}

\noindent The quantities $\mu_{E}$ and $\mu_{I}$ give the mean firing threshold of each subpopulation, and the gain parameters $a_{E}$ and $a_{I}$ set the spread of the firing thresholds for the two groups. 

Dynamics of the excitatory ensemble are driven by \textit{(1)} the local interaction strength within the excitatory subpopulation $c_{EE}$, \textit{(2)} the interaction strength from the inhibitory subpopulation to the excitatory subpopulation $c_{IE}$, \textit{(3)} constant, non-specific inputs $P_{E,j}$, and also \textit{(4)} interactions $W_{ij}$ corresponding to long-range excitatory inputs from different populations $i$ that link to unit $j$ via white matter connectivity. Following \cite{Gollo2015:DwellingQuietly,Roberts2019:Metastable,Demirtas2018:Hierarchical,Hlinka2012:UsingComputationalModels}, we let $W_{ij} = \frac{A_{ij}}{\sum_{i}A_{ij}}$, which is simply the anatomical connection strength from unit $i$ to unit $j$, normalized by the total input to region $j$. Furthermore, $C$ is a global coupling that tunes the overall interaction strength between different brain areas, and $\tau_{ij}$ is a time delay between regions $i$ and $j$ that arises due to the spatial embedding of the brain network and the fact that signal transmission speeds are finite \cite{Deco2009:KeyRoleofCoupling,Muldoon2016:StimulationBased,Abeysuriya2018:ABiophysicalModel,Cabral2011:RoleOfLocal,Ritter2013:TheVirtual,Cabral2014:Exploring}. We set the time delay to $\tau_{ij} = \frac{D_{ij}}{v}$, where $D_{ij}$ is the Euclidean distance between regions $i$ and $j$, and where $v$ is a constant signal conduction speed. Activity in the inhibitory ensemble depends on \textit{(1)} the interaction strength $c_{EI}$ from the excitatory subpopulation, \textit{(2)} the local interaction strength within the inhibitory subpopulation $c_{II}$, and \textit{(3)} other possible non-specific inputs $P_{I}$. Finally, to increase biological plausibility and incorporate the stochastic nature of neural dynamics, we add a term $\sigma_{E}\xi(t)$ to Eq.~\ref{eq:WC_network_E} and a term $\sigma_{I}\xi(t)$ to Eq.~\ref{eq:WC_network_I}, which correspond to Gaussian white noise with zero mean and standard deviations $\sigma_{E}$ and $\sigma_{I}$, respectively \cite{Muldoon2016:StimulationBased,Deco2009:KeyRoleofCoupling}. In what follows, we will take the excitatory neuronal population activities $E_{i}(t)$ of each brain area as the observables of interest \cite{Muldoon2016:StimulationBased,Abeysuriya2018:ABiophysicalModel,Roberts2019:Metastable,Murray2018:BiophysicalModeling,Hlinka2012:UsingComputationalModels}.

\subsubsection{Model parameters}

Under an appropriate choice of parameters, the WC model can give rise to oscillatory firing rate dynamics \cite{Wilson1972:ExcitatoryandInhibitory}. Such rhythmic activity is ubiquitous in large-scale neural systems \cite{Buzsaki2006:RhythmsOfTheBrain} and is the dynamical behavior of interest for this investigation. While the frequencies of oscillations observed in neural systems can span orders of magnitude \cite{Buzsaki2006:RhythmsOfTheBrain}, local neuronal ensembles and individual brain areas often generate oscillations in the gamma frequency band (30-90Hz) as a result of feedback between coupled excitatory and inhibitory neurons \cite{Wang2010:Neurophysiological,Borgers2003:Synchronization,Kopell2010:GammaAndTheta}. Furthermore, gamma oscillations and synchronization of gamma activity between distributed brain areas are associated with the flow of information between neuronal ensembles \cite{Fries2015:Rhythms,Palmigiano2017:FlexibleInformation,Battaglia2012:DynamicEffective}, are modulated by stimuli \cite{Henrie2005:LFPPowerSpectra,Brosch2002:Stimulus}, and are thought to underlie a number of cognitive processes \cite{Pascal2009:NeuronalGamma}. Motivated by the fact that gamma oscillations are robustly observed in local excitatory-inhibitory circuits, we set the parameters of the phenomenological WC model such that individual brain regions oscillate in the gamma band \cite{Deco2009:KeyRoleofCoupling} (see Table ~\ref{t:WC_params}). We note that it may also be interesting in future work to investigate other frequency bands or multiple frequency bands simultaneously \cite{Mejias2016:FeedforwardAndFeedback}.

As discussed further in Sec. SI, the non-specific background input $P_{E}$ is the typical control parameter used to tune the behavior of an isolated WC unit. In particular, at low values of $P_{E}$, a single WC unit flows towards a low-activity steady-state (Fig. S1A), and at high values of $P_{E}$, the system reaches a stable high-activity steady-state (Fig. S1C). At intermediate values of the excitatory drive, an isolated unit -- with the parameters given in Table ~\ref{t:WC_params} --  will undergo a bifurcation and exhibit rhythmic activity in the gamma frequency band (Fig. S1B). Up to a point, increasing $P_{E}$ within this intermediate region leads to oscillations with increasing amplitude and frequency (Fig. S1D--F).

The situation becomes more complex when multiple WC units are coupled via the structural connectome. In this scenario, an individual region's dynamics are determined by a combination of the constant background drive $P_{E}$ and the strengths of delayed inputs from other parts of the network, which are modulated by the coupling $C$ and the structural connectivity $\mathbf{A}$. To account for these two influences, we consider both $P_{E}$ and $C$ as tuning parameters, and examine working points of the model at which the combination of $P_{E}$ and $C$ generate oscillatory activity in individual brain areas. Finally, we set the signal propagation speed to a fixed value of $v = 10m/s$, which is in the range of empirical observations and previous large-scale modeling efforts \cite{Muldoon2016:StimulationBased}.    

\begin{table}[h!]
	\caption{Parameter values for the large-scale Wilson-Cowan neural mass model and for the numerical simulations.} 
	\begin{center}
		\begin{tabular}{ c c c } 
			\hline
			Parameter & Description & Value \\ 
			\hline
			\hline
			$v$ 						& propagation speed		& 		10m/s 											\\
			$C$							& global coupling strength					&		0--5					    				\\
			$\tau_{E}$				& excitatory time constant						&		2.5ms 											\\
			$\tau_{I}$				& inhibitory time constant						&		3.75ms 											\\	   				
			$a_{E}$					& excitatory gain						&		1.5 											\\
			$a_{I}$					& inhibitory gain						&		1.5 											\\
			$\mu_{E}$			& excitatory firing threshold					    	&		3.0 											\\
			$\mu_{I}$			& inhibitory firing threshold							&		3.0 											\\
			$c_{EE}$			& local E to E coupling 							&		16 												\\
			$c_{IE}$			& local I to E coupling						&		12 												\\
			$c_{EI}$			& local E to I coupling								&		15 												\\
			$c_{II}$			 & local I to I coupling								&		3 												\\
			$P^{\mathrm{base}}_{E}$	& baseline excitatory background drive						&		0.45--0.9	   									\\
			$\Delta P_{E}$ 	& perturbation strength		    					&  	    0.1 									    \\
			$P_{I}$				& inhibitory background drive							&		0 									            \\
			$\sigma_{E}$	& excitatory noise strength								&		$5\times 10^{-5}$										\\
			$\sigma_{I}$	& inhibitory noise strength							&		$5\times 10^{-5}$										\\
			$dt$				& integration time step							&		$7\times10^{-5}$s 											\\
			$dt_{\mathrm{ds}}$	& downsampled time step							&		$1\times 10^{-3}$s 											\\ 
			\hline
			\hline
		\end{tabular}
	\end{center}
	\label{t:WC_params}
\end{table}

\subsubsection{Incorporating the effects of local excitations on network activity}
\label{s:model_perturbation}

We wish to investigate how local perturbations of population activity affect brain-wide dynamics. In particular, we will examine the effects of increased excitation to a single brain area. Such heightened regional activation is incorporated into the model by increasing the level of drive to the excitatory subpopulation of the perturbed neural mass unit, while leaving the excitatory input at the original baseline value for all other brain areas \cite{Muldoon2016:StimulationBased}. We note that, phenomenologically, excitation of a local brain region could occur through a number of mechanisms, including sensory input, brain stimulation, or, alternatively, due to internal processes that regulate inputs to or excitability levels of specific neuronal populations. The goal of this work is to study the effects of localized excitations generally, rather than to design a detailed model of a specific type of perturbation. 

In the model, a selective increase in drive to region $j$ is implemented by increasing the constant background excitation to region $j,$ $P_{E,j}$, by an amount $\Delta P_{E} > 0$ (see Fig.~\ref{f:methods}E for a schematic). Thus, we take $P_{E,j} \rightarrow P_{E,j} + \Delta P_{E}$, where $\Delta P_{E}$ denotes the strength of the activation. The baseline condition corresponds to the situation in which all brain areas receive the same level of  background input, such that $P_{E,i} = P_{E}^{\mathrm{base}}$ for all $i$. The dynamics of the system in the baseline state can then be compared to the case of specific additional drive to a single unit $j$, where we have $P_{E,j} = P_{E}^{\mathrm{base}} + \Delta P_{E}$ and $P_{E,i} = P_{E}^{\mathrm{base}}$ for all $i \neq j$. 
 
\subsection{Numerical methods and simulations}

The equations governing the time evolution of the excitatory and inhibitory sub-population activity of each brain area form a system of coupled stochastic, delayed differential equations. We numerically integrate this system of equations using the Euler-Mayurama method with a time step of $dt=7\times 10^{-5}$s. For the time delays, we round each $\tau_{ij}$ to the nearest multiple of the integration time step \textit{dt}, such that each delay corresponds to an integer number of time steps. For the initial conditions, we assume a constant history for each unit's activity of length equal to the longest delay in the system. After running a simulation, we discard the first $t_{\mathrm{burn}} = 1$ second so that our analysis is not biased by transients or the specific choice of initial conditions. Each time-series is then downsampled to a resolution of $dt_{\mathrm{ds}} = 1\times 10^{-3}$s. We often report quantities averaged over several realizations of the simulations with different instantiations of the initial conditions and noise. We specify when reported measures refer to ensemble averages, and when they depict data derived from a single simulation. The parameters for the numerical simulations are shown in Table~\ref{t:WC_params}.

\subsection{Power spectra}
\label{s:power_spectra}

Useful characteristics of the simulated brain activity are apparent in the frequency domain (see Fig.~\ref{f:methods}F). Here, we use Welch's method (as implemented in MATLAB R2019b) to estimate the power spectral density (psd) of the excitatory population activities. We use window sizes of 1 second with 99\% overlap, and subtract the mean of each time-series before computing the psd. 

\subsection{Quantifying interareal phase-locking}
\label{s:quantifying_coherence}

To quantify the strength of temporal coupling between different brain areas, we use the phase-locking value (PLV) \cite{Lachaux1999:MeasuringPhase}. This measure is commonly used to assess the level of coherence between pairs of oscillatory time-series in a given frequency band. Importantly, because the state variables in the WC model are real-valued signals corresponding to population firing rates, with possibly multiple spectral components, we compute PLVs for a given frequency band by \textit{(1)} filtering raw excitatory time-series within the specified frequency range, and \textit{(2)} extracting instantaneous phases for the given frequency band using the Hilbert transform (see Fig.~\ref{f:methods}F). In the following two sections, we first describe the estimation of instantaneous phases using the Hilbert transform, and we then describe the computation of the phase-locking value in more detail.

\subsubsection{Instantaneous phases from the Hilbert transform}
\label{s:hilbert_transform}

Given a real-valued signal $X(t)$, it is possible to define instantaneous phase and amplitude variables that describe the signal using the Hilbert transform. Importantly, although the Hilbert transform can theoretically be computed for an arbitrary signal $X(t)$, the instantaneous amplitude $A(t)$ and phase $\theta(t)$ are only physically meaningful for relatively narrowband signals \cite{Pikovsky:2003a}. It is therefore necessary to bandpass filter a signal before taking the Hilbert transform. Here, raw time-series were bandpass filtered in a frequency range $f_{o} \pm \Delta f$ Hz using a 6th-order Butterworth filter in the forward and backward directions. In the results section, we describe how $f_{o}$ and $\Delta f$ are determined during the presentation of various findings that depend on computing the Hilbert phase. Filtering was carried out in MATLAB using the `butter' and `filtfilt' functions. After filtering the simulated neural activity, the Hilbert transform was applied in order to extract instantaneous phases for the given frequency band. The Hilbert transform was implemented using the `hilbert' function in MATLAB. More details on the Hilbert transform can be found in Sec. SVIII.

\subsubsection{Functional connectivity from the phase-locking value}
\label{s:phase_locking_value}

The outputs of the filtering and Hilbert transform process described in the previous section are instantaneous phases $\theta_{i}(f_{o},t)$ derived from the excitatory activity $E_{i}(t)$ of each brain region $i$ at a given central frequency $f_{o}$ and time $t$ (Fig.~\ref{f:methods}F). Equipped with these phase variables, we can quantify functional interactions among brain areas in a given frequency band using measures of phase-locking. A frequently used measure of the coherence between distinct brain areas is the PLV. The PLV -- which we will denote symbolically as $\rho_{ij}$ -- between two phase time-series $\theta_{i}(t)$ and $\theta_{j}(t)$ is given by

\begin{equation}
\rho_{ij} = \Bigg | \frac{1}{T_s} \sum_{t=1}^{T_s} e^{i[\theta_{i}(t)-\theta_{j}(t)]} \Bigg |,
\label{eq:plv}
\end{equation}

\noindent where $T_s$ is the number of sample time points over which we wish to compute the phase locking. The PLV thus quantifies the consistency of the relation between two phases $\theta_{i}(t)$ and $\theta_{j}(t)$. If the phase difference $\Delta \theta_{ij}(t) = \theta_{i}(t) - \theta_{j}(t)$ is constant over a given time window, $\rho_{ij}$ will be equal to 1. On the other hand, if the phase differences $\Delta \theta_{ij}(t)$ are distributed uniformly across a given time segment, $\rho_{ij}$ will be approximately 0. In this study, we will use the PLV to quantify phase coherence between the simulated firing rate activity of pairs of brain regions within a particular frequency band (Fig.~\ref{f:methods}F).  

We would also like to ensure that the PLV reflects the consistency of phase relations that arise from interactions (direct or indirect), and not locking arising from the fact that two regions happen to have the same frequency, but, possibly, a different phase relation in every trial. We therefore concatenate phase time-series from different trials before computing the PLV \cite{Lowet2016:Quantifying}, where each trial is a simulation run with a different random initial condition and different noise realization. In this way, a high PLV between two brain regions indicates that, across time \textit{and} trials, the activity of the two regions exhibits a consistent phase relationship within a particular frequency band.

As with structural connectivity, it is useful to think of a given $N \times N$ matrix of PLV values as a network where the element (edge) $\rho_{ij}$ is the phase coherence between region (node) $i$ and region (node) $j$. In contrast to the structural network, this network represents the presence of functional associations (the PLV) between brain regions' activity. Following common terminology, we will thus often refer to phase-locking as ``functional connectivity'', and we will refer to phase-locking matrices as ``functional connectivity matrices'' or ``functional networks''.

\subsubsection{Statistical analyses}
\label{s:statistical_analyses}

All data and statistical analysis was performed in MATLAB release R2019a. Statistical dependencies between two variables were assessed via the Spearman rank correlation, using the built-in MATLAB function `corr'. Throughout the text, we denote the Spearman correlation coefficient as $r_{s}$. Rank correlations are considered statistically different from zero if the corresponding $p$-value is less than 0.05.

\section{Results}

\subsection{Baseline dynamical regimes of the brain network model}
\label{s:baseline_state}

As described in Sec.~\ref{s:model}, we model whole-brain dynamics as a system of coupled WC units, where the coupling topology is informed by empirical measurements of anatomical brain connectivity. Depending on the values of various parameters, the brain network model can exhibit different qualitative behaviors. Importantly, different baseline states may in turn result in distinct modulations of brain-wide activity patterns in response to local perturbations. Thus, in this section, we first characterize the behavior of the system for the baseline condition in which all areas receive the same amount of excitatory drive. This initial study will allow us to focus our subsequent investigations on relevant parts of the parameter space, and to study how the effects of excitations depend upon the system's baseline state.

We focus on two parameters of interest: \textit{(1)} the level of generic background input received by the excitatory populations of all brain areas $P_{E}^{\mathrm{base}}$ and \textit{(2)} the global coupling strength $C$. Recall that for an isolated WC unit, $P_{E}^{\mathrm{base}}$ is a control parameter that induces a bifurcation \cite{Wilson1972:ExcitatoryandInhibitory,Hoppensteadt2012:Weakly} in the firing rate activity. However, when examining a network of coupled neural masses, the dynamics of each component are also dictated by inputs from other units in the system. The parameter $C$ is a second control parameter that globally scales the interaction strength between pairs of brain areas by tuning how much input a given region receives from its neighbors in the network. This coupling can hence be thought of as modulating the overall influence of network activity on regional dynamics. The nature of both local and network-wide dynamical behaviors will thus change depending on the combination of $P_{\mathrm{E}}^{\mathrm{base}}$ and $C$, allowing the system to exist in markedly different states.

To quantify how model behavior varies as a function of $P_{E}^{\mathrm{base}}$ and $C$, we perform a sweep over a broad range of these two quantities, considering values of $P_{E}^{\mathrm{base}} \in [0.45,0.9]$ in steps of $\Delta P_{E}^{\mathrm{base}} = 0.05$, and values of $C \in [0,5]$ in steps of $\Delta C = 0.1$. These parameter ranges were chosen to allow for the exploration of multiple dynamical regimes of the system while ensuring that the model exhibits oscillatory dynamics. For each parameter combination, we run five 2-second-long simulations. The values of all other parameters are defined in Table \ref{t:WC_params}, with the exception that, for these phase-space analyses, we run noiseless simulations in order to more precisely demarcate the boundaries between different dynamical regimes of the model.

\subsubsection{Long-range coupling strength and background drive tune baseline dynamical state} 

\begin{figure*}
	\centering
	\includegraphics[width=0.95\textwidth]{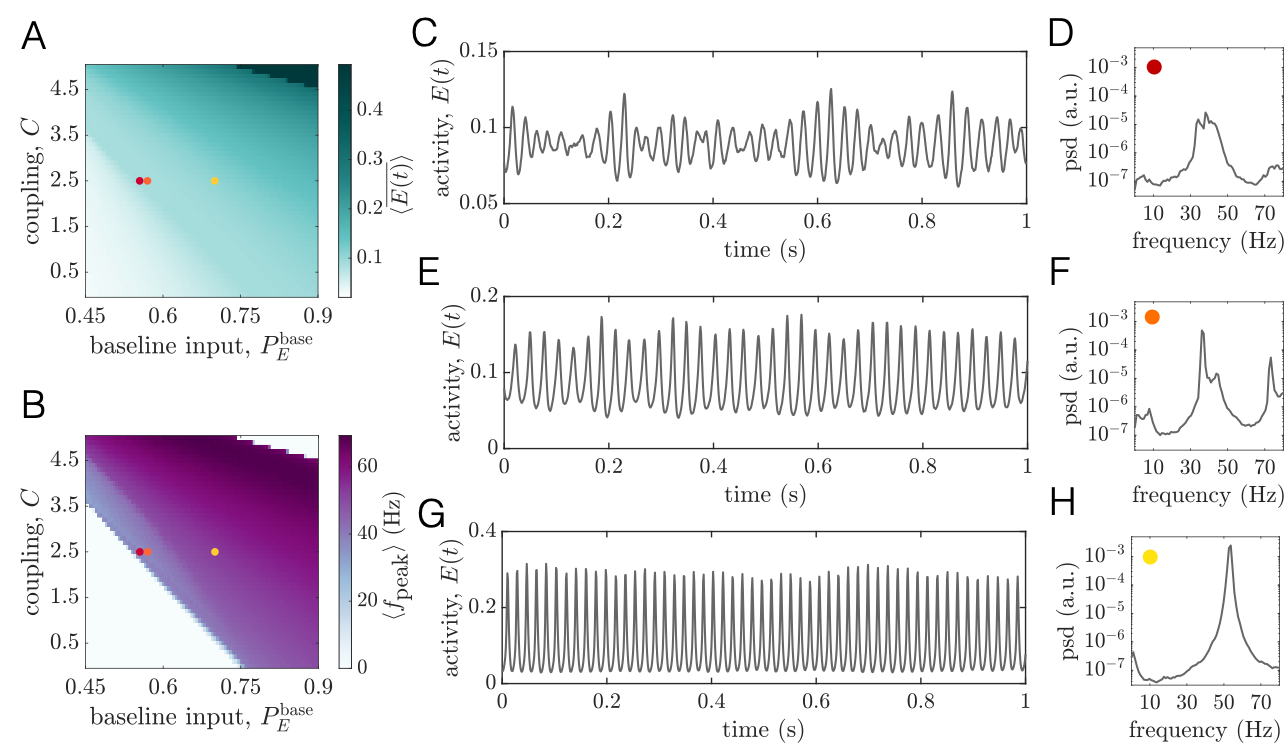}
	\caption{\textbf{Long-range coupling strength $C$ and background drive $P_{E}^{\mathrm{base}}$ modulate regional firing rates and oscillation frequencies at baseline}. \textit{(A)} The time- and network-averaged population firing rate $\langle \overline{E(t)} \rangle$ as a function of $C$ and $P_{E}^{\mathrm{base}}$ (units are arbitrary). \textit{(B)} The network-averaged peak frequency of regional activity $\langle f_{\mathrm{peak}}\rangle$ as a function of $C$ and $P_{E}^{\mathrm{base}}$. \textit{(C)} A segment of the activity of one brain area and \textit{(D)} the corresponding power spectra of the same area at the baseline working point denoted by the red dot in panels \textit{A} and \textit{B} ($P_{E}^{\mathrm{base}} = 0.553$, $C = 2.5$).  \textit{(E)} A segment of the activity of one brain area and \textit{(F)} the corresponding power spectra of the same area at the baseline working point denoted by the orange dot in panels \textit{A} and \textit{B} ($P_{E}^{\mathrm{base}} = 0.57$, $C = 2.5$). \textit{(G)} A segment of the activity of one brain area and \textit{(H)} the corresponding power spectra of the same area at the baseline working point denoted by the yellow dot in panels \textit{A} and \textit{B} ($P_{E}^{\mathrm{base}} = 0.7$, $C = 2.5$).}
	\label{f:baseline_state1}
\end{figure*}

We begin by computing two measures that quantify regional dynamics: \textit{(1)} the time-averaged firing rate of a region $\overline{E(t)}$, and \textit{(2)} the frequency at maximum power (peak frequency) of a region $f_{\mathrm{peak}}$. To obtain summary measures characterizing the state of the system as a whole, we compute network-averages of these quantities, denoted by angled-brackets. In studying $\langle \overline{E(t)} \rangle$ as a function of $P_{E}^{\mathrm{base}}$ and $C$, we observe three principal regimes (Fig.~\ref{f:baseline_state1}A). When both $P_{E}^{\mathrm{base}}$ and $C$ are low, the system settles to a state of low average firing rate (white region); this state corresponds to a non-oscillatory, low-activity equilibrium. In contrast, when $P_{E}^{\mathrm{base}}$ and $C$ are both high, the average firing rate saturates at a high level (dark green region); this state corresponds to a non-oscillatory, high-activity equilibrium. Finally, at intermediate values of these parameters, the mean firing rate varies between the low and high extremes, and the regional activity is oscillatory; because we wish to consider the rhythmic nature of neuronal population activity, this is the relevant portion of parameter space.

Next we seek to understand how $\langle f_{\mathrm{peak}} \rangle$ varies in the $P_{E}^{\mathrm{base}}$ -- $C$ plane (Fig.~\ref{f:baseline_state1}B). A clear wedge-shaped area marks parameter combinations that give rise to network-averaged peak frequencies in the gamma range (i.e., the frequencies range from approximately 30 -- 65 Hz in the colored region). As with the firing rate, the main oscillation frequency tends to increase (decrease) with either increasing (decreasing) background excitation or coupling strength within this parameter regime. By comparing Fig.~\ref{f:baseline_state1}B to Fig.~\ref{f:baseline_state1}A, we see that the white areas surrounding the purple wedge correspond to the regions of phase space where the firing saturates at a fixed low or high value. In Sec. SII, we describe a systematic method for determining boundaries in the 2D space spanned by $C$ and $P_{\mathrm{E}^{\mathrm{base}}}$ that indicate the onset or disappearance of oscillatory activity (see Fig. S2). In what follows, we use $P_{\mathrm{E}}^{\mathrm{*}}(C)$ to denote the level of background drive at which oscillations begin to emerge for a fixed coupling strength $C$.

To provide further intuition for how dynamics vary within parameter space, we study example time-series and power spectra for three different baseline states (colored dots in Figs.~\ref{f:baseline_state1}A,B). Note that these working points correspond to an intermediate coupling value of $C=2.5$, but varying levels of input $P_{E}^{\mathrm{base}}$. We begin with the working point $P_{E}^{\mathrm{base}} = 0.553$, which sits just beyond the boundary indicating the transition to sustained rhythmic activity. From the time-series, we observe that the firing rate is oscillating (Fig.~\ref{f:baseline_state1}C), and from the spectra, we observe a peak frequency of $\approx$40Hz on a broadband background (Fig.~\ref{f:baseline_state1}D). We next consider the working point $P_{\mathrm{E}}^{\mathrm{base}} = 0.57$. In this state, each unit receives a slightly higher level of excitatory drive, leading to higher-amplitude oscillations (Fig.~\ref{f:baseline_state1}E,F). However, although peak spectral power increases, complex amplitude modulations can still be seen in the corresponding time-series (Fig.~\ref{f:baseline_state1}D). Finally, we consider the working point $P_{\mathrm{E}}^{\mathrm{base}} = 0.7$. Here, the activity is characterized by regular, high-amplitude oscillations (Fig.~\ref{f:baseline_state1}G). Furthermore, inspection of the power spectra indicates a single, narrow peak at a slightly higher frequency than the previous working point (Fig.~\ref{f:baseline_state1}H).

\subsubsection{Network-averaged phase coherence is non-monotonically modulated by coupling strength and background drive}
\label{s:baseline_PLVs}

\begin{figure*}
	\centering
	\includegraphics[width=0.95\textwidth]{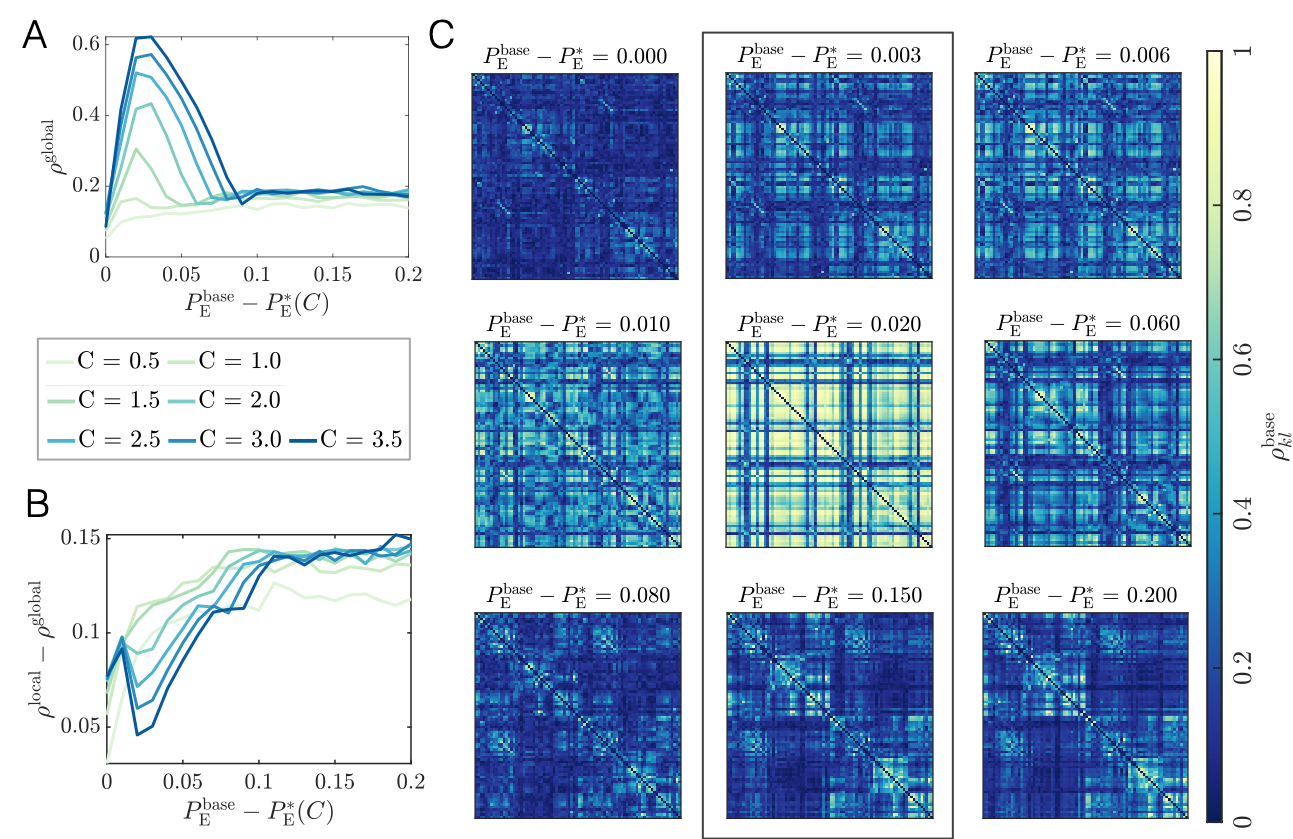}
	\caption{\textbf{Long-range coupling strength $C$ and background drive $P_{E}^{\mathrm{base}}$ modulate network phase-coherence at baseline}. \textit{(A)} The global order parameter $\rho^{\mathrm{global}}$ \textit{vs.} $P_{\mathrm{E}}^{\mathrm{base}} - P_{\mathrm{E}}^{\mathrm{*}}(C)$, for different fixed values of $C$. \textit{(B)} The difference between the global and local order parameters, $\rho^{\mathrm{global}} - \rho^{\mathrm{local}}$, \textit{vs.} $P_{\mathrm{E}}^{\mathrm{base}} - P_{\mathrm{E}}^{\mathrm{*}}(C)$, for different fixed values of $C$. \textit{(C)} Region-by-region PLV matrices for various values of $P_{\mathrm{E}}^{\mathrm{base}} - P_{\mathrm{E}}^{\mathrm{*}}(C)$ at fixed $C = 2.5$. The boxed matrices (from top to bottom) correspond to the red, orange, and yellow working points in Fig.~\ref{f:baseline_state1}.}
	\label{f:baseline_state2}
\end{figure*}

Both the firing rate and the power spectra are local measures that quantify the behavior of individual regions' dynamics. However, in network systems, it is also imperative to define order parameters that capture information about interactions between components or the amount of order in the network. Indeed, for systems composed of coupled units, the system ``state" is defined not only by the dynamical behavior of individual units, but also by how those dynamics are interrelated between units. Here, we are interested in the degree to which regional dynamics are coherent, which we quantify via the PLV between regions' firing rates (Sec.~\ref{s:phase_locking_value}). To compute the PLV between brain areas for baseline conditions, we first determined the peak frequency of each unit at a given set of parameters. We then filtered regional activity in a frequency band spanning 10Hz above the maximum peak frequency and 10Hz below the minimum peak frequency across all brain areas, and extracted Hilbert phases from the filtered signals (see Sec.~\ref{s:quantifying_coherence} for details). Finally, PLVs between all pairs of brain areas were computed as described in Sec.~\ref{s:phase_locking_value} and Eq.~\ref{eq:plv}, using 50 different simulations (trials) of 5 seconds each (with noise included).

To summarize how the overall level of coherence in the network varies as a function of the background drive and coupling strength, we defined a macroscopic phase-locking order parameter: the average of the pairwise PLVs across all pairs of units in the network, 
\begin{equation}
\rho^{\mathrm{global}} = \langle \rho_{ij} \rangle.
\label{eq:global_order_parameter}
\end{equation} 
\noindent This quantity ranges between 0 and 1, where large values indicate a more dynamically ordered state and where low values indicate a more incoherent state of network activity. In general, we find that the level of background input and the coupling strength interdependently tune the global level of coherence in the system (Fig.~\ref{f:baseline_state2}A). At low coupling strengths, brain areas cannot coordinate their dynamics and $\rho^{\mathrm{global}}$ remains at a relatively low value for a range of excitatory drives. Hence, for very weak coupling, system dynamics are in a relatively disordered state. In contrast, as the coupling is increased, we begin to see a qualitative change in behavior. For higher values of $C$, we observe that $\rho^{\mathrm{global}}$ varies non-monotonically as a function of the background drive: $\rho^{\mathrm{global}}$ first increases and then decreases as a function of $P_{\mathrm{E}}^{\mathrm{base}}$. For a given coupling $C$, there appears to be a ``critical" value at relatively small but non-zero $P_{E}^{\mathrm{base}} - P_{E}^{\mathrm{*}}(C)$ where the system develops a well-defined peak in global coherence. As the background drive is increased further, the global order parameter begins to decrease and then eventually plateaus, albeit with some fluctuations. More specifically, at levels of background drive well beyond the ``critical state'' of peak coherence, $\rho^{\mathrm{global}}$ relaxes to an intermediate value between its peak value and its value at the lowest background input. In this regime, the system resides in a state of partial order. Increasing the interareal coupling strength has the effect of amplifying the peak value of $\rho^{\mathrm{global}}$ (although $\rho^{\mathrm{global}}$ remains below 1 for all couplings considered), but does not appear to affect the global phase-locking level to the right side of the peak.

To provide further intuition for this behavior, we focus on an intermediate coupling of $C = 2.5$, and calculate the pairwise coherence patterns $\rho_{ij}$ for several values of the background drive (Fig.~\ref{f:baseline_state2}C). At the lowest level of excitatory input ($P_{\mathrm{E}}^{\mathrm{base}} - P_{\mathrm{E}}^{\mathrm{*}}(C) = 0$), some organization can be seen in the coherence matrix, but the system is very weakly phase-locked overall. In this state, units exhibit relatively low amplitude oscillations, and are therefore more influenced by noise. It is thus reasonable to expect relatively low levels of global phase-locking at low background drive. However, with only a small increase in the level of background drive (e.g., $P_{\mathrm{E}}^{\mathrm{base}} - P_{\mathrm{E}}^{\mathrm{*}}(C) = 0.003$), we observe distributed increases in the coherence values and a large spread of high, medium, and low coherence pairs dispersed throughout the network. Increasing the background drive slightly more (e.g., $P_{\mathrm{E}}^{\mathrm{base}} - P_{\mathrm{E}}^{\mathrm{*}}(C) = 0.02$) leads to the emergence of large, highly-coherent blocks that span the network. This input level corresponds to the peak in $\rho^{\mathrm{global}}$ and represents a highly ordered state of the system. As $P_{\mathrm{E}}^{\mathrm{base}}$ is increased further, though, phase-locking begins to decrease widely throughout the network and the coherence pattern markedly changes. In particular, for high $P_{\mathrm{E}}^{\mathrm{base}} - P_{\mathrm{E}}^{\mathrm{*}}(C)$, we observe the emergence of localized groups of intermediate-to-high coherence (Fig.~\ref{f:baseline_state2}C, Row 3). For these high input levels, the phase-locking matrix exhibits distinct block-like architecture that more closely resembles the structural connectivity network. To understand this shift in the dynamic state of the system, it is important to note that the increased amplitude of local excitatory activity due to increasing $P_{\mathrm{E}}^{\mathrm{base}}$ alters the extent to which units' dynamics are locally generated rather than induced by network coupling. In particular, as $P_{\mathrm{E}}^{\mathrm{base}}$ increases, regions are pushed further towards their individual limit cycles and dynamics become more locally driven. The increased influence of local dynamics seems to eventually hinder the ability of individual units to adjust their rhythms and exhibit consistent phase relationships. Note that phase-locking is also made especially difficult by the large variance in the distribution of interareal delays imposed by the connectome's spatial embedding, and indeed, for high background drive states, it appears that more strongly connected and spatially nearby units are able to maintain stronger coherence.

In general, our observations point to complex behavior in which the macroscopic order parameter (global phase coherence) varies non-monotonically as a function of the background drive and network coupling strength (Fig.~\ref{f:baseline_state2}A,C). Therefore, a variety of qualitatively different regimes exist, beyond just a simple binary separation between an asynchronized state and a synchronized state. It is also important to note that, while a similar level of $\rho^{\mathrm{global}}$ can occur for input strengths below and above the coherence peak, these two states of partial coherence are qualitatively different from one another (compare the first and third rows of Fig.~\ref{f:baseline_state2}C). To more quantitatively distinguish network states before and after the point of peak coherence, we also considered a local order parameter, $\rho^{\mathrm{local}}$:
\begin{equation}
\rho^{\mathrm{local}} = \frac{\sum_{i,j=1}^{N}\mathrm{A}_{ij}\rho_{ij}}{\sum_{i,j=1}^{N}\mathrm{A}_{ij}}.
\label{eq:local_order_parameter}
\end{equation}
\noindent From Eq.~\ref{eq:local_order_parameter}, we see that $\rho^{\mathrm{local}}$ is a weighted average of $\rho_{ij}$, with weights equal to the strength of structural brain network connections. In this way, $\rho^{\mathrm{local}}$ will be larger when more strongly structurally connected brain areas are more phase-locked. In Fig.~\ref{f:baseline_state2}B, we show $\rho^{\mathrm{local}} -\rho^{\mathrm{global}}$ \textit{vs.} $P_{\mathrm{E}}^{\mathrm{base}} - P_{\mathrm{E}}^{\mathrm{*}}(C)$ for different values of the coupling strength $C$. These plots exhibit a clear upward trend, indicating that with increasing background input, the amount of local coherence increases relative to the amount of global coherence. Hence, the macroscopic state of the system becomes increasingly constrained by structure as local excitatory drive increases. Thus, even though the global phase-locking value can be similar on either side of peak $\rho^{\mathrm{global}}$, the system is in qualitatively different dynamical modes.

It is crucial to remark that the behavior we observe here contrasts starkly with the behavior that occurs in simpler phase-oscillator models, where synchronization typically increases monotonically with coupling strength. A critical difference between phase-based models and the more realistic Wilson-Cowan system considered here is that, for the latter case, unit dynamics are described and coupled not by a phase variable, but rather by an oscillatory, real-valued signal that represents fluctuating population firing rates. Hence, changes in the amplitude or stability of a region's local dynamics can affect the dynamical state of the system as a whole. Indeed, the preceding analyses show that modulations of the network interaction strength, as well as the level of background input, push the system into very different dynamical regimes. Crucially, this capacity of the model will allow us to examine how the effects of local changes in neural activity depend not only on which area in the network is excited, but also on the baseline working point of the system as a whole, as determined by the level of background drive to the network. 

\subsection{Effects of regional excitations on spectra and phase-locking} 
\label{s:stimulation_effects}

Having characterized the brain network model under baseline conditions, we now investigate how local enhancements of neural activity modulate brain network dynamics. Because we are interested in how the macroscopic state of the network may influence the impact of a localized perturbation, we study three qualitatively distinct dynamical regimes, and examine in detail the effects of local excitations for each case. In particular, we focus on a fixed intermediate coupling strength $C = 2.5$ for which the system exhibits a clear peak in $\rho^{\mathrm{global}}$ (Fig.~\ref{f:baseline_state2}A), and we select three different values of the background input $P_{\mathrm{E}}^{\mathrm{base}}$ that place the system either in a state preceding, at, or following the global coherence peak. For each working point (WP1, WP2, and WP3), we examine how regional perturbations of a fixed strength affect the spectra of individual regions and influence interareal phase-locking. Note that stimulation of a single brain region $j$ is always introduced by increasing the excitatory input to region $j$ by an amount $\Delta P_{\mathrm{E,}j} = 0.1$, while keeping all other regions at their working-point-specific baseline drive (see Sec.~\ref{s:model_perturbation} for details). We also remark that our goal is not to exhaustively analyze all possible working points, but rather to demonstrate that different dynamical regimes can produce qualitatively distinct behaviors. Finally, we note that in Sec. SV, we verify that results hold for different values of $P_{\mathrm{E}}^{\mathrm{base}}$ in the vicinity of those studied in the main text, and in Sec. SVII, we examine an alternative value of the global coupling ($C = 2$).

\subsection{Working point 1: Pre-global coherence peak}
\label{s:wp1}

We begin by considering a working point (WP1) located at $C = 2.5$ and $P_{\mathrm{E}}^{\mathrm{base}} = 0.553$, below the point of peak coherence (Fig.~\ref{f:baseline_state2}C, Row 1, Column 2). In this state, the system is perched just past the boundary marking the transition between the quiescent state of low activity and the commencement of rhythmic dynamics. Hence, regional activity is oscillatory but of relatively low amplitude (see Fig.~\ref{f:baseline_state1}C), and the power spectra is broad (see Fig.~\ref{f:baseline_state1}D). 

\subsubsection{Local excitations induce distinct modifications to power spectra}
\label{s:powSpec_wp1}

\begin{figure*}
	\centering
	\includegraphics[width=\textwidth]{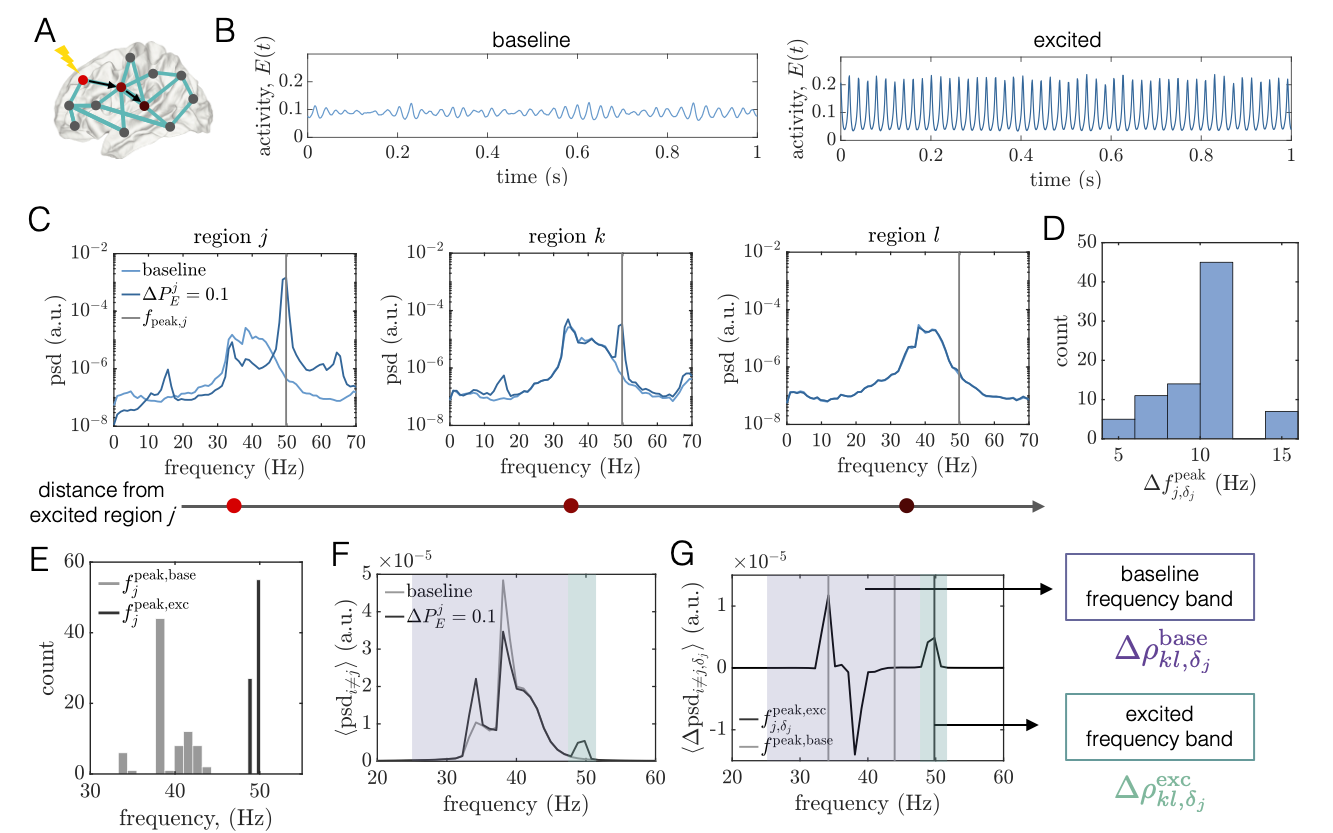}
	\caption{\textbf{Regional excitation causes local and downstream changes to brain areas' power spectra in different frequency bands at WP1.} \textit{(A)} Schematic of a brain network depicting the site $j$ of selective excitation in brightest red. The black arrows point to two other regions $k$ and $l$ that lie progressively further away from the perturbed area. Note that in this figure, regions $j$, $k$, and $l$ correspond to brain areas 1 (R--Lateral Orbitofrontal), 4 (R--Medial Orbitofrontal), and 10 (R--Precentral), respectively. \textit{(B)} Left: A segment of region $j$'s activity time-course in the baseline condition. Right: A segment of region $j$'s activity time-course when it is perturbed with additional excitatory input $\Delta P_{\mathrm{E},j}$. \textit{(C)} Power spectra of the excited area $j$ and two other downstream regions $k$ and $l$. In all three panels, the lighter curves correspond to the baseline condition, and the darker curves correspond to the state in which $j$ is driven with increased current. The gray vertical lines indicate the peak frequency of region $j$, $f_{\mathrm{peak},j}$, in the excited condition. \textit{(D)} Histogram of the shift in peak frequency $\Delta f^{\mathrm{peak}}_{j,\delta_j}$ induced by exciting unit $j$, plotted over all choices of the perturbed area. \textit{(E)} Distribution of peak frequencies of all units in the baseline condition (light gray) and distribution of the peak frequency units acquire when directly excited (dark gray). \textit{(F)} Average power spectra $\langle \mathrm{psd}_{i} \rangle$ over all units $i\neq j$ at baseline (light gray) and when unit $j$ is perturbed with additional input (dark gray).  \textit{(G)} The average difference $\langle {\Delta} \mathrm{psd}_{i\delta_j} \rangle$ of the spectra of unit $i\neq j$ when unit $j$ is excited and in the baseline condition, where the average is over all units $i\neq j$.  For reference, the two light gray vertical lines denote the minimum and maximum peak frequency of units in the baseline state, and the dark gray line indicates the peak frequency acquired by the excited region $j$. Shaded boxes denote two frequency bands of interest: \textit{(1)} the \textit{baseline} band (purple) consisting of the main oscillation frequencies of brain areas under baseline conditions, and \textit{(2)} the \textit{excited} band (green) centered around the peak frequency that perturbed regions inherit under increased excitatory drive. Note that $\langle {\Delta} \mathrm{psd}_{i\delta_j} \rangle$ exhibits modulations in both bands. In subsequent analyses, we assess perturbation-induced changes in functional interactions between brain areas by computing the difference in the PLV between the baseline state and the state when unit $j$ is excited. These modulations are computed separately for the baseline frequency band, $\Delta \rho^{\mathrm{base}}_{kl,\delta_j}$ (purple), and for the excited band $\Delta \rho^{\mathrm{exc}}_{kl,\delta_j}$ (green).}
	\label{f:ts_ps_perturb}
\end{figure*}

To begin, we examine the effects of a localized increase in drive to a single brain area on regions' time series and power spectra (Fig.~\ref{f:ts_ps_perturb}A-C). In agreement with a number of past experimental and modeling studies \cite{Lowet2015:InputDependentFrequency,Mejias2016:FeedforwardAndFeedback,Henrie2005:LFPPowerSpectra,Jia2013:NoConsistent}, increased drive to the excitatory pool of region $j$ increases the firing rate and frequency of oscillations in the activated region (Fig.~\ref{f:ts_ps_perturb}C, Left). In particular, stimulation causes an increase in the power, narrowing of the spectra (associated with an increase in periodicity of regional activity), and a shift of the peak frequency from $\approx$ 40Hz at baseline to $\approx$ 50Hz when excited. We also note the appearance of modulation sidebands \cite{Lowet2016:Quantifying} in the excited spectra to the left and right of the peak frequency. The left sideband occurs near the main frequencies of the baseline spectral peak ($\approx$ 34Hz) and the right sideband is nearly a mirror image. These sidebands arise due to the modulation of the excited region's time-series by the lower-frequency input it receives from other areas in the network \cite{Lowet2016:Quantifying}. This modulation also results in a spectral peak at $\approx$ 16Hz -- which is the difference between the new, excited frequency and the sideband peaks, and is a marker of quasiperiodic amplitude modulation in the time-series. To more carefully quantify the effects of an excitation to region $j$, we consider the shift in the peak frequency of unit $j$, $\Delta f^{\mathrm{peak}}_{j,\delta_j}= f^{\mathrm{peak}}_{j,\delta_j} - f^{\mathrm{peak}}_{j,o}$, between its baseline and excited states (Fig.~\ref{f:ts_ps_perturb}D). We calculate these differences for all choices of the activated brain area, and find that they range from about 6Hz to 16Hz, with an average value of $\langle \Delta f^{\mathrm{peak}}_{j,\delta_j} \rangle = \langle f_{\mathrm{peak},j}^{\delta_j} - f_{\mathrm{peak},j}^{o} \rangle \approx 10$Hz. We also find that these perturbation-induced shifts in the peak frequency yield excited frequencies that are well-separated from the range of peak frequencies in the baseline state (Fig.~\ref{f:ts_ps_perturb}E). 

We next consider the power spectra of two other units $k$ and $l$ located at increasing distances from the excited region (Fig.~\ref{f:ts_ps_perturb}C, Middle, Right). We observe that unit $k$ maintains its initial frequency content, but also develops new clearly defined peaks centered at the frequency of the excited region and at the difference of the excited frequency and the baseline peak. In contrast, the spectra of unit $l$ -- which is further away from the excited site -- is relatively unchanged. Hence, depending on the network coupling, increasing the drive to region $j$ can also modulate the spectra of a downstream region. To summarize more generally how the spectra of downstream brain areas are altered by driving area $j$ with additional input, we examine the average power spectral density $\langle \mathrm{psd}_{i} \rangle$ over all units $i\neq j$ both at baseline and when unit $j$ is excited (Fig.~\ref{f:ts_ps_perturb}F). At baseline, the network-averaged spectra at this working point is relatively broad and complex. It contains multiple peaks -- a main one at 38Hz and two smaller peaks around 34Hz and 41Hz. In addition, a local excitation in the network produces complex and broadband alterations in power, as expected in a scenario of quasiperiodic entrainment between nonlinear oscillators \cite{Schuster2006:Deterministic}. For this example, we observe the appearance of an entirely new peak at 50Hz, but also an enhancement of the lowest baseline peak and a depression of the highest baseline peak. These modulations are perhaps more apparent in Fig.~\ref{f:ts_ps_perturb}G, which shows the average difference $\langle {\Delta} \mathrm{psd}_{i,\delta_j} \rangle$ of the spectra of unit $i\neq j$ between when unit $j$ is excited and the baseline condition, where the average is over all units $i\neq j$. In sum, we see that a regional enhancement of neural activity causes non-local modulations in power both at the frequency of the directly excited brain area, as well at the system's baseline oscillation frequencies. These analyses suggest that there are two relevant frequency bands to consider for subsequent analysis: \textit{(1)} a relatively broad band containing the main frequencies of brain areas in the baseline state, and \textit{(2)} a band centered around the peak frequency of the excited unit. In what follows, we will denote these two bands as ``baseline" and ``excited", and consider the change in phase-locking, $\Delta \rho^{\mathrm{base}}_{kl,\delta_j}$ and $\Delta \rho^{\mathrm{exc}}_{kl,\delta_j}$, in each band due to a local perturbation.

\subsubsection{Excitations of regional activity induce or alter interareal phase-locking in excited and baseline frequency bands}
\label{s:PLV_wp1}

We are now prepared to study how local excitations of a single brain area alter the coordination of network-wide dynamics. To do so, we examine changes in interregional phase-locking. We separate our analysis into two frequency bands -- baseline and excited -- by filtering the activity time-series in each frequency band, extracting Hilbert phases from the filtered signals, and then calculating the PLV for each pair of regions within each band (see Fig.~\ref{f:methods}E). Details regarding the filtering process, the extraction of phases via the Hilbert Transform, and the calculation of the PLV can be found in Secs.~\ref{s:hilbert_transform} and ~\ref{s:phase_locking_value}. Since spectra are relatively broad at baseline, the baseline frequency band is determined by finding the peak frequency of all units in the baseline state, and setting the band to encompass frequencies 10Hz below the smallest peak frequency and 10Hz above the largest peak frequency. For the much narrower excited band, we extracted the peak frequency of the activated region (which is well-defined), and set the band to range from 1.5Hz below to 1.5Hz above that frequency. This range was chosen so as to contain the majority of the excited band peak, while including as little of the original baseline band as possible. If the peak frequency of the stimulated area was not more than 3.5Hz above the largest baseline peak frequency, then we only examined phase-locking changes in the baseline frequency band. Our choices are motivated by the following observation: the notion of an excited frequency band is only meaningful when a perturbation introduces a new frequency into the system that is well separated from the frequencies present in the baseline condition. Also, because the excited frequency band is induced by the perturbation, we consider only increases in the PLV within the excited band. On the other hand, since units are already oscillating at frequencies in the baseline band prior to a perturbation, we consider the absolute change in the PLV within the baseline band.

\begin{figure*}
	\centering
	\includegraphics[width=0.9\textwidth]{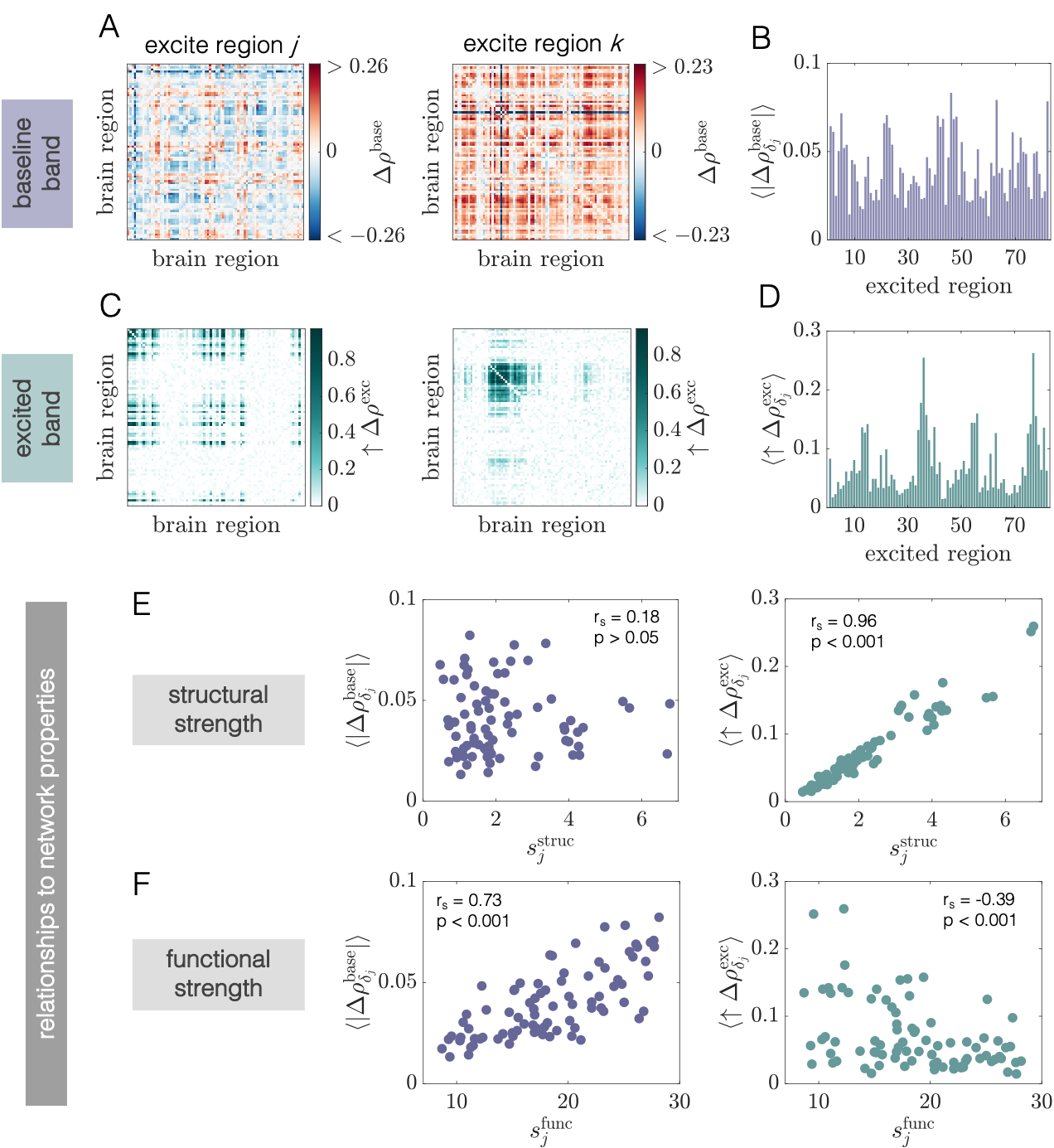}
	\caption{\textbf{Phase coherence modulations at WP1 are driven by local excitations of neural activity, differ between excited and baseline frequency bands, and are differentially related to structural and functional network properties.} \textit{(A)} Pairwise changes in the PLV inside the baseline band $ \Delta \rho^{\mathrm{base}}$ when region $j$ (Left) or region $k \neq j$ (Right) is perturbed. Note that in this figure, regions $j$ and $k$ correspond to regions 4 (R--Medial Orbitofrontal) and 23 (R--Lateral Occipital), respectively. \textit{(B)} Network-averaged absolute PLV changes in the baseline band $ \langle | \Delta \rho^{\mathrm{base}}_{\delta_j} | \rangle $ caused by excitation to different brain areas. \textit{(C)} Pairwise increases in PLV inside the excited band $ \Delta \uparrow \rho^{\mathrm{exc}} $ when region $j$ (Left) or region $k \neq j$ (Right) is perturbed. \textit{(D)} Network-averaged positive PLV changes in the excited band $ \langle \uparrow \Delta \rho^{\mathrm{exc}}_{\delta_j} \rangle $ caused by excitation to different brain areas. \textit{(E)} The quantity $\langle | \Delta \rho^{\mathrm{base}}_{\delta_j} | \rangle $ \textit{vs.} structural node strength $s^{\mathrm{struc}}_{j}$ (Left), and the quantity $\langle \uparrow \Delta \rho^{\mathrm{exc}}_{\delta_j} \rangle $ \textit{vs.} structural node strength $s^{\mathrm{struc}}_{j}$ (Right). \textit{(F)} The quantity $\langle | \Delta \rho^{\mathrm{base}}_{\delta_j} | \rangle $ \textit{vs.} functional node strength $s^{\mathrm{func}}_{j}$ (Left), and the quantity $\langle \uparrow \Delta \rho^{\mathrm{exc}}_{\delta_j} \rangle $ \textit{vs.} functional node strength $s^{\mathrm{func}}_{j}$ (Right). In panels \textit{(E)} and \textit{(F)}, insets indicate Spearman correlation coefficients between the plotted quantities, and their associated $p$-values).}
	\label{f:phaseCoh_modulations_wp1}
\end{figure*}

To provide intuition about how interareal phase-locking is altered upon a local excitation, we consider the effect of exciting two different brain areas (Fig.~\ref{f:phaseCoh_modulations_wp1}A,C). These examples first show that excitation of a local brain area can induce phase-locking at the excited frequency (Fig.~\ref{f:phaseCoh_modulations_wp1}C), but can also cause reconfigurations of coherence in the frequency band containing the original oscillatory activity of the system (Fig.~\ref{f:phaseCoh_modulations_wp1}A). Further, the patterns induced in the excited band are clearly distinct from the modulations in coherence that occur in the baseline band, for both choices of the stimulation region. Second, we note that the changes in baseline band phase-locking $\Delta \rho_{ij}^{\mathrm{base}}$ (Fig.~\ref{f:phaseCoh_modulations_wp1}A) and the increases of excited band phase-locking $\uparrow \Delta \rho_{ij}^{\mathrm{exc}}$ (Fig.~\ref{f:phaseCoh_modulations_wp1}C) are markedly different between stimulation of region $j$ (left panels) and stimulation of region $k \neq j$ (right panels). Thus, depending on the frequency band considered and on the excited area's location within the large-scale brain network, stimulation induces different modifications to phase-locking relationships across the system as a whole. 

To summarize the overall diversity of the network-wide response to a regional excitation, we calculate the average change in the PLV induced by driving each brain area with additional input. We use the notation $\langle | \Delta \rho_{\delta_j}^{\mathrm{base}} | \rangle$ and $\langle \uparrow \Delta \rho_{\delta_j}^{\mathrm{exc}} \rangle$ to denote the network-average of absolute changes in baseline band PLV and positive changes in excited band PLV induced by activating region $j$, respectively. Using these quantities, we observe that the global response to an excitation of a specific brain area exhibits a large degree of variability (Fig.~\ref{f:phaseCoh_modulations_wp1}B,D). Activating some areas induces much greater system-wide modulations of interareal phase locking than activating others. Furthermore, regions that cause the most noticeable enhancement of the PLV inside the excited frequency band are not necessarily those that cause the largest absolute alterations of the PLV in the baseline band. This observation suggests that two different mechanisms may be responsible for the distinct effects observed at the baseline and excited frequencies.

Another important consideration is how phase-locking changes are spread throughout the network. Are PLV modifications distributed locally -- concentrated near the excited brain region -- or are the changes dispersed more widely across the system? By quantifying the spatial extent of phase-coherence modulations (see Sec. SIII), we find that the phase-locking reconfigurations inside the baseline frequency band are distributed more widely than the phase-locking enhancements inside the excited band (Fig. S3A). This observation indicates that alterations to functional interactions induced in the baseline band tend to be more widely distributed, whereas alterations induced in the excited band are more focused near the site of the perturbation.

\subsubsection{Structural and functional connectivity are linked to different types of phase-locking modulations}
\label{s:struc_func_PLV_WP1}

The results presented in the previous section suggest that enhancements of neural activity in a localized brain area can cause distinct types of alterations to the temporal coordination between brain areas, and also suggest that different regions in a large-scale brain network may play variable roles in their ability to induce or modulate such interactions. These findings beg the question: What aspects of the system may drive or predict these diverse responses? In this section, we explore the extent to which properties of structural or baseline functional network connectivity are related to changes in phase-locking patterns induced by regional excitations.

We begin by considering relationships between the structural connectivity of a given brain area and the changes in PLV induced by regional excitation. Because the network of anatomical connections couples different brain areas and allows them to directly interact, it is reasonable to hypothesize that the organization of this network should play a role in guiding the influence of a perturbation. A common measure of a brain area's importance in an anatomical network is its structural strength $s_{j}^{\mathrm{struc}} = \sum_{i=1}^{N} A_{ij}$ \cite{Rubinov2010:Complex}, where larger strengths correspond to more strongly anatomically connected brain areas. We study $ \langle |\Delta \rho_{\delta_j}^{\mathrm{base}}| \rangle$ and $ \langle \uparrow \Delta \rho_{\delta_j}^{\mathrm{exc}} \rangle $ as functions of $s_{j}^{\mathrm{struc}}$ (Fig.~\ref{f:phaseCoh_modulations_wp1}E). Interestingly, at this working point we find that global changes in the PLV at frequencies corresponding to regions' naturally-emerging oscillations are \textit{not} strongly predicted by the structural strength of the brain area that receives additional drive (scatter plot of purple data points). In contrast, we do observe a strong relationship (Spearman correlation $r_{s} = 0.96$, $p<0.001$) between structural strength and the induced phase-locking in the excited band (scatter plot of green data points). This result indicates that more structurally connected units generate larger overall enhancements of phase-locking at the frequency of the directly stimulated area.

Although the organization of the brain's structural network is crucial in determining the evolution of macroscale brain dynamics, it is not the only governing factor. Intuitively, we may also expect functional interactions to play a role in guiding the effects of a perturbation, since the presence of a functional connection between two brain areas implies an interdependence of their dynamics -- enforced by the current collective oscillatory state -- that can occur even in the absence of a direct structural connection. Given this reasoning, we proceeded to examine relationships between units' baseline coherence and changes in phase-locking induced by local excitations. More specifically, we consider the baseline \textit{functional} strength of node $j$, defined as $s_{j}^{\mathrm{func}} = \sum_{i=1}^{N}\rho_{ij}$ (where $\rho_{ij}$ is phase-coherence between units $i$ and $j$ at baseline). This measure is equal to the total phase-coherence between region $j$ and all other brain areas, and quantifies how dynamically integrated a given region is to the network as a whole.

We study $ \langle |\Delta \rho_{\delta_j}^{\mathrm{base}}| \rangle$ and $ \langle \uparrow \Delta \rho_{\delta_j}^{\mathrm{exc}} \rangle $ as functions of $s_{j}^{\mathrm{func}}$ (Fig.~\ref{f:phaseCoh_modulations_wp1}F). Curiously, in contrast to structural strength, we observe a consistently stronger relationship between baseline band phase-coherence modulations and functional strength (Fig.~\ref{f:phaseCoh_modulations_wp1}F, Left). In particular, the average absolute change in phase-locking in the baseline frequency band, $ \langle | \Delta \rho_{\delta_j}^{\mathrm{base}}| \rangle $, is strongly positively correlated with functional node strength ($r_{s} = 0.73$, $p<0.001$). Thus, at this working point, perturbing more functionally interconnected regions drives larger absolute changes in interareal phase-locking in a frequency band containing the baseline oscillations. This result should be contrasted to the results from structural node strength, for which there were no strong relationships regarding the reconfiguration of coherence at the baseline oscillation frequencies. Finally, we consider $ \langle \uparrow \Delta \rho_{\delta_j}^{\mathrm{exc}} \rangle $ as a function of the functional strength $s_{j}^{\mathrm{func}}$ (Fig.~\ref{f:phaseCoh_modulations_wp1}F, Right), and find that these two quantities exhibit a weaker but still significant negative Spearman correlation ($r_{s} = -0.39$; $p<0.001$). This observation indicates that regions which make strong functional associations to the rest of the network actually tend to produce weaker enhancements of phase-locking in the excited band when driven with increased input. Put another way, regions that tend to have stronger coherence at baseline are less flexible in their ability to induce phase-locking at the new, higher frequency that they acquire when perturbed. 

The set of results discussed in this section highlights the importance of considering both structural network topology and the organization of functional interactions -- the latter being driven not only by structure, but also by the dynamical state of the system. For the model working point considered in this section, we find that structural and functional network properties are related to distinct types of perturbation-induced changes. These findings also make clear that two different mechanisms govern the PLV changes induced in the excited \textit{vs.} baseline frequency bands. The phase-locking that arises in the excited band reflects the transmission and replication of periodic input from the directly excited area to and in downstream regions. In particular, if the anatomical connection between the stimulated site and a downstream area is strong enough, then the stimulated area's activity will induce a new spectral component in the receiving area (see, e.g., Fig.~\ref{f:ts_ps_perturb}C), and subsequently, the two regions will exhibit phase-locking at the excited frequency. In addition, even two areas that are not directly linked can display a high PLV in the excited band due to strong common input from the stimulated region, or due to the propagation of the periodic stimulation signal along alternative paths in the network. In sum, because the spreading of the stimulated area's activity is highly constrained by the presence of structural connections, it is sensible that regions with stronger structural connectivity to other areas will more effectively drive downstream regions and lead to larger excited band effects. Perhaps more interesting are the reconfigurations of phase locking that occur at the spontaneous frequencies of the network. Modulations of phase coherence in the baseline frequency band arise not due to a direct transmission of input, but rather via a modification of the ongoing entrainment between units’ spontaneous rhythms. In turn, changes in the mutual entrainment between brain areas' activity induces complex, nonlinear reorganizations of coherent interactions. For WP1, these changes in functional coupling are more related to the stimulated region's coherence with the network at baseline, rather than its anatomical connectivity. Intuitively, this may in part be due to the fact that perturbing a particular area decouples it from other areas at the original oscillation frequency, such that stimulating regions that are strongly coherent to begin with leads to a reconfiguration of existing interactions in the baseline frequency band. 

\subsection{Working point 2: Global coherence peak}
\label{s:wp2}

In this section, we consider the effects of perturbations to neural activity at a second working point (WP2), where the baseline state of the system corresponds to the ``critical'' state at the peak of $\rho^{\mathrm{global}}$ (see Fig.~\ref{f:baseline_state2}A). More specifically, we use the same coupling $C = 2.5$ as in WP1 but consider a slightly higher level of background drive $P_{\mathrm{E}}^{\mathrm{base}} = 0.57$ (see Figs.~\ref{f:baseline_state1}E and F for example time-series and power spectra at WP2). At this working point, we observe blocks of relatively strong phase-locking in the interregional baseline PLV matrix (Fig.~\ref{f:baseline_state2}C, Row 2, Column 2). However, the network-averaged PLV is only $\approx$ 0.56, which is still significantly less than the maximum possible value of $\rho^{\mathrm{global}} = 1$. 

\subsubsection{Spectral modifications in baseline and excited bands persist at state of peak coherence}
\label{s:powSpec_critical}

\begin{figure*}
	\centering
	\includegraphics[width=\textwidth]{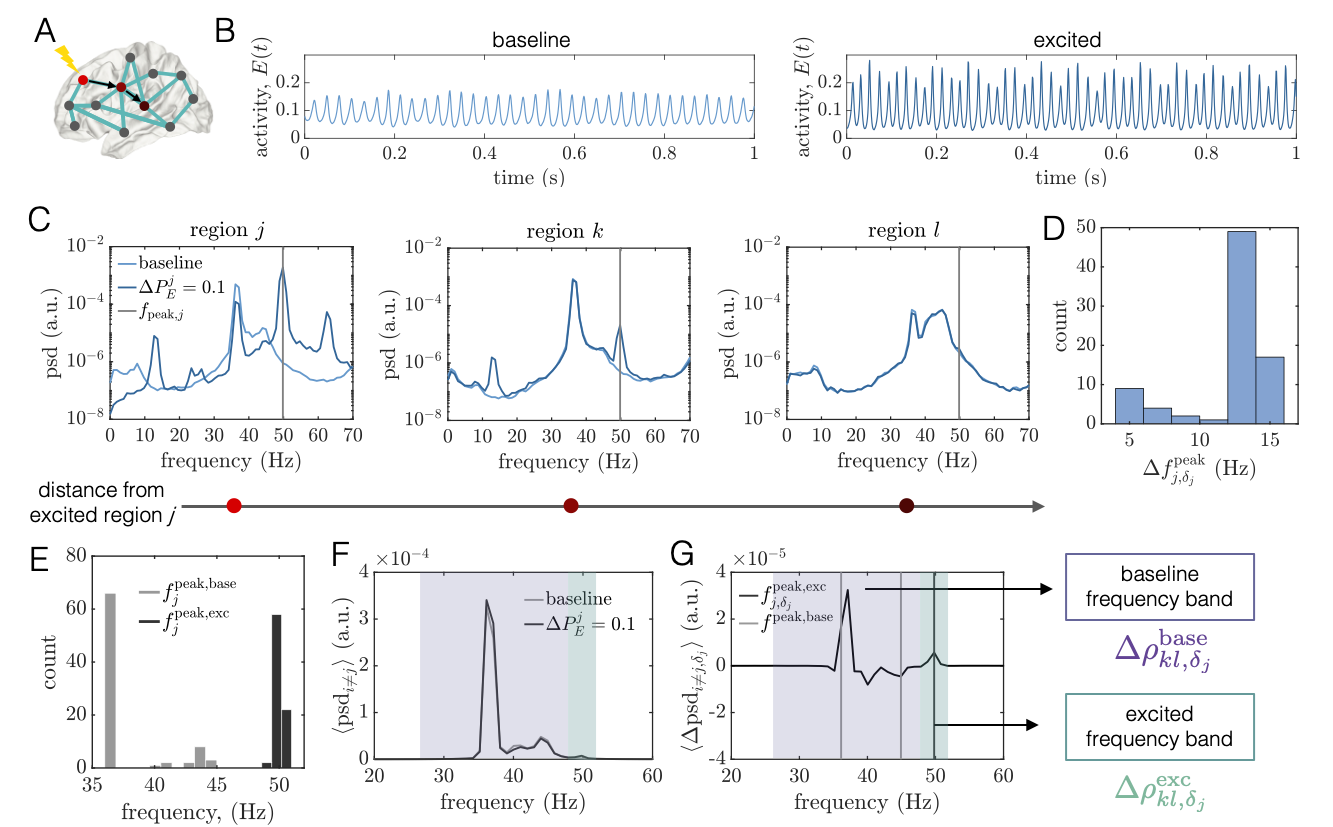}
	\caption{\textbf{Alterations to power spectra induced by local stimulation of neural activity in the state of peak baseline coherence (WP2).} \textit{(A)} Schematic of a brain network depicting the site $j$ of selective excitation in brightest red. The black arrows point to two other regions $k$ and $l$ that lie progressively further away from the perturbed area. Note that in this figure, regions $j$, $k$, and $l$ correspond to brain areas 1 (R--Lateral Orbitofrontal), 4 (R--Medial Orbitofrontal), and 10 (R--Precentral), respectively. \textit{(B)} Left: A segment of region $j$'s activity time-course in the baseline condition. Right: A segment of region $j$'s activity time-course when it is perturbed with additional excitatory input $\Delta P_{\mathrm{E},j}$. \textit{(C)} Power spectra of the excited area $j$ and two other downstream regions $k$ and $l$. In all three panels, the lighter curves correspond to the baseline condition, and the darker curves correspond to the state in which region $j$ is driven with increased current. The gray vertical lines indicate the peak frequency of region $j$, $f_{\mathrm{peak},j}$, in the excited condition. \textit{(D)} Histogram of the shift in peak frequency $\Delta f^{\mathrm{peak}}_{j,\delta_j}$ induced by exciting unit $j$, plotted over all choices of the perturbed area. \textit{(E)} Distribution of peak frequencies of all units in the baseline condition (light gray) and distribution of the peak frequency that units acquire when directly excited (dark gray). \textit{(F)} Average power spectra $\langle \mathrm{psd}_{i} \rangle$ over all units $i\neq j$ at baseline (light gray) and when unit $j$ is perturbed with additional input (dark gray).  \textit{(G)} The average difference $\langle {\Delta} \mathrm{psd}_{i,\delta_j} \rangle$ of the spectra of unit $i\neq j$ when unit $j$ is excited and in the baseline condition, where the average is over all units $i\neq j$.  For reference, the two light gray vertical lines denote the minimum and maximum peak frequency of units in the baseline state, and the dark gray line indicates the peak frequency acquired by the excited region $j$. Shaded boxes denote two frequency bands of interest: \textit{(1)} the \textit{baseline} band (purple) consisting of the main oscillation frequencies of brain areas under baseline conditions, and \textit{(2)} the \textit{excited} band (green) centered around the peak frequency that perturbed regions inherit under increased excitatory drive. Note that $\langle \Delta \mathrm{psd}_{i\delta_j} \rangle$ exhibits modulations in both bands. In subsequent analyses, we assess perturbation-induced changes in functional interactions between brain areas by computing the difference in the PLV between the baseline state and the state when unit $j$ is excited. These modulations are computed separately for the baseline frequency band, $\Delta \rho^{\mathrm{base}}_{kl,\delta_j}$ (purple) and for the excited band $\Delta \rho^{\mathrm{exc}}_{kl,\delta_j}$ (green).}
	\label{f:wpCritical_powSpec}
\end{figure*}

Similar to our observations at WP1, stimulating a particular region with increased excitatory drive causes an increase in the amplitude and frequency of the neural activity (Fig.~\ref{f:wpCritical_powSpec}B and Fig.~\ref{f:wpCritical_powSpec}C,Left). In Fig.~\ref{f:wpCritical_powSpec}D, we show a histogram of the shift in peak frequency $\Delta f^{\mathrm{peak}}_{j,\delta_j}$ across all choices of the activated region, which range between approximately 5 and 15 Hz. In addition to an increase in overall amplitude, the activity of the perturbed region under excited conditions acquires more complex temporal structure, exhibiting enhanced amplitude modulations that are indicative of quasiperiodic dynamics. Due to interactions with the network, the excited area's spectra also shows sidebands to the left and right of its main peak, and a bump at a frequency equal to the difference in the sideband and peak frequencies, which is related to the amplitude modulations observed in the time-series. We also examine the power spectra of two downstream regions $k$ and $l$, located at increasing distances from the perturbed area $j$ (Fig.~\ref{f:wpCritical_powSpec}C,Middle,Right). As for WP1, downstream region $k$ develops a new spectral component at the peak frequency of the excited region -- and also at a lower frequency approximately equal to the difference of its baseline peak and the excited peak -- whereas unit $l$ does not.

Next we consider the distribution of units' peak frequencies in the baseline condition and when excited with additional input (Fig.~\ref{f:wpCritical_powSpec}E). We again observe a clear separation between the two histograms, which indicates that there are two frequency ranges of interest for further analysis. To probe this behavior, we first study the average spectra $\langle \mathrm{psd}_{i} \rangle$ over all units $i \neq j$ in the baseline state and in the state when unit $j$ is selectively excited (Fig.~\ref{f:wpCritical_powSpec}F). We then consider the average difference $\langle {\Delta} \mathrm{psd}_{i,\delta_j} \rangle$ of the spectra of unit $i\neq j$ between when unit $j$ is excited and in the baseline condition (Fig.~\ref{f:wpCritical_powSpec}G). We observe clear power modulations in both the baseline frequency band -- which contains the original oscillation frequencies -- and in an excited frequency band centered around the peak frequency of the directly stimulated region. In the next section, we proceed to examine changes in the PLV in these two bands -- $\langle \rho_{kl,\delta_j}^{\mathrm{base}} \rangle$ and $\langle \rho_{kl,\delta_j}^{\mathrm{exc}} \rangle$ -- induced by excitations of regional activity.

\subsubsection{Structural and functional network connectivity continue to predict overall changes in excited and baseline band phase-locking at the peak-coherence working point}

\begin{figure*}
	\centering
	\includegraphics[width=0.9\textwidth]{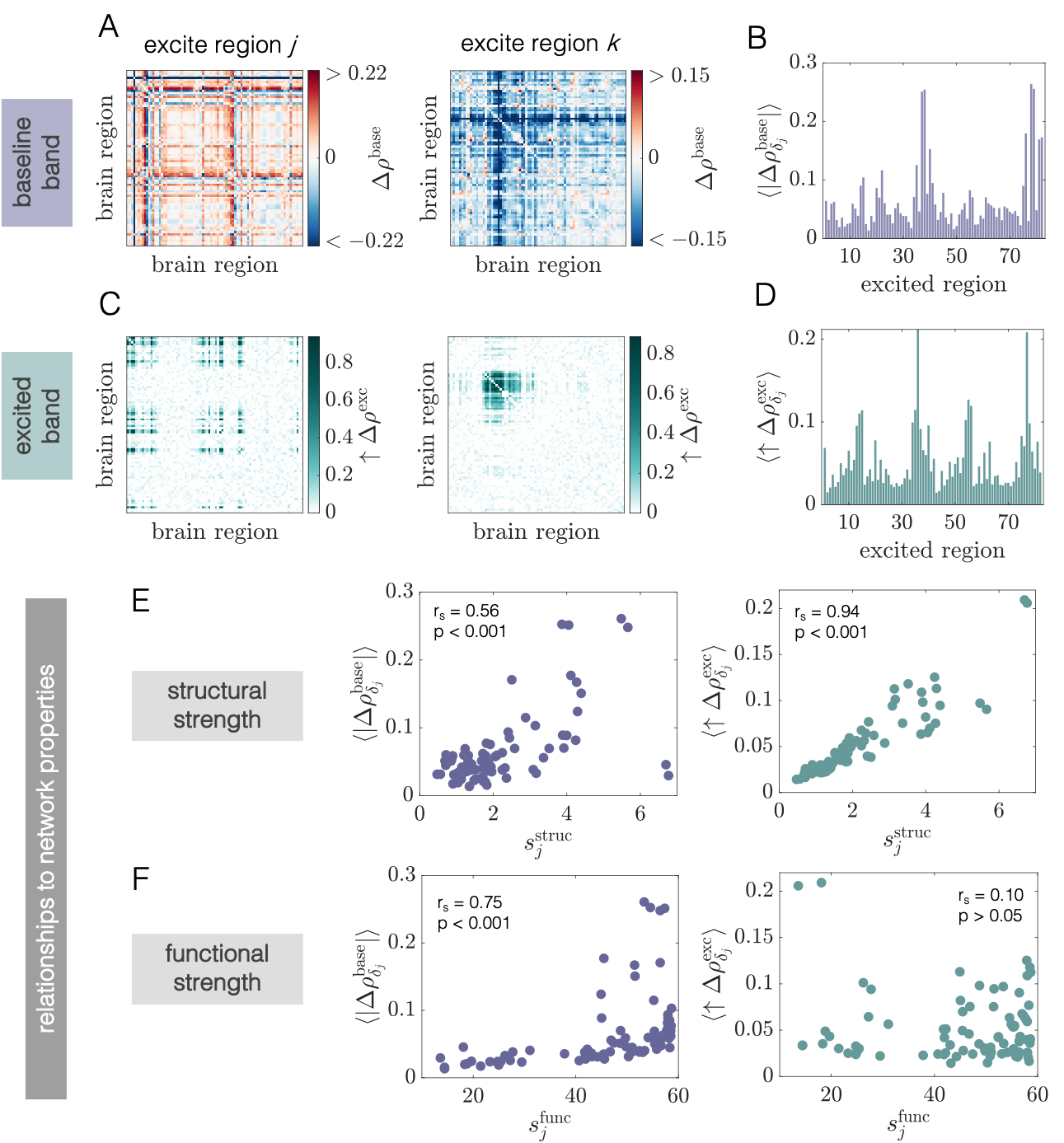}
	\caption{\textbf{Phase-locking modulations at WP2 are driven by local excitations of neural activity, differ between excited and baseline frequency bands, and are differentially related to structural and functional network properties.}  \textit{(A)} Pairwise changes in phase-locking inside the baseline band $ \Delta \rho^{\mathrm{base}} $ when region $j$ (Left) or region $k \neq j$ (Right) is perturbed. Note that in this figure, regions $j$ and $k$ correspond to regions 4 (R--Medial Orbitofrontal) and 23 (R--Lateral Occipital), respectively. \textit{(B)} Network-averaged absolute PLV changes in the baseline band $ \langle | \Delta \rho^{\mathrm{base}}_{\delta_j} | \rangle $ caused by excitation to different brain areas. \textit{(C)} Pairwise increases in phase-locking inside the excited band $ \Delta \rho^{\mathrm{exc}} $ when region $j$ (Left) or region $k \neq j$ (Right) is perturbed. \textit{(D)} Network-averaged positive PLV changes in the excited band $ \langle \uparrow \Delta \rho^{\mathrm{exc}}_{\delta_j} \rangle $ caused by excitation to different brain areas. \textit{(E)} The quantity $\langle | \Delta \rho^{\mathrm{base}}_{\delta_j} | \rangle $ \textit{vs.} structural node strength $s^{\mathrm{struc}}_{j}$ (Left) and the quantity $\langle \uparrow \Delta \rho^{\mathrm{exc}}_{\delta_j} \rangle $ \textit{vs.} structural node strength $s^{\mathrm{struc}}_{j}$ (Right). \textit{(F)} The quantity $\langle | \Delta \rho^{\mathrm{base}}_{\delta_j} | \rangle $ \textit{vs.} functional node strength $s^{\mathrm{func}}_{j}$ (Left) and the quantity $\langle \uparrow \Delta \rho^{\mathrm{exc}}_{\delta_j} \rangle $ \textit{vs.} functional node strength $s^{\mathrm{func}}_{j}$. In panels \textit{(E)} and \textit{(F)}, insets indicate Spearman correlation coefficients between the plotted quantities and their associated $p$-values).}
	\label{f:phaseCoh_modulations_wpCritical}
\end{figure*}

Similar to the working point below peak synchrony (WP1), local stimulation at WP2 induces phase-locking changes that differ depending on which part of the network is activated and which frequency band is examined (Fig.~\ref{f:phaseCoh_modulations_wpCritical}A--D). To appreciate this fact, it is helpful to study examples that display the changes in the PLV in both the baseline band and the excited frequency band when either region $j$ or region $j\neq k$ is excited (Fig.~\ref{f:phaseCoh_modulations_wpCritical}A,C). Upon examination of the network-averages of the PLV modulations (Figs.~\ref{f:phaseCoh_modulations_wpCritical}B,D), we again find significant variability in the global responses to regional stimulation. Actually, in comparing the overall baseline band changes at WP1 (Fig.~\ref{f:phaseCoh_modulations_wp1}B) to the changes at WP2 (Fig.~\ref{f:phaseCoh_modulations_wpCritical}B), we find that the dispersion and the mean value of the global responses are both larger at the second working point. We quantified and compared the heterogeneity of the responses by computing the coefficient of variation (CoV) of the distribution of $\langle | \Delta \rho_{\delta_j}^{\mathrm{base}} | \rangle $ (across the choice of the stimulated region) for the two baseline conditions. For WP1, we find that the CoV for $\langle |\Delta \rho_{\delta_j}^{\mathrm{base}} | \rangle $ is equal to 0.43, whereas for WP2 it is equal to 0.86. The mean response across all choices for the excited region is also higher at WP2 (0.065) than at WP1 (0.040). We additionally note, however, that the phase-locking induced in the excited frequency band, $\langle | \Delta \rho_{\delta_j}^{\mathrm{base}} | \rangle $, decreases from 0.07 at WP1 to 0.05 at WP2. Hence, when the system operates at the state of peak $\rho^{\mathrm{global}}$, the stimulation-induced responses in the baseline frequency band are larger -- on average -- and also more heterogeneously distributed relative to the responses at the working point below the state of peak $\rho^{\mathrm{global}}$, while the average response in the excited band decreases. We return to these points again in Sec.~\ref{s:state_dependence}, where we more generally examine the state-dependence of perturbation-induced changes in phase-locking. As for WP1, we also continue to find that the phase-coherence modulations in the baseline frequency band are more spatially distributed throughout the system than those induced in the excited frequency band (see Sec. SIII, Fig. S3B).

To conclude this section, we consider the relationships between changes in phase-coherence induced by stimulating a given region and the structural or functional node strength of the stimulated area (see Sec.~\ref{s:struc_func_PLV_WP1} for the definitions of these two quantities). A number of the associations that exist at WP2 (Fig.~\ref{f:phaseCoh_modulations_wpCritical}E,F) were also observed at WP1 (see Fig.~\ref{f:phaseCoh_modulations_wp1}E,F). For example, the structural strength $s_{j}^{\mathrm{struc}}$ remains strongly positively correlated with the average phase-locking induced in the excited frequency band $\langle \uparrow \Delta \rho_{\delta_j}^{\mathrm{exc}} \rangle$ (Fig.~\ref{f:phaseCoh_modulations_wpCritical}E, Right). Furthermore, functional strength $s_{j}^{\mathrm{func}}$ retains the strongest positive correlation with the absolute change in baseline band phase-coherence $\langle | \Delta \rho_{\delta_j}^{\mathrm{base}} | \rangle$ (Fig.~\ref{f:phaseCoh_modulations_wpCritical}F, Left). Thus, the phase-locking induced in the new, excited frequency band continues to be strongly predicted by anatomical connectivity, whereas the reconfigurations of phase coherence in the baseline frequency band continue to be best predicted by the strength of regions' initial functional interactions with the system as a whole. The main difference between WP1 and WP2 is that for the second working point, a positive correlation also emerges between $\langle | \Delta \rho_{\delta_j}^{\mathrm{base}} | \rangle$ and $s_{j}^{\mathrm{struc}}$ (Fig.~\ref{f:phaseCoh_modulations_wpCritical}E, Left). The existence of this relationship indicates that when the system state is at peak $\rho^{\mathrm{global}}$, structural connectivity also becomes partially indicative of stimulation-induced changes in baseline band phase-locking. Nonetheless, the level of regions' initial coherence with the network remains the strongest predictor of the total response at units' baseline frequencies.

\subsection{Working point 3: Post-global coherence peak}
\label{s:wp3}

In the previous sections, we examined how localized excitations of neural activity modulated brain network dynamics when the system was operating in either \textit{(1)} a regime of relatively low background drive and below the point of peak global coherence (WP1), or \textit{(2)} in a state of intermediate background drive located right at the point of peak global coherence (WP2). In this section, we examine the effects of perturbations at a third and final dynamical configuration, where the system receives high background drive such that it resides well beyond the point of maximal coherence. In particular, we consider the working point located at $C = 2.5$ and $P_{\mathrm{base}} = 0.7$ (WP3). Relative to WP1 and WP2, the activity at WP3 is characterized by regular, high-amplitude oscillations of slightly higher frequency (see Fig.~\ref{f:baseline_state1}G,H). Moreover, the phase-locking matrix for the system at WP3, in contrast to the previous working points, is characterized by a higher degree of local as opposed to global phase-locking (Fig.~\ref{f:baseline_state2}C, Row 3, Column 2).
 
\subsubsection{Spectral changes are more restrained at the high-background drive working point}
\label{s:powSpec_wp2}

\begin{figure*}
	\centering
	\includegraphics[width=\textwidth]{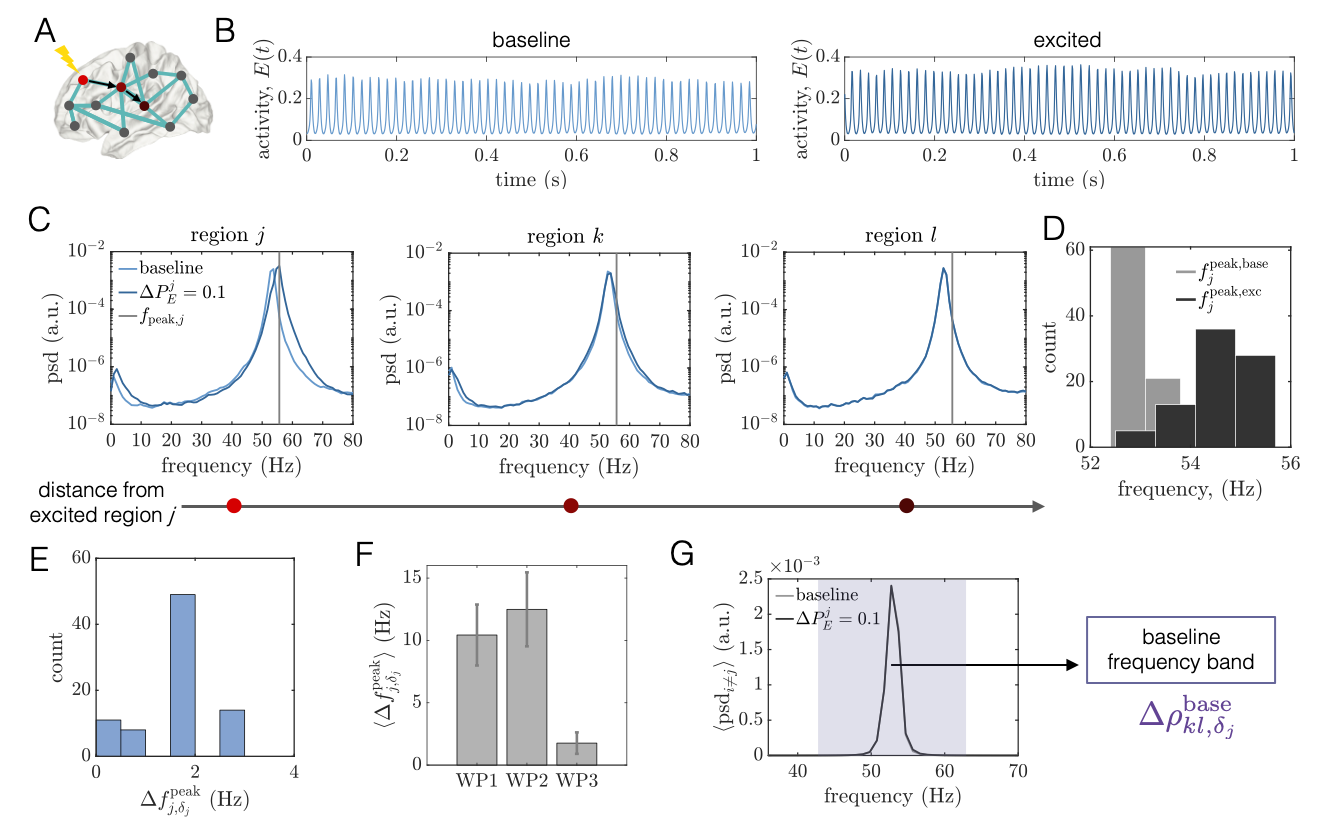}
	\caption{\textbf{Effects of local excitations on power spectra are more constrained at the high background-drive working point (WP3).} \textit{(A)} Schematic of a brain network depicting the site $j$ of selective excitation in brightest red. The black arrows point to two other regions $k$ and $l$ that lie progressively further away from the perturbed area. Note that in this figure, regions $j$, $k$, and $l$ correspond to brain areas 1 (R--Lateral Orbitofrontal), 4 (R--Medial Orbitofrontal), and 10 (R--Precentral), respectively. \textit{(B)} Left: A segment of region $j$'s activity time-course in the baseline condition. Right: A segment of region $j$'s activity time-course when it is perturbed with additional excitatory input $\Delta P_{\mathrm{E},j}$. \textit{(C)} Power spectra of the excited area $j$ and two other downstream regions $k$ and $l$. In all three panels, the lighter curves correspond to the baseline condition, and the darker curves correspond to the state in which region $j$ is driven with increased current. The gray vertical lines indicate the peak frequency of region $j$, $f_{\mathrm{peak},j}$, in the excited condition. \textit{(D)} Distribution of peak frequencies of all units in the baseline condition (light gray) and distribution of the peak frequency that units acquire when directly excited (dark gray).  \textit{(E)} Histogram of the shift in peak frequency $f^{\mathrm{peak}}_{j,\delta_j}$ induced by exciting unit $j$, plotted over all choices of the perturbed area. \textit{(F)} The average shift in the peak frequency of the excited region $\langle \Delta f^{\mathrm{peak}}_{j,\delta_j} \rangle$ for WP1, WP2, and WP3 (error bars indicate the standard deviation over all choices of the excited unit). \textit{(G)} Average power spectra $\langle \mathrm{psd}_{i} \rangle$ over all units $i\neq j$ at baseline (light gray) and when unit $j$ is perturbed with additional input (dark gray). Because the spectra are relatively robust to excitations at this working point, phase coherence modulations induced by a local excitation to region $j$, $\Delta \rho^{\mathrm{base}}_{kl,\delta_j}$, are computed only for a single baseline frequency band (purple area).}
	\label{f:powSpec_WP3}
\end{figure*}

Upon examination of the time-series corresponding to baseline and excited conditions, we find noticeable differences between how additional input changes local activity at WP3 versus at either WP1 or WP2. Specifically, when the system operates in the high-drive state, an excitation of the same strength appears to have a much less drastic effect on the stimulated region's firing rate activity, inducing relatively small changes to its amplitude and frequency (Fig.~\ref{f:powSpec_WP3}B). 

In fact, stimulation generally leaves weaker effects on regional spectra at WP3. We arrive at this conclusion by first considering the baseline spectra and the spectra under excitation to unit $j$ itself and two downstream areas $k$ and $l$ located at increasing distances from the excited region $j$ (Fig.~\ref{f:powSpec_WP3}C). As in the other two dynamical regimes, the peak frequency and power of the excited region shifts to higher values, but at WP3, the increase is modest relative to the shifts that occur at either WP1 or WP2, and no modulation sidebands arise in the spectra. Consequently, we also observe that an excitation to unit $j$ seems to have relatively little impact on the spectra of downstream regions, especially when compared to the effects at WP1 and WP2. In particular, no second peak emerges in the spectra of region $k$ or region $l$ (Fig.~\ref{f:powSpec_WP3}C, Middle, Right). To more generally quantify the effects of localized excitations on the perturbed regions' power spectra, we examine the distribution of the differences in peak frequency of the perturbed unit between its excited and baseline states, $\Delta f^{\mathrm{peak}}_{j,\delta_j}$. The largest of these shifts is only about 3Hz (Fig.~\ref{f:powSpec_WP3}E). Hence, relative to WP1 and WP2, the average shift in peak frequency $\langle \Delta f^{\mathrm{peak}}_{j,\delta_j} \rangle$ is greatly reduced at WP3 (see Fig.~\ref{f:powSpec_WP3}F). Furthermore, due to the stiffness of individual regions' dynamics at WP3, units do not exhibit a large enough increase in their peak frequency upon excitation to produce a new, well-defined excited frequency band. Unlike the previous two working points, the baseline and excited peak frequency distributions significantly overlap at WP3 (Fig.~\ref{f:powSpec_WP3}D), precluding the notion of separate baseline and excited frequencies. For this reason, in our subsequent analyses we only consider phase-locking changes inside a single frequency band (Fig.~\ref{f:powSpec_WP3}G). While we refer to this set of frequencies as the ``baseline band", we note that it still contains the peak frequency of the directly excited unit, since its frequency shift is so small.

The results presented in this section indicate that regional dynamics are more robust to excitations at WP3 compared to WP2 or WP1. This robustness can in part be understood by considering the effects of the baseline level of background input, $P_{\mathrm{E}}^{\mathrm{base}}$, which is the parameter that is tuned to move from WP1 to WP2, and from WP2 to WP3. In particular, the increase in background drive at WP3 relative to WP1 and WP2 pushes each unit in the system further towards the stable limit cycle it would approach in the case of zero coupling, in turn resulting in robust endogenous oscillations at each brain area. These regular and high-amplitude oscillations that arise in the high-drive state are more difficult to perturb, which leads to minimal changes in the power spectra under local excitations of neural activity. For the same reasons, it is also more difficult for a local change in activity to propagate and influence the dynamics of remote areas. In contrast, when the system operates at either WP1 or WP2, the baseline oscillations at each brain area are weaker; they are of lower amplitude and give rise to a broader spectra in comparison to WP3. In those states, the system is more plastic and more susceptible to local perturbations. This flexibility of dynamics at WP1 and WP2 is reflected by clear modifications to regional spectra upon local stimulation and the signatures of the stimulation effect in downstream regions (Figs.~\ref{f:ts_ps_perturb} and~\ref{f:wpCritical_powSpec}).

\subsubsection{Selective excitations yield a distinct and more homogenous set of phase-coherence modulations at the high-drive working point}

\begin{figure*}
	\centering
	\includegraphics[width=0.9\textwidth]{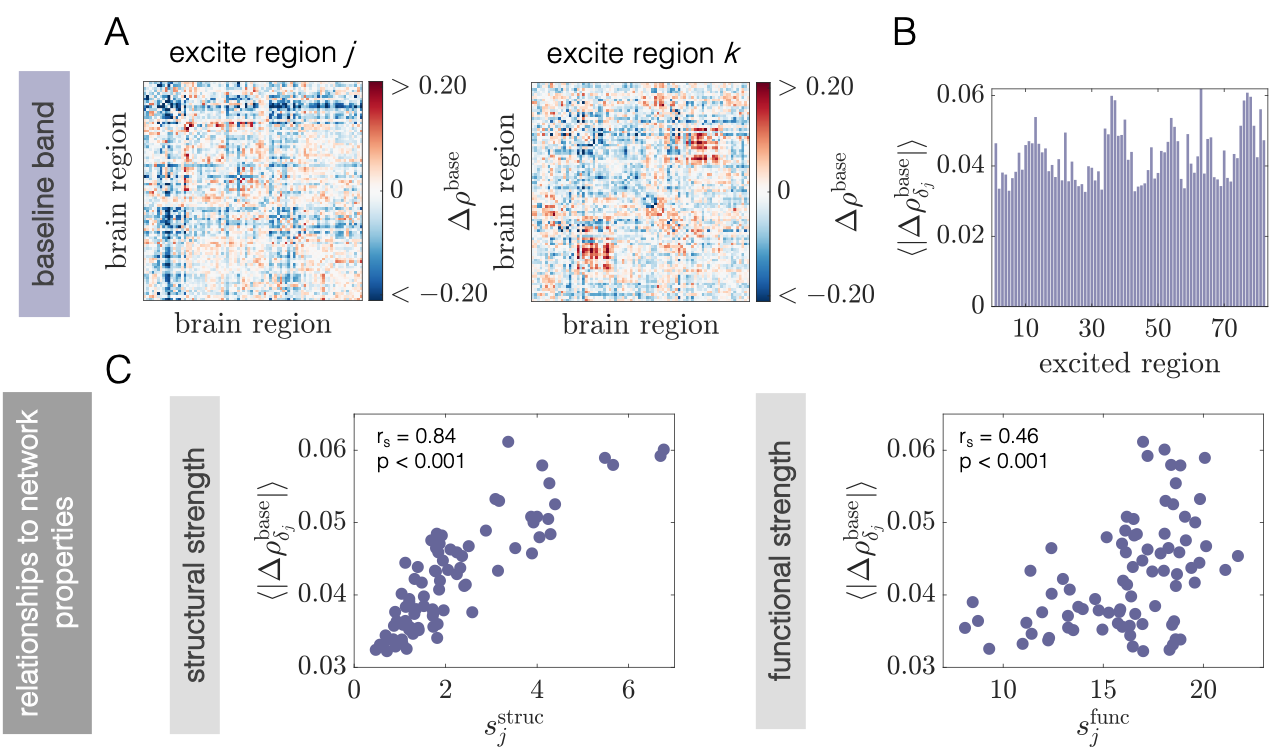}
	\caption{\textbf{Phase coherence modulations induced by regional stimulation at WP3 and their associations with network properties.} \textit{(A)} Pairwise changes in phase-locking inside the baseline band $ \Delta \rho^{\mathrm{base}} $ when region $j$ (Left) or region $k \neq j$ (Right) is perturbed. Note that in this figure, regions $j$ and $k$ correspond to regions 10 (R--Precentral) and 15 (R--Isthmus), respectively. \textit{(B)} Network-averaged absolute PLV changes in the baseline band $ \langle | \Delta \rho^{\mathrm{base}}_{\delta_j} | \rangle$ caused by excitation to different brain areas. \textit{(C)} The quantity $\langle | \Delta \rho^{\mathrm{base}}_{\delta_j} | \rangle $ \textit{vs.} structural node strength $s^{\mathrm{struc}}_{j}$ (Left) and \textit{vs.} functional node strength $s^{\mathrm{struc}}_{j}$ (Right). Insets indicate Spearman correlation coefficients between these quantities, and their associated $p$-values).}
	\label{f:WP3_PLVmodulations}
\end{figure*}

While the increased rigidity of the system at WP3 prevents the emergence of two well-separated frequency bands, we can still assess the effects of selective excitations on PLVs in the single, baseline frequency band. We carry out such an analysis in this section, focusing on contrasting the results obtained at WP3 to those obtained previously at WP1 and WP2. To begin, we show examples of the pairwise changes in PLV induced by excitation of two different brain areas $j$ and $k$ (Fig.~\ref{f:WP3_PLVmodulations}A). In each case, both positive and negative modulations occur, but the response clearly differs depending on the stimulated area. In Fig.~\ref{f:WP3_PLVmodulations}B, we visualize the average absolute changes $\langle | \Delta \rho_{\delta_j}^{\mathrm{base}} | \rangle$ in phase-locking caused by excitations of different brain areas. As with the previously examined working points, we observe that particular regions in the network have more or less influence in their ability to alter interareal coherence. 

However, despite the variability in the global response across the choice of the excited region, a visual comparison of the responses at WP3 (Fig.~\ref{f:WP3_PLVmodulations}B) to those at either WP1 (Fig.~\ref{f:phaseCoh_modulations_wp1}B) or WP2 (Fig.~\ref{f:phaseCoh_modulations_wpCritical}B) suggests that the overall modulations may be less variable across regions when the system operates in the high-drive state. Indeed, although the mean of the absolute response across all choices of the excited region is approximately the same at WP1 and WP3 ($\approx$ 0.04 in both cases), when we compare the CoV of $\langle | \Delta \rho_{\delta_j}^{\mathrm{base}} | \rangle $, we find that the CoV for WP3 is 0.18, while for WP1 it is 0.43. The reduced CoV for the responses at WP3 indicate more homogenous -- and therefore less region-specific -- global responses upon stimulation of different brain areas in the high background drive regime. Because we are particularly interested in how perturbations may differentially affect network dynamics depending on the state of the system at baseline, we also asked if the overall phase-coherence modulations induced by exciting different brain areas were correlated between WP1 and WP3. We found that this relationship was not statistically significant (Fig. S4), indicating that distinct sets of regions produce the largest or smallest overall changes in network-wide coherence at the two different baseline states. 

Given the observed differences in the effects of dynamical state on the propagation of local excitations, we next considered the associations between global changes in phase-locking induced by regional stimulation and structural or functional properties of network nodes (Fig.~\ref{f:WP3_PLVmodulations}C). At WP3, we find a strong positive correlation between $\langle | \Delta \rho_{\delta_j}^{\mathrm{base}} | \rangle$ and structural strength $s_{j}^{\mathrm{struc}}$ ($r_{s} = 0.84$, $p<0.001$; Fig.~\ref{f:WP3_PLVmodulations}C, Left). Hence, modulations in coherence at the high-drive working point are dependent on the presence of strong anatomical connections emanating from the stimulated region. This result may be a consequence of the fact that brain areas exhibit relatively strong spontaneous rhythms at WP3, which are less easily adjusted by non-local changes in activity. In turn, it will be the more strongly connected regions -- which have more direct influence on other areas of the network -- that most effectively alter ongoing baseline dynamics and lead to greater overall disruptions of interareal coherence. In terms of the associations between functional strength and phase-locking modulations, we also found a significant positive correlation between $s_{j}^{\mathrm{func}}$ and $\langle | \Delta \rho^{\mathrm{base}}_{\delta_j} | \rangle$ ($r_{s} = 0.46$, $p<0.001$; Fig.~\ref{f:WP3_PLVmodulations}C, Right). This result indicates that there is some overlap between how structural and functional connectivity are related to changes in phase-coherence at WP3, but that structural connectivity is a more robust predictor of these effects. 

To conclude this section, we note that an equally important observation is that, depending on the dynamical state of the system, different mechanisms guide how perturbations alter functional interactions. As a consequence, in the low-drive state, the emergent dynamical coordination between units is more strongly related to the reconfiguration of phase-coherence at regions' spontaneous frequencies, whereas structure seems to govern the effects of focal perturbations in the high-drive regime. Because the collective state of the system can change even when anatomical connectivity is fixed, both network structure as well as dynamical interactions are important to consider.

\subsection{Effects of local excitations on functional interactions vary with the dynamical state}
\label{s:state_dependence}

\begin{figure*}
	\centering
	\includegraphics[width=\textwidth]{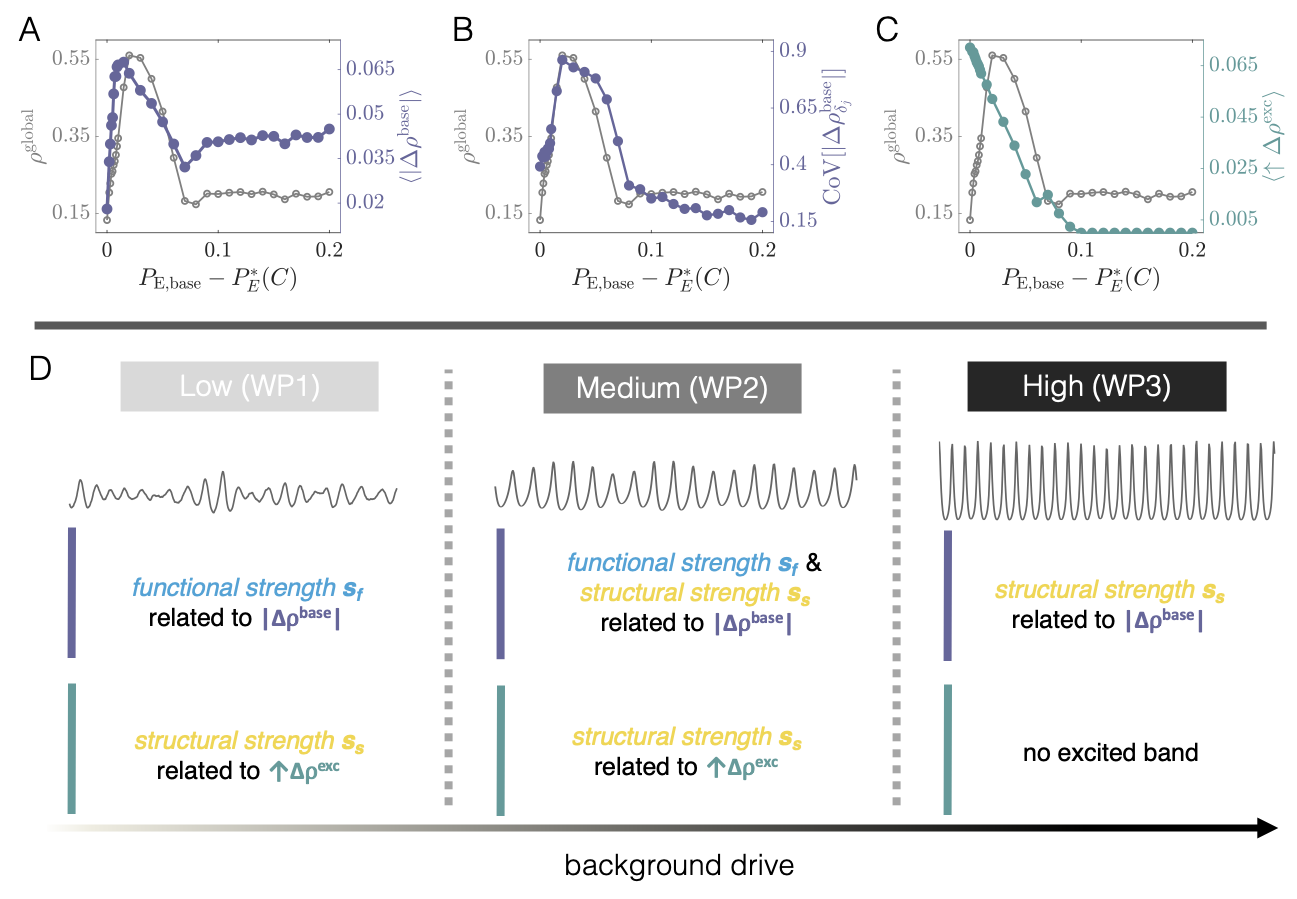}
	\caption{\textbf{Dependence of phase-locking modulations on the baseline state of the system and a schematic showing relationships to structural or functional network properties for WP1, WP2, and WP3.} \textit{(A)} The left axis (gray) shows the global PLV $ \rho^{\mathrm{global}} $ (at baseline) as a function of the background input level $P_{\mathrm{E,base}} - P_{\mathrm{E}}^{*}(C)$. The right axis (purple) shows the network- and region-averaged absolute change in baseline band phase coherence due to regional excitations $\langle | \Delta \rho^{ \mathrm{base}} | \rangle$ as a function of $P_{\mathrm{E,base}} - P_{\mathrm{E}}^{*}(C)$. \textit{(B)} The left axis (gray) shows the global PLV $ \rho^{\mathrm{global}} $ (at baseline) as a function of the background input level $P_{\mathrm{E,base}} - P_{\mathrm{E}}^{*}(C)$. The right axis (purple) shows the coefficient of variation (computed across the choice of the excited brain area) of the network-averaged absolute change in baseline band phase coherence due to a regional excitation $\langle | \Delta \rho^{ \mathrm{base}}_{\delta_j} | \rangle$ as a function of $P_{\mathrm{E,base}} - P_{\mathrm{E}}^{*}(C)$.  \textit{(C)} The left axis (gray) shows the global PLV $ \rho^{\mathrm{global}} $ (at baseline) as a function of the background input level $P_{\mathrm{E,base}} - P_{\mathrm{E}}^{*}(C)$. The right axis (green) shows the network- and region-average of the positive changes in excited band phase coherence due to regional excitations $\langle \uparrow \Delta \rho^{\mathrm{exc}} \rangle$ as a function of $P_{\mathrm{E,base}} - P_{\mathrm{E}}^{*}(C)$. \textit{(D)} A schematic depicting how structural $s^{\textrm{struc}}$ or functional $s^{\textrm{func}}$ strength are related to either baseline or excited band PLV changes, for the three studied working points. As the background drive varies from low (WP1) to medium (WP2) to high (WP3), the dynamical state of the system changes, and so does the association of different phase-locking modulations to structural or functional network properties.}
	\label{f:phaseCoh_modulations_varyPeB}
\end{figure*}

In Secs.~\ref{s:wp1}, \ref{s:wp2}, and \ref{s:wp3} we examined in detail how additional excitation of a single brain area induced or modulated interareal phase-locking for three different baseline working points of the model. In particular, the level of background drive -- which regulates the influence of locally-generated dynamics on network activity -- was tuned to move the system between qualitatively distinct operating points. These three regimes differed both in terms of the nature of regional activity, as well as the collective dynamical coordination of that activity (see Sec.~\ref{s:baseline_state}). Crucially, we found that local excitations had markedly different effects on system dynamics and associations to network connectivity depending on the baseline state. In this section, we more generally examine and summarize the relations between the dynamical state and the effects of excitations on interareal coherence. 

To examine the effects of selective increases in drive on the coordination of network dynamics, we consider the overall PLV changes in the baseline and excited frequency bands as functions of $P_{\mathrm{E,base}} - P_{\mathrm{E}}^{*}(C)$. As before, we fix $C = 2.5$ and vary only $P_{\mathrm{E,base}}$. We begin by studying the average absolute modulations of phase-coherence in the baseline frequency band $\langle | \Delta \rho^{\mathrm{base}} | \rangle$ (Fig.~\ref{f:phaseCoh_modulations_varyPeB}A), where the angled brackets indicate a mean first over all pairs of brain areas and then across all $N$ choices of the excited region. We find that the general shape of $\langle | \Delta \rho^{\mathrm{base}} | \rangle$ as a function of $P_{\mathrm{E,base}} - P_{\mathrm{E}}^{*}(C)$ tends to mimic that of the global baseline phase-coherence $\rho^{\mathrm{global}}$, but with a slight shift to the left. In particular, $\langle | \Delta \rho^{\mathrm{base}} | \rangle$ is at a minimum value for the smallest background drive state, then undergoes a steep rise to a well-defined peak at still a relatively low value of $P_{\mathrm{E,base}} - P_{\mathrm{E}}^{*}(C) = 0.01$, followed by a slower decline and then a slight increase to a plateau as $P_{\mathrm{E,base}} - P_{\mathrm{E}}^{*}(C)$ increases further.

The peak in the phase-locking modulation curve signifies a distinct working point at which regional excitations produce the largest overall changes to functional interactions. In this state, the system is highly flexible in that local amplifications of neural activity can generate large modifications to overall levels of interareal coherence patterns in a frequency band encompassing units' baseline rhythms. It is important to note that this working point occurs just prior to the global PLV peak, which is located at $P_{\mathrm{E,base}} - P_{\mathrm{E}}^{*}(C) = 0.02$. Thus, the system exhibits the largest response to localized excitations when it is perched between the disordered and ordered phase (as measured by the level of $\rho^{\mathrm{global}}$). At lower values of $P_{\mathrm{E,base}} - P_{\mathrm{E}}^{*}(C)$, activation of a single brain area has weaker effects on functional interactions. This behavior likely occurs in part because there is less initial coherence for local excitations to disrupt. As $ \rho^{\mathrm{global}} $ declines after peaking with increasing $P_{\mathrm{E,base}} - P_{\mathrm{E}}^{*}(C)$, so too does the overall response to regional excitation. The phase-coherence change approaches a local minimum at $P_{\mathrm{E,base}} - P_{\mathrm{E}}^{*}(C) = 0.07$, after which it exhibits a brief rise as the system transitions into states dominated by enhanced local baseline coherence (see Figs.~\ref{f:baseline_state2}B,C). Finally, at working points well beyond peak $\rho^{\mathrm{global}}$ (e.g., at WP3), $\langle | \Delta \rho^{\mathrm{base}} | \rangle$ settles to an intermediate value. 

In addition to understanding how the average response of the system varies depending on the dynamical state, it is also of interest to examine how the variability of the responses (over excitation of different brain areas) changes as a function of state. In other words, is the overall stimulation-induced change in phase coherence specific to the unit that received additional input, or do different units tend to yield similar amounts of functional reconfiguration? To address this question, we considered the coefficient of variation (CoV) of the network-averaged absolute PLV change, $\mathrm{CoV}[\langle | \Delta \rho_{\delta_j}^{\mathrm{base}}  | \rangle]$, where the CoV was computed across the different choices for the excited brain area $j$. In general, we find that $\mathrm{CoV}[\langle | \Delta \rho_{\delta_j}^{\mathrm{base}}  | \rangle]$ does exhibit a state dependence. In considering $\mathrm{CoV}[\langle \Delta \rho_{\delta_j}^{\mathrm{base}} \rangle]$ \textit{vs.} $P_{\mathrm{E,base}} - P_{\mathrm{E}}^{*}(C)$ (Fig.~\ref{f:phaseCoh_modulations_varyPeB}), we observe that $\mathrm{CoV}[\langle \Delta \rho_{\delta_j}^{\mathrm{base}} \rangle]$ begins at an intermediate value of $\approx$ 0.4 for the state of lowest background drive, and then steadily increases to a global maximum at $P_{\mathrm{E,base}} - P_{\mathrm{E}}^{*}(C) = 0.02$, which coincides with the peak of $ \rho^{\mathrm{global}} $. This behavior suggests that the responses to local stimulation are most heterogeneously distributed when the system operates at the point of peak global coherence. After peaking, $\mathrm{CoV}[\langle \Delta \rho_{\delta_j}^{\mathrm{base}} \rangle]$ first rapidly and then more slowly declines as the level of background drive increases. Importantly, for high-drive states well beyond the peak of $\rho^{\mathrm{global}}$ (such as WP3), the variability of the global responses decreases to values even lower than its initial level at $P_{\mathrm{E,base}} - P_{\mathrm{E}}^{*}(C) = 0$. This result signifies that when the network operates in a regime where local dynamics have increased influence, the system's responses to regional stimulation are more homogenous, and therefore less region-specific. In contrast, for low-drive working points below and up to the peak $\rho^{\mathrm{global}}$, there is greater dispersion in $\langle | \Delta \rho_{\delta_j}^{\mathrm{base}}| \rangle$, such that the responses are more dependent on the precise location in the network that is excited.

For our final analysis of state-dependence, we assess how the network working point influences the average amount of phase-locking induced in the excited frequency band, when it exists (see Sec.~\ref{s:PLV_wp1} for the criteria). We find that as the level of generic background input increases, the overall response in the excited band decays and eventually vanishes. As discussed in Sec.~\ref{s:wp3}, this behavior is due to the amplification of local baseline oscillations with increasing background drive. In particular, the heightened rigidity of the system's spontaneous dynamics is more difficult for local perturbations to override, and thus prevents the emergence of widespread phase-locking at the excited frequency of the stimulated region.

Before concluding this section, we reiterate once again that the role of structural and functional network connectivity in determining how stimulation reorganizes dynamical, interareal interactions qualitatively varies depending on the operating point of the model. In Fig.~\ref{f:phaseCoh_modulations_varyPeB}D, we schematically summarize the most robust associations between quantities, separating the background drive $P_{\mathrm{E,base}} - P_{\mathrm{E}}^{*}(C)$ into three main regimes: ``low", ``medium", and ``high" (corresponding to WP1, WP2, and WP3, respectively). Although we have not comprehensively detailed the effects of perturbations for all working points of the model, the results discussed throughout the text highlight the critical influence of a brain network's collective dynamical state in dictating the outcomes of localized stimulation.

\section{Discussion}

In this study, we set out to explore the relations between large-scale brain connectivity, dynamics, and the local and widespread impacts of regional perturbations to neural activity. We began by building a reduced computational model of whole-brain network dynamics. Following the efforts of past work \cite{Muldoon2016:StimulationBased}, brain areas were represented as Wilson-Cowan neural masses with long-range coupling between regions constrained by empirical diffusion tractography measurements. We then aimed to investigate how the stimulation of a particular brain area affects network dynamics, and to determine whether and how the collective dynamical state of the system plays a role in modulating such effects. Here, we chose to examine state-dependence by changing the combination of generic background drive and interareal coupling strength, which vary the influences of local oscillatory activity and network interactions on collective dynamics, respectively. By tuning these parameters, we identified qualitatively distinct dynamical regimes of the model, and at these different working points, we assessed how local excitations of fixed strength induced changes to measures of interareal phase-locking. We found that, depending on the baseline dynamical regime of the system, the network exhibits different responses to regional perturbations. Furthermore, altering the working point of the model also qualitatively altered the relationships between stimulation-induced changes in phase-locking and properties of structural and functional network connectivity. To the best of our knowledge, these points have yet to be investigated at the whole-brain scale via computational modeling.

In order to frame the remainder of the discussion, we begin here with a summary of our main results. First, we found that in states of low-to-intermediate background drive (e.g., WP1 and WP2), local excitations had two main effects. Locally, stimulation caused a significant shift in the excited area's power spectra. This led to the emergence of a new,``excited" frequency in the system that was well-separated from the main oscillation frequencies at baseline. Second, there were modifications in downstream regions' spectra within a frequency band corresponding to the spontaneous, ``baseline" oscillations. Beyond altering brain areas' spectra, a focal perturbation also induced or reconfigured temporal coordination between regional activity in both the excited and baseline frequency bands. At WP1 (Sec.~\ref{s:wp1}) and WP2 (Sec.~\ref{s:wp2}) we found that induced phase-locking in the excited frequency band was tightly constrained by anatomical connectivity, demonstrating the fact that the excited band effects arise due to a direct propagation of strong, periodic activity from the stimulated region to downstream areas along structural connections. In contrast, we observed that the absolute modulations in phase-locking in the baseline frequency band were more strongly related to the stimulated region's functional strength, rather than its structural strength. In other words, network-wide changes in the levels of coherence at units' baseline frequencies were better predicted by the perturbed region's initial coherence with the system as a whole. In contrast to the propagation mechanism underlying the induced phase-locking in the excited band, changes in baseline band coherence arise from modifications of the dynamical interactions between the ongoing rhythms of different brain areas, which is a function of not only network structure, but also the collective oscillatory state of the system. At WP2, we additionally found that anatomical strength started to become predictive of baseline band coherence changes (albeit at a slightly weaker level than the relationship with functional strength). Such an association between structural strength and baseline band alterations in coherence did not arise at WP1, and thus distinguishes the two working points. For the high background drive state (WP3; Sec.~\ref{s:wp3}), endogenous oscillations are significantly less responsive to perturbations. Specifically, at this working point, stimulation no longer generates well-separated baseline and excited frequency bands, and, furthermore, the system-wide changes in phase-locking are more homogenous across different choices of the excited brain area. At WP3, we also found that the overall responses to perturbations become most strongly related to the stimulated region's structural strength. This reflects the idea that -- due to the enhanced role of locally-generated activity relative to network interactions at WP3 -- areas with stronger anatomical connections are those able to produce larger global responses when stimulated. In sum, these results indicate that depending on the collective state of network activity, the widespread impacts of stimulation can differ, and may be driven by distinct mechanisms. 

Before moving on to a further discussion of the results, broader implications, and limitations of this work, it is critical to state that this investigation is certainly not the first to examine the impacts of perturbations using computational models of brain dynamics. In fact, our investigation was motivated and inspired by a number of previous large-scale modeling studies that have uncovered key insights into either the network-wide \cite{Gollo2017:MappingHowLocal,Muldoon2016:StimulationBased,Spiegler2016:SelectiveActivation,Kunze2016:Transcranial} or state-dependent \cite{Alagapan2016:Modulation,Lefebvre2017:Stochastic,Li2017:Unified} influences of stimulation or perturbations to neural activity. For example, \cite{Gollo2017:MappingHowLocal} modeled slow blood-oxygen-level dependent signals using a simplified oscillator model, and by incorporating a realistic hierarchy of timescales, uncovered novel interplays between brain areas' intrinsic dynamics, structural connectivity, and the effects of regional stimulation on systems-level functional interactions. In another study published around the same time \cite{Spiegler2016:SelectiveActivation}, the authors employed neural mass modeling to investigate the transient responses of brain networks to brief stimulation, focusing in particular on the roles of both long-range and short-range anatomical connectivity in shaping the organization of stimulation-induced functional subnetworks. Because our study builds on its methodology, we also specifically emphasize the investigation by \textcite{Muldoon2016:StimulationBased}, where (as here) the authors used interconnected Wilson-Cowan units to simulate brain activity, and then utilized recent advances in network control theory \cite{Tang2018:Control} to make predictions about the overall functional effects of regional stimulation. Finally, we bring particular attention to the study conducted by \textcite{Alagapan2016:Modulation}, where the authors built a simplified model to examine the state-dependent effects of alternating current stimulation on local cortical oscillations. As part of their study, they found that the strength of endogenous oscillations altered the susceptibility of local dynamics to external stimulation. In the present investigation, we vary the background drive (the main tuning parameter in the WC model), which also modulates the amplitude of regional activity. Therefore, although our study is focused on the network-wide effects of focal perturbations, it is important to heed and connect our results to those previously reported in \cite{Alagapan2016:Modulation}. We try to point out such similarities in the ensuing discussion.

Although our work was heavily influenced by and builds upon these and other past efforts, it is also important to highlight some key distinctions and extensions of our analysis. First, in contrast to \cite{Gollo2017:MappingHowLocal}, we opted to use the more complex but in some ways more biophysically-motivated Wilson-Cowan system as the fundamental dynamical unit in the model, rather than a pure phase oscillator. Studying models that incorporate both phase and amplitude dynamics (for example, the WC model) will likely be critical for a more complete understanding of oscillatory neural activity, since variations in amplitude can affect functional couplings in the brain, are modulated by behavioral conditions, and are also elicited by various forms of stimulation \cite{Tewarie2018:Relationships,Canolty2010:TheFunctional,Jia2013:NoConsistent,Daffertshofer2011:OnTheInfluence,Brookes2011:Changes}. A second key difference between our study and those conducted in \cite{Muldoon2016:StimulationBased} and \cite{Spiegler2016:SelectiveActivation} is that the previously mentioned studies analyzed a situation in which brain activity was assumed to be in a state of low, non-oscillatory activity prior to any perturbation or stimulation. Here, however, we were interested particularly in baseline dynamical states corresponding to ongoing oscillatory activity, and the interaction between those baseline rhythms and increased excitation to a particular brain area. Finally, in an extension of prior work that has begun to examine the state-dependent impacts of stimulation on single cortical areas \cite{Alagapan2016:Modulation}, we specifically wished to consider a large-scale network model that allows for an analysis of how stimulation can spread to induce or modulate dynamical interactions between widespread brain areas. Hence, the key contribution of this study is its simultaneous investigation of \textit{(1)} not only the focal, but also the \textit{distributed} impacts of regional stimulation, and \textit{(2)} how the \textit{collective} regime of system activity as a whole (i.e. that arising from a combination of local dynamical properties and network coupling) influences such network-wide effects. To the best of our knowledge, it remains an open question how the collective state of brain activity alters the way focal stimulation impacts functional relations between distant brain areas.

In the firing-rate model implemented here, the general focal effect of stimulating a particular brain area was to increase both the amplitude and frequency of that region. Such changes in neural activity are broadly consistent with the effects of natural stimulation of certain cortical areas, such as visual stimuli impinging on visual cortex \cite{Jia2013:NoConsistent,Henrie2005:LFPPowerSpectra} and the effects of increased excitatory drive to neuronal populations generally \cite{Lowet2015:InputDependentFrequency}. We also found, though, that an excitation of the same strength had significantly stronger effects on local activity (i.e., the induced shift in power and frequency was larger) for states of lower background drive. Though the analogy is not perfectly direct, this result is undoubtedly akin to and consistent with the findings reported in \cite{Alagapan2016:Modulation}, wherein alternating current stimulation induced the strongest effects at the stimulation frequency in the absence of strong endogenous oscillations. However, moving beyond local responses, it is also critical to acknowledge that a given neuronal population exists in the context of a larger network of areas, such that local changes in activity can induce distributed effects \cite{Luft2014:Best,To2018:Changing,Shafi2012:Exploration,Polania2018:Studying}. In this regard, as perhaps one might expect, we found that in states of low background drive, downstream areas situated structurally close to the activated region developed peaks in their spectra at the frequency of the stimulated region. Hence, activity from the excited area is transferred along anatomical connections and becomes oscillatory input to other parts of the network, where it ``excites" a new spectral component in those regions' signals. In fact, the strong and regular oscillatory input from the directly driven area leaves spectral signatures in downstream regions that are qualitatively similar to those induced by rhythmic stimulation, where activity in an area stimulated at a particular frequency shows enhanced power at the stimulation frequency \cite{Ahrens2002:SpectralMixing,Alagapan2016:Modulation}. Furthermore, due to this direct propagation mechanism, the activity induced at the excited frequency was phase-locked across different units of the network.

More interesting, though, was that in states of relatively low background drive, selectively exciting a given brain area could also leave spectral fingerprints at downstream units' baseline frequencies. Generic broad-band alterations of spectra, beyond effects at the main spectral peak, can occur in coupled neuronal populations with long-range excitatory interactions \cite{Battaglia2007:TemporalDecorrelation}. Indeed, in one of only a few possible scenarios, mutually interacting nonlinear oscillators engage first into quasiperiodic dynamics, and then develop chaos, which is associated with spectral changes over continuous ranges \cite{Schuster2006:Deterministic}. Here, we also found that focal stimulation can lead to complex modulations of dynamical interactions between brain areas' ongoing rhythms, that in turn modify the temporal coordination of regional activity in a band containing units' baseline frequencies. Furthermore, in contrast to the excited band effects (which were highly constrained by anatomical connectivity), we found that for some working points of the model, the excited region's initial functional connectivity strength was more predictive of the baseline band effects. Generally, the notion that inputting additional energy to a single brain area can modify the interactions between ongoing oscillations of other units in the network is a critical point that highlights the complex, non-linear, and non-local effects of altering activity in one part of a larger system. Although the field has not reached a clear consensus on the distributed effects of stimulation, the general finding in our study that enhancing the input to a single brain area can induce or reorganize functional interactions is consistent with previous modeling work \cite{Muldoon2016:StimulationBased,Spiegler2016:SelectiveActivation}. We also make some new, specific predictions about different types of changes that can occur and their potential underlying mechanisms. Perhaps most interesting about these results is the suggestion that both network structure and coordinated dynamical organization may play a role in guiding the non-local effects of perturbations to neural activity. While network neuroscience has traditionally focused on how structure can predict function \cite{Honey2009:Predicting,Hermundstad2013:StructuralFoundations,Shen2015:NetworkStructure,Shen2015:StableLong,Avena-Koenigsberger2018:Communication}, the collective dynamics of a system need not be completely constrained by structure alone \cite{Bargmann2013:FromTheConnectome,Kopell2014:Beyond,Kirst2016:Dynamic,Battaglia2012:DynamicEffective,Bargmann2012:Beyond,Vazquez-Rodriguez2019:Gradients,Battaglia2020:Functional,Gutierrez2014:Modulation}. Our findings indicate that, as a consequence, the collective oscillatory state of the system could also be important in determining how a perturbation can lead to distributed changes in functional interactions. Specifically, we found here that in certain cases, functional connectivity was actually a better predictor of the overall response to perturbations than structural connectivity.

Another key finding of this study is that a local amplification of neural activity manifests differently -- both at the site of the excitation itself and across the broader brain network -- depending on the collective state of the system. Although we utilized simplified models to describe brain network dynamics and perturbations to neural activity, our results add to a growing body of literature on state-dependent stimulation \cite{Alagapan2016:Modulation,Lefebvre2017:Stochastic,Li2017:Unified,Bergmann2018:BrainState,Thut2017:Guiding,Silvanto2008:StateDependency,Neuling2013:Orchestrating,Ruhnau2016:EyesWide}. Brain state-dependent stimulation is an important but only fairly recently examined idea recognizing that the effects of an exogenous perturbation to a particular brain area can be conditional on the endogenous rhythmic or spontaneous activity of the system at the time of stimulation. For example, empirical studies have shown that outcomes of various types of stimulation can differ as a function of cognitive state (e.g., task \textit{vs.} rest) \cite{Thut2017:Guiding,Silvanto2008:StateDependency,Neuling2013:Orchestrating,Ruhnau2016:EyesWide,Li2019:BrainState}. Importantly, a few recent modeling studies have also begun to systematically investigate and provide mechanistic explanations for state-dependent responses to alternating current stimulation \cite{Alagapan2016:Modulation,Lefebvre2017:Stochastic,Li2017:Unified}. However, those studies presented dynamical models of only a single or a few coupled cortical and/or subcortical regions, and therefore focused mainly on the impacts of stimulation directly on the perturbed area. Here, in contrast, our goal was to consider the notion of state-dependence in the context of a whole-brain model, wherein the ``state" of the system is defined not only by the activity pattern or parameters of a single brain area, but also by the collective dynamics of the system as a whole. Our approach allows us to consider both the influence of large-scale anatomical connectivity as well as the dynamical regime of the network in shaping the local and widespread effects of regional alterations to neural activity. 

The idea that brain network responses to stimulation depend on the collective dynamical state also has implications for the control of brain dynamics. In neuroscience, stimulation has often been used to tease apart the function of a particular neural component and unveil its role in a larger context or understand its influences on other elements in a system. Going a step further, stimulation may also be used to intervene when brain dynamics go awry -- for example, in neurological diseases such as Parkinson's and epilepsy \cite{Schulz2013:Noninvasive,Johnson2013:Neuromodulation,Fisher2014:Electrical}. More generally, stimulation holds promise for controlling the brain out of and into specific dynamical configurations \cite{Tang2018:Control,Stiso2019:WhiteMatter,Khambhati2019:FunctionalControl}. However, realizations that both local and network-wide responses to stimulation can vary as a function of the endogenous state also imply that the dynamical regime of the brain should be considered when attempting to control network dynamics via stimulation. While there has been some recent progress \cite{Bergmann2018:BrainState}, the notion of state-dependent control of brain activity is a relatively nascent area of study, and therefore still lacks models that can provide insight into existing empirical results and aid our ability to make further predictions. Nonetheless, there have been efforts in disciplines outside of neuroscience to develop models and theory for the control of network dynamics that take into account context, or system state \cite{Skardal2015:Control,Lynn2016:Maximizing}. Although these ideas may not be immediately transferrable to the study of brain network dynamics, they could provide a useful foundation, and perhaps be tuned and adapted for application to the problem of controlling networked neural systems specifically.

Given that our study is based on computational modeling and simulation, it is also important to comment on the general utility of such approaches. Following a number of past efforts \cite{Honey2009:Predicting,Gollo2015:DwellingQuietly,Deco2009:KeyRoleofCoupling,Cabral2011:RoleOfLocal,Hlinka2012:UsingComputationalModels,Abeysuriya2018:ABiophysicalModel,Bansal2018:Personalized,Bansal2019:Cognitive,Ritter2013:TheVirtual,Sanz-Leon2015:Mathematical,Glomb2017:RestingState,Cabral2014:Exploring,Vuksanovic2015:Dynamic}, we constructed a whole-brain model as a network of neural masses coupled according to empirically-derived large-scale brain architecture. In these setups, reduced mean-field descriptions of neuronal population activity (of which a classic example is the Wilson-Cowan model) are adopted over detailed, biophysically-precise models of individual neurons \cite{Breakspear2017:DynamicModels,Coombes2019:NextGeneration}. The motivation behind this procedure is that, without dimensionality reduction, it would be essentially impossible to build and analyze a model of whole-brain dynamics due to the sheer number of components and interactions that would need to be included. However, while macroscale models of brain activity based on interconnected neural masses constitute a major simplification of the true system (see below for more on limitations), they still maintain the potential to posit, clarify, or rule out the operating principles behind empirical observations, make new predictions about untested phenomena, and provide general intuitions about the complexities of brain structure and dynamics. Indeed, there is now a relatively long history demonstrating that simplified models of large-scale brain activity, which necessarily abstract away many details, can nevertheless provide important insights into brain (dys)function \cite{Breakspear2017:DynamicModels,Murray2018:BiophysicalModeling}. For example, these models have been successfully applied to discern structure-function relationships in brain networks broadly \cite{Honey2009:Predicting}, describe mechanisms that could underlie specific phenomena such as the behavior of time-varying neural synchrony \cite{Hansen2015:FunctionalConnectivity} or the emergence of brain waves \cite{Roberts2019:Metastable}, and to also understand pathological activity patterns such as seizure spread \cite{Kameneva2017:NeuralMass,Baier2018:Design}. 

In the case of the present study, we used a biophysically-motivated neural mass model to make predictions about the focal and diffuse effects of stimulation to a particular brain area. While conjectures based on simplified models are useful and important, they must ultimately be tested empirically in order to substantiate whether or not they provide biologically meaningful predictions and explanations. In order to test the overall effects of perturbing different regions, multiple brain areas would need to be individually stimulated and their global responses compared. While these experiments would be a major undertaking, recent advances combining non-invasive brain stimulation techniques with measurement modalities like EEG and MEG \cite{Thut2009:NewInsights,Witkowski2016:Mapping,Antal2004:Oscillatory,Neuling2015:Friends,Siebner2009:Consensus} do make it possible to target multiple sites while also recording brain activity from distributed regions. These types of experiments -- which have already begun to be performed \cite{Thut2009:NewInsights,Witkowski2016:Mapping,Antal2004:Oscillatory,Neuling2015:Friends,Siebner2009:Consensus} -- are crucial in validating results from computational models concerning how activity changes not only in the directly activated region, but also how stimulation induces or alters functional couplings across different parts of the brain. Perhaps one of the most exciting testable predictions of the model is that focal stimulation propagates to cause downstream modulations of power and phase-locking at both the dominant frequency acquired by the activated area, but also at regions' baseline frequencies. Validating this prediction would require an experiment in which stimulation forces the perturbed area to oscillate at a well-defined frequency -- which could perhaps be achieved by alternating current stimulation \cite{Thut2012:TheFunctional} -- and a simultaneous measurement of other brain areas' dynamics. Second, it would be interesting to test how stimulation may differentially alter network dynamics when applied during qualitatively different brain states, such as during states of lower \textit{vs.} higher coherence between certain parts of the brain. It is very important to note that some studies have begun to test how behavioral state \cite{Alagapan2016:Modulation,Lefebvre2017:Stochastic,Li2017:Unified,Bergmann2018:BrainState,Thut2017:Guiding,Silvanto2008:StateDependency,Neuling2013:Orchestrating,Ruhnau2016:EyesWide} affects the focal outcomes of stimulation. However, it remains unclear how the dynamical state of the brain as whole mediates the widespread, network effects of stimulation, and especially, the influence of stimulation on functional interactions.

\subsection{Methodological limitations and future work}

There are a number of methodological considerations to comment on regarding this work. First, we used a relatively coarse-grained parcellation ($N = 82$ regions) to construct human structural brain networks. This resolution is roughly consistent with several other whole-brain modeling studies \cite{Muldoon2016:StimulationBased,Abeysuriya2018:ABiophysicalModel,Cabral2011:RoleOfLocal,Tewarie2018:Relationships,Cabral2014:Exploring}, and, importantly, allows us to run faster, and hence more, simulations. However, the parcellation we used also represents a significant simplification of the underlying anatomy, since the regions represent large pieces of neural tissue and hence remain agnostic to potentially important structural heterogeneities at finer scales. In addition, the limitations of human brain imaging and tractography preclude a perfect reconstruction of interareal connections \cite{Lazar2010:Mapping}. One primary drawback is that these methods cannot resolve the directedness of interareal connections, which could impact subsequent results \cite{Kale2018:Estimating}. Finally, we used a group-averaged connectome in this study. On the one hand, this simplification allowed us to focus on general trends and behaviors, but on the other hand, it leaves no room for examining how individual differences may affect certain findings. In future work, it will be interesting and important to explore how various results generalize to both higher-resolution and higher-quality brain data, and to understand how variability in brain structure across different human subjects \cite{Bansal2018:Personalized,Bansal2018:DataDriven} or even across different species \cite{Schmidt2018:MultiScale} may relate to differences in how perturbations of neural activity are expressed in system dynamics.

A second limitation of this study concerns the dynamical model used to simulate brain activity. We opted to use the Wilson-Cowan model \cite{Wilson1972:ExcitatoryandInhibitory}, which is a canonical neural mass system that embodies a tradeoff between biological realism and tractability. A key aspect of the WC model is that it generates oscillations, which are characteristic of and ubiquitously found in large-scale empirical brain recordings \cite{Buzsaki2006:RhythmsOfTheBrain}. Furthermore, in the WC model, oscillations are generated via the interaction of excitatory and inhibitory neuronal populations, which is considered a biophysically-plausible and likely mechanism underlying observed rhythmicity in brain dynamics. However, while the WC model does constitute a realistic improvement over simpler linear models, phase-oscillators, or generic Hopf bifurcation models, it is still phenomenological in nature. Future work should test whether results generalize to other neural mass models and whether various findings are mechanism-dependent. For simplicity and in line with past work \cite{Muldoon2016:StimulationBased,Abeysuriya2018:ABiophysicalModel,Hlinka2012:UsingComputationalModels,Roberts2019:Metastable,Honey2009:Predicting,Gollo2015:DwellingQuietly,Glomb2017:RestingState}, we also only considered long-range couplings between the excitatory populations of different brain areas. However, interareal connections likely target both excitatory and inhibitory neurons, and it would therefore be interesting to include long-range excitatory-to-inhibitory links. In this study we also used a fixed value for the signal propagation velocity. Although we attempted to choose an empirically constrained value, it is known that delays can have significant consequences on brain dynamics \cite{Deco2009:KeyRoleofCoupling,Petkoski2019:Transmission}, and so our results should be interpreted accordingly. 

Another important assumption of the model is that each unit has identical parameters. While this is a reasonable and useful setup to analyze first, there is a large body of literature detailing very specific heterogeneities across brain areas. For example, different regions may operate at different intrinsic time-scales \cite{Murray2014:AHierarchy}, have different intrapopulation architectures \cite{Murray2018:BiophysicalModeling} (e.g., excitatory or inhibitory coupling strengths), or exhibit rhythmic activity in several distinct frequency bands \cite{Bastos2015:Visual}. Recent modeling efforts have begun to incorporate some of these additional complexities, finding that doing so can lead to more realistic baseline dynamics and can explain certain empirically-observed behaviors not accounted for by simpler models \cite{Demirtas2018:Hierarchical,Mejias2016:FeedforwardAndFeedback}. In future work, it will be important to ask how the impacts of perturbations are affected when additional details about the heterogeneity of the underlying anatomy or dynamics of brain areas are included in models \cite{Gollo2017:MappingHowLocal}. Importantly, this would also allow for making more concrete statements about how different regions influence large-scale activity patterns. However, we also stress that the model implemented in the present work does not prevent the ensuing dynamics from being complex and rich, and actually allows us to appreciate how dynamical -- in additional to structural -- complexity can be a key driver of stimulation-induced effects. Finally, we note that we examined only a few representative working points of the model, out of the many that exist. Though our goal in this work was to illustrate that the collective state of the system can influence the effects of local perturbations, we do not claim to have provided an exhaustive description of all possible behaviors.

In addition to investigating alternative and/or more detailed models, forthcoming studies could also consider different ways of operationalizing perturbations to neural activity. Here, we built upon past work \cite{Muldoon2016:StimulationBased} and opted to invoke ``stimulation" in the simplest possible manner. Specifically, we increased the level of excitatory drive to the perturbed area, which has the consequence of raising the amplitude and frequency of the perturbed region. While these changes are generally consistent with the effects of certain types of stimulation \cite{Jia2013:NoConsistent,Henrie2005:LFPPowerSpectra}, one could also imagine modeling alternating current stimulation, for example. In this case, the stimulated area receives oscillatory input of a particular amplitude and frequency \cite{Reato2013:Effects}, both of which can be independently tuned. While this kind of stimulation has been examined in models of single cortical regions \cite{Alagapan2016:Modulation,Lefebvre2017:Stochastic,Li2017:Unified}, it would be exciting to scale up to systems-level models. One could also investigate alternative ways of modulating brain state. In this study, we varied the baseline dynamical regime in perhaps the most straightforward way possible, by tuning the level of excitatory input globally for all network elements. This variation changed the local dynamics of each brain area, and in turn, the macroscopic state of the system as a whole. However, brain state could also be modulated by tuning a different physiologically-interpretable parameter, such as the gain in the sigmoidal activation functions. Indeed, recent modeling studies have shown that altering neural gain can lead to dynamical regimes in which functional integration and segregation are balanced \cite{Shine2017:TheModulation}. It would also be interesting to understand how widespread changes of neural gain affect the way focal perturbations materialize. 

\section{Outlook}

Our goal in this work was to conduct an idealized investigation into the effects of focal stimulation on brain network dynamics, and, especially, how the collective state of the system may influence the distributed impacts of such perturbations. To the best of our knowledge, this latter point has only recently begun to be examined in the context of large-scale brain networks, and therefore warrants investigation via simple but biophysically-principled models. However, as noted already, the results of this study must be interpreted cautiously as they are yet to be compared against or validated by empirical data. As with any modeling study, this is an imperative step that must be taken before conclusive statements can be made. While there are now a handful of studies that interrogate the impacts of regional stimulation on brain network dynamics, a particularly challenging aspect for future work will be to test the question of state dependence. In this study, we varied the baseline dynamical regime of the model by smoothly tuning interpretable parameters in the same way for all network elements. Although this methodology made practical sense in the context of the model, the question of how to best translate this procedure to experiments remains unanswered. Future studies would benefit from a tight feedback loop between experiment and theory. In particular, one could carefully fit model parameters separately for different empirically-observed brain states, such that various aspects of simulated brain activity match experimental results. The effects of stimulation could then be re-investigated in the context of the highly-constrained biophysical model, and compared against empirical observations. Although here we conducted a more abstract and theoretical investigation, working towards increasingly realistic and experimentally testable models is an exciting direction for forthcoming studies.

\section{Funding}
This work was primarily supported by NSF BCS-1631550, ARO W911NF-18-1-0244, and NIH R01-MH -116920. DSB, LP, and CWL would like to acknowledge additional support from the John D. and Catherine T. MacArthur Foundation, the Alfred P. Sloan Foundation, the ISI Foundation, the Paul Allen Foundation, the Army Research Laboratory (W911NF-10-2-0022), the Army Research Office (Bassett-W911NF-14-1-0679, Grafton-W911NF-16-1-0474, DCIST-W911NF-17-2-0181), the Office of Naval Research, the National Institute of Mental Health (2-R01-DC-009209-11, R01-MH112847, R01-MH107235, R21-M MH-106799), the National Institute of Child Health and Human Development (1R01HD086888-01), National Institute of Neurological Disorders and Stroke (R01 NS099348), and the National Science Foundation (BCS-1441502, BCS-1430087, and NSF PHY-1554488). LP was also supported by a Graduate Research Fellowship from the National Science Foundation for part of this work. DB acknowledges support from the EU i-Innovative Training Network “i-CONN” (H2020 MSCA ITN 859937) and the French Agence Nationale pour la Recherche (“ERMUNDY”, ANR-18-CE37-0014-02). The content is solely the responsibility of the authors and does not necessarily represent the official views of any of the funding agencies.
\label{s:funding}

\section{Acknowledgments}
The authors would like to thank Richard Betzel and John Medaglia for providing assistance with the structural brain data. LP also thanks Panos Fotiadis and Keith Wiley for feedback on prior versions of this manuscript, as well as Jason Kim, Tatyana Gavrilchenko, Ann Sizemore Blevins, and Tanner Kaptanoglu for many helpful discussions.
\label{s:acknowledgments}

\section{Citation diversity statement}

Recent work in neuroscience and other fields has identified a bias in citation practices such that papers
from women and other minorities are under-cited relative to the number of such papers in the field \cite{Dworkin2020:TheExtent,Maliniak2013:TheGender,Caplar2017:QuantitativeEvaluation,Chakravartty2018:Communication,Dion2018:GenderedCitation,Thiem2018:JustIdeas}. Here we sought to proactively consider choosing references that reflect the diversity of the field in thought, form of contribution, gender, race, geographic location, and other factors. We used automatic classification of gender based on the first names of the first and last authors \cite{Dworkin2020:TheExtent}, with possible combinations including male/male, male/female, female/male, female/female. Excluding self-citations to the senior authors of our current paper, the references contain 58\% male/male, 10\% male/female, 25\% female/male, 7\% female/female. We look forward to future work that could help us to better understand how to support equitable practices in science.

\bibliography{WC_relations.bib}

\end{document}


\begin{flushleft}
	{\huge{SUPPLEMENTARY TEXT} \newline\newline\newline\newline
		\Large{\bf{``Relations between large-scale brain connectivity and effects of regional stimulation depend on collective dynamical state"}} 
	}
\newline
\\
Lia Papadopoulos\textsuperscript{1},
Christopher W. Lynn\textsuperscript{1},
Demian Battaglia\textsuperscript{2},
Danielle S. Bassett\textsuperscript{1,3,4,5,6,7*},
\\
\bigskip
\textbf{1} Department of Physics \& Astronomy, University of Pennsylvania, Philadelphia, PA 19104, USA
\\
\bigskip
\textbf{2} Institute for Systems Neuroscience, University Aix-Marseille, Boulevard Jean Moulin 27, 13005 Marseille, France
\\
\bigskip
\textbf{3} Department of Bioengineering, University of Pennsylvania, Philadelphia, PA 19104, USA
\\
\bigskip
\textbf{4} Department of Electrical \& Systems Engineering, University of Pennsylvania, Philadelphia, PA 19104, USA
\\
\bigskip
\textbf{5} Department of Neurology, University of Pennsylvania, Philadelphia, PA 19104, USA
\\
\bigskip
\textbf{6} Department of Psychiatry, University of Pennsylvania, Philadelphia, PA 19104, USA
\\
\bigskip
\textbf{7} Santa Fe Institute, Santa Fe, NM 87501, USA
\\
\bigskip
* dsb@seas.upenn.edu

\end{flushleft}

\newpage

\section{Dynamics of an isolated Wilson-Cowan unit}
\label{s:isolatedWC}

In this section, we briefly describe the behavior of a single Wilson-Cowan unit, which forms the fundamental dynamical component of the whole-brain computational model. An isolated Wilson-Cowan unit evolves according to Eqs. 1 and 2, with the coupling term set to $C = 0$. Here we take all other model parameters to be those displayed in Table 1, with the exception that we consider noiseless simulations for the purpose of demonstration.

\begin{figure*}
	\centering
	\includegraphics[width=\textwidth]{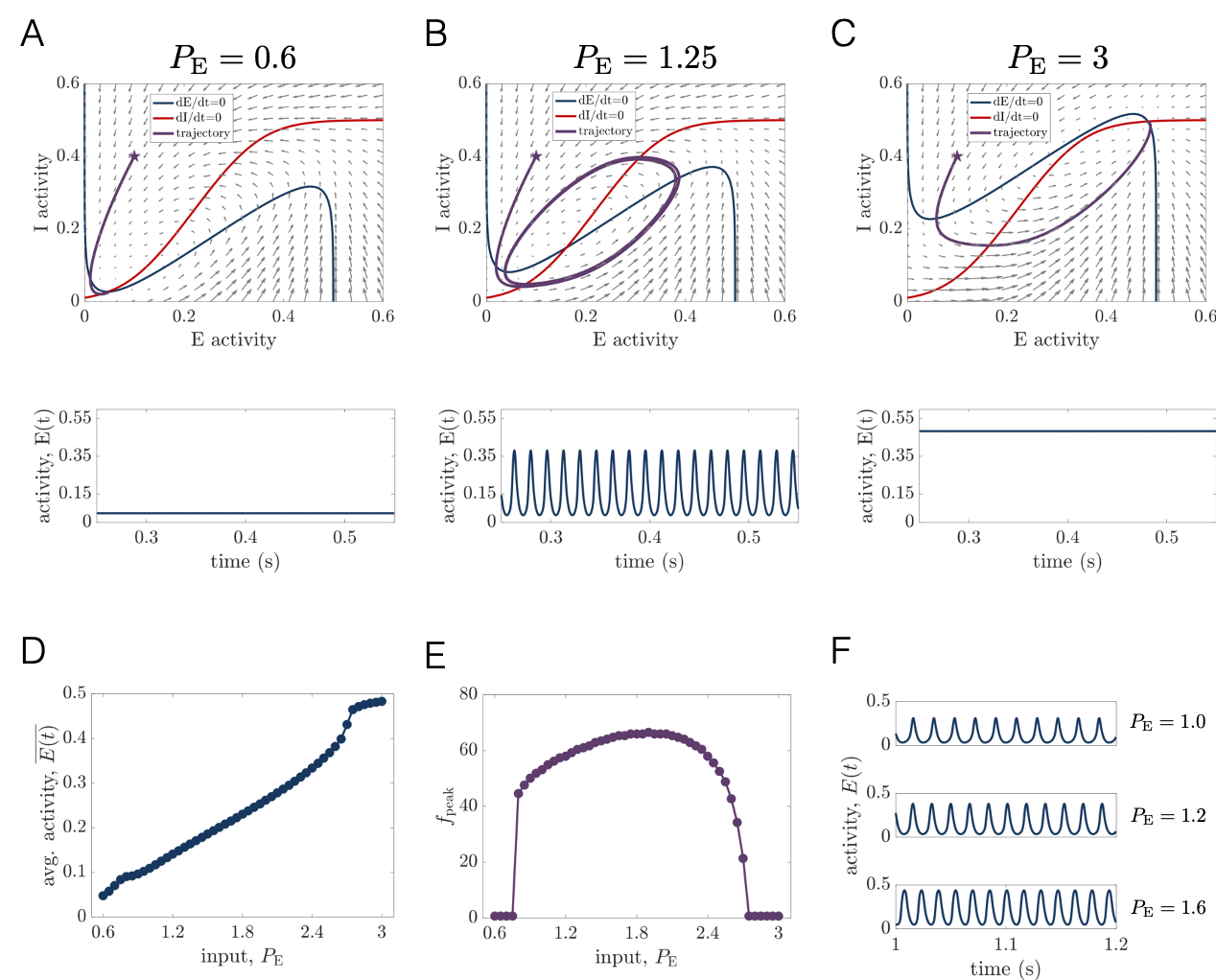}
	\caption{\textbf{Behavior of an isolated Wilson-Cowan unit.} \textit{(A-C)} Phase plane representations (top) and excitatory time series (bottom) for an isolated Wilson-Cowan unit subject to varying levels of excitatory drive: $P_{\mathrm{E}} = 0.6$ (\textit{A}), $P_{\mathrm{E}} = 1.25$ (\textit{B}), and $P_{\mathrm{E}} = 3.0$ (\textit{C}). \textit{(D)} The time-average of the excitatory firing-rate $\overline{E(t)}$ as a function of the input $P_{\mathrm{E}}$ for a single Wilson-Cowan unit. \textit{(E)} The peak frequency $f_{\mathrm{peak}}$ of the excitatory activity as a function of the input $P_{\mathrm{E}}$ for a single Wilson-Cowan unit. \textit{(F)} Examples of excitatory firing-rate activity for three different values of the input ($P_{\mathrm{E}} = \{1.0, 1.2, 1.6\}$ from top to bottom) that place the system in the oscillatory regime.}
	\label{f:isolated_WC_unit}
\end{figure*}

A typical bifurcation parameter for the Wilson-Cowan model is the excitatory drive $P_{\mathrm{E}}$. When other parameters are appropriately tuned, varying $P_{\mathrm{E}}$ can induce oscillatory activity in the $E$ and $I$ neuronal sub-populations. The top row of Figs. \ref{f:isolated_WC_unit}A-C show phase plane representations of a single Wilson-Cowan unit for three different levels of the excitatory input $P_{\mathrm{E}}$, and the bottom row of each panel shows the time-evolution of the excitatory activity $E(t)$ for the given parameter value. In the phase planes, the blue lines correspond to the excitatory variable nullcline ($dE/dt = 0$), the red lines correspond to the inhibitory variable nullcline ($dI/dt = 0$), and the purple lines show an example trajectory that begins at the point denoted by the star. For a low input level of $P_{\mathrm{E}} = 0.6$ (panel \textit{A}), the system has a single stable fixed-point corresponding to a low activity steady-state. For an intermediate drive of $P_{\mathrm{E}} = 1.25$ (panel \textit{B}), the system exhibits a stable limit cycle and the firing-rate activity oscillates in time. For a high input of $P_{\mathrm{E}} = 3$ (panel \textit{C}), the system again exhibits a single stable fixed-point, but corresponding to a high activity steady-state. 

To summarize the effect of the external input $P_{\mathrm{E}}$ on the behavior of an isolated Wilson-Cowan unit, we first plot the time-average of the excitatory firing-rate $\overline{E(t)}$ as a function of the input $P_{\mathrm{E}}$ (Fig. \ref{f:isolated_WC_unit}D). Note that $\overline{E(t)}$ increases monotonically with increasing drive. Second, we consider how the peak frequency $f_{\mathrm{peak}}$ of the excitatory activity varies with the input level $P_{\mathrm{E}}$ (Fig. \ref{f:isolated_WC_unit}E). The peak frequency was determined by finding the frequency at which the Welch's power spectral density of the excitatory time-series was maximum (see Sec. IIE for details). For low inputs, $f_{\mathrm{peak}}$ is approximately zero; the system resides in the low activity steady-state and there are no intrinsic oscillations. As the input is increased, though, oscillations emerge with frequencies in the gamma range (30 - 80 Hz). Increasing the input $P_{\mathrm{E}}$ in the oscillatory regime first raises the peak frequency from its initial value up to $\sim$65Hz. Such an increase in oscillatory frequency with increasing drive has also been found experimentally \cite{Henrie2005:LFPPowerSpectra,Lowet2015:InputDependentFrequency,Roberts2013:Robust,Jia2013:NoConsistent}. However, beyond a certain point, further increasing the excitatory drive causes the peak frequency to decline back to zero as the system approaches the high activity fixed point where oscillations again cease completely. In Fig. \ref{f:isolated_WC_unit}F, we show excitatory activity time-series for three different values of the input $P_{\mathrm{E}}$ that place the system in the oscillatory regime. It is clear by eye that for these parameters, increasing the excitatory drive increases the amplitude and frequency of the oscillations.

\section{Determining the onset of oscillatory activity in the whole-brain model}
\label{s:phaseSpace_boundary}

\begin{figure*}
	\centering
	\includegraphics[width=0.5\textwidth]{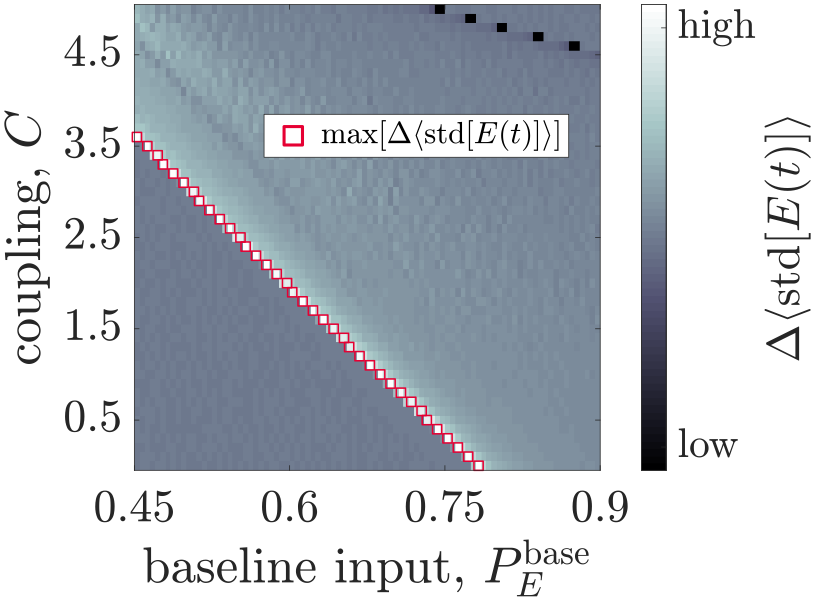}
	\caption{\textbf{The change in the network-averaged standard deviation of excitatory activity $\Delta \langle\mathrm{std}[E_{i}(t)]\rangle$ as a function of the global coupling $C$ and the level of non-specific baseline input $P_{\mathrm{E}}^{\mathrm{base}}$.} The change in $\langle\mathrm{std}[E_{i}(t)]\rangle$ is computed by taking the difference of this quantity between consecutive values of the input $P_{\mathrm{E}}^{\mathrm{base}}$, while holding the coupling $C$ fixed. The red squares denote the value of $P_{\mathrm{E}}^{\mathrm{base}}$ (for a given coupling $C$) at which $\Delta \langle\mathrm{std}[E_{i}(t)]\rangle$ is largest. These points serve to delineate a transition in regional brain dynamics from a state of low, non-oscillatory activity to a state of oscillatory activity.}
	\label{f:boundary_phaseSpace}
\end{figure*}

In order to systematically determine the boundary marking the onset of oscillatory activity as a function of the background drive $P_{\mathrm{E}}^{\mathrm{base}}$ and the coupling $C$, we examined the network-averaged standard deviation of the firing rate, $\langle \mathrm{std}[E_{i}(t)] \rangle$. This quantity captures the strength of fluctuations of the excitatory population activities around their mean values. Thus, by noting when $\langle \mathrm{std}[E_{i}(t)] \rangle$ jumps from a value near zero to a higher, positive value, we can qualitatively determine the transition from the state of low, non-oscillatory firing-rates to the onset of rhythmic dynamics in regional activity. Here, we are interested in finding the level of background excitation $P^{*}_{\mathrm{E}}(C)$ that is needed to induce oscillations at each brain area for a given coupling $C$. To determine these ``boundary" points $P^{*}_{\mathrm{E}}(C)$, we thus hold $C$ fixed, and consider the difference in $\langle \mathrm{std}[E_{i}(t)] \rangle$ between consecutive values of $P_{\mathrm{E}}^{\mathrm{base}}$. We plot this difference $\Delta\langle \mathrm{std}[E_{i}(t)] \rangle$ as a function of $P_{\mathrm{E}}^{\mathrm{base}}$ and $C$ in Fig. ~\ref{f:boundary_phaseSpace}, where we indeed observe a sharp boundary (i.e., a large increase in the network-averaged standard deviation as the background input is increased). We define the boundary points by finding the value $P^{*}_{\mathrm{E}}(C)$ at which the difference $\Delta\langle \mathrm{std}[E_{i}(t)] \rangle$ is maximized, and repeating this proceess for each value of the coupling $C$. We mark the boundary corresponding to the onset of oscillatory activity with red squares in Fig. ~\ref{f:boundary_phaseSpace}. 

\section{Weighted distances for quantifying the spatial spread of phase-locking changes}
\label{s:weighted_distances}

\begin{figure*}
	\centering
	\includegraphics[width=0.75\textwidth]{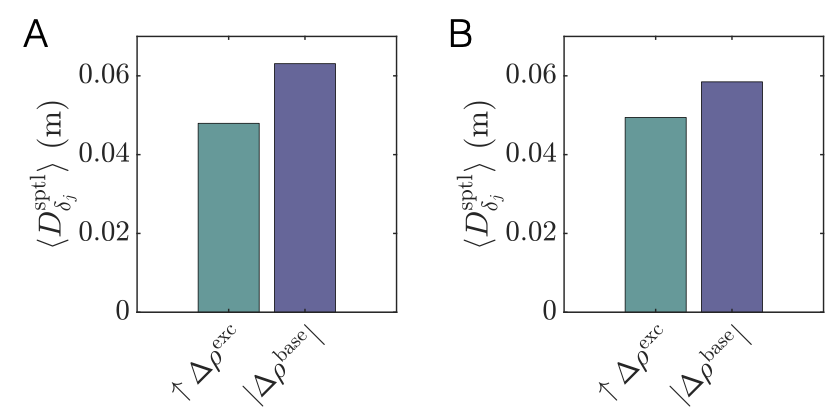}
	\caption{\textbf{Comparison of the spatial spread of phase-locking changes in the excited and baseline frequency bands.} \textit{(A)} Network-average of the mean spatial distance from the location of the excited region $j$ to regions $k$ and $l$, weighted by the normalized PLV change that region $j$ induces between region $k$ and region $l$ at WP1; bars represent the mean of the weighted average of the distance across all choices of the perturbed unit, and different colors represent weighting the distance by different phase-locking modulations. \textit{(B)} Network-average of the mean spatial distance from the location of the excited region $j$ to regions $k$ and $l$, weighted by the normalized PLV change that region $j$ induces between region $k$ and region $l$ at WP2; bars represent the mean of the weighted average of the distance across all choices of the perturbed unit, and different colors represent weighting the distance by different phase-locking modulations.}
	\label{f:distance_PLVmodulations}
\end{figure*}

In this section, we investigate how phase-locking changes are spatially-distributed for modulations induced either in the baseline or in the excited frequency band. Accordingly, we examine a weighted-distance measure $D^{\mathrm{sptl}}_{\delta_j}$ that captures the average spatial extent of phase-coherence modulations relative to the site of the perturbation. This quantity is defined explicitly as

\begin{equation}
D^{\mathrm{sptl}}_{\delta_j} = \frac{\sum_{kl} (D_{kl,\delta_j}) (\Delta \rho_{kl,\delta_j})}{\sum_{kl} \Delta \rho_{kl,\delta_j}},
\label{eq:PLVweighted_distance}
\end{equation}

\noindent where $D_{kl,\delta_j}$ is the average Euclidean distance from the excited area $j$ to regions $k$ and $l$, and where $\Delta \rho_{kl,\delta_j}$ is the change in phase-locking induced between regions $k$ and $l$ due to an excitation of region $j$. Note, then, that $D^{\mathrm{sptl}}_{\delta_j}$ is simply a weighted average of the mean spatial distance from the excited region $j$ to a pair of regions $k$ and $l$, where the weights are given by the phase-coherence modulation between $k$ and $l$ induced by driving region $j$. Here we separately consider the distance measure $D^{\mathrm{sptl}}_{\delta_j}$ using two different types of phase-locking changes: $\uparrow \Delta \rho_{kl,\delta_j}^{\mathrm{exc}}$ and $| \Delta \rho_{kl,\delta_j}^{\mathrm{base}} |$. More specifically, in Fig.~\ref{f:distance_PLVmodulations} we show the average distance $\langle D^{\mathrm{sptl}}_{\delta_j} \rangle$ computed over all $N$ choices of the excited brain area; panel \textit{(A)} shows the two distances for WP1 and panel \textit{(B)} shows the distances for WP2. For both working points, this weighted-distance measure suggests that phase-coherence modulations that occur in the baseline frequency band tend to be more spatially-distributed than those arising in the excited frequency band.

\section{Average responses to perturbations in the baseline frequency band at WP1 vs. WP3}
\label{s:WP1_vs_WP3}

\begin{figure*}
	\centering
	\includegraphics[width=0.45\textwidth]{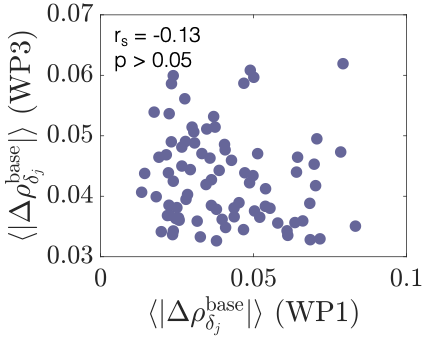}
	\caption{\textbf{Network-averaged absolute change in baseline band phase-coherence $\langle | \Delta \rho_{\delta_j}^{\mathrm{base}} | \rangle$ plotted at WP1 \textit{vs.} WP3.} Each point corresponds to a different choice of the stimulated region.}
	\label{f:WP1_vs_WP3}
\end{figure*}

To investigate how the dynamical state of the brain network model influences the effects of focal stimulation, we consider the relationship between the global responses to stimulation at two different working points. In particular, we examine the average absolute change in phase-locking in the baseline frequency band, $\langle | \Delta \rho_{\delta_j}^{\mathrm{base}} | \rangle$, at WP3 \textit{vs.} at WP1 (Fig.~\ref{f:WP1_vs_WP3}). Recall that WP1 corresponds to a low background drive working point preceding peak baseline coherence (Sec. IIIC), whereas WP3 corresponds to a high background drive working point following peak baseline coherence (Sec. IIIE). It is clear upon visual inspection of Fig.~\ref{f:WP1_vs_WP3} that there is no consistent relationship between these two quantities. Hence, regions that induce a large response at the system's spontaneous frequencies at WP1 are not necessarily those that induce a large response at WP3. This finding supports the notion that the dynamical state of network activity impacts how local perturbations manifest and alter functional interactions.

\section{Verification of relationships between phase-locking modulations and structural or functional connectivity at different working points in the low, medium, and high background drive regimes}
\label{s:alternate_WPs}

In the main text, we studied the effects of focal stimulation at three different working points of the brain network model (WP1, WP2, and WP3). At each of those states, we also considered the associations between the average change in phase-locking induced by regional perturbations (within both the baseline and excited frequency bands) and structural or dynamical (functional) network properties of the stimulated region. We found that, depending on the baseline state of the system, different relationships emerged between the perturbation-induced responses and structural or functional node strengths (see Figs. 5, 7, and 9). In this section, we verify that qualitatively similar relationships hold for other working points in the immediate vicinity of those studied in the main text. Note that for each alternative working point, we consider the same excitation strength used originally (i.e., $\Delta P_{\mathrm{E},j} = 0.1$). 

\begin{figure*}
	\centering
	\includegraphics[width=0.9\textwidth]{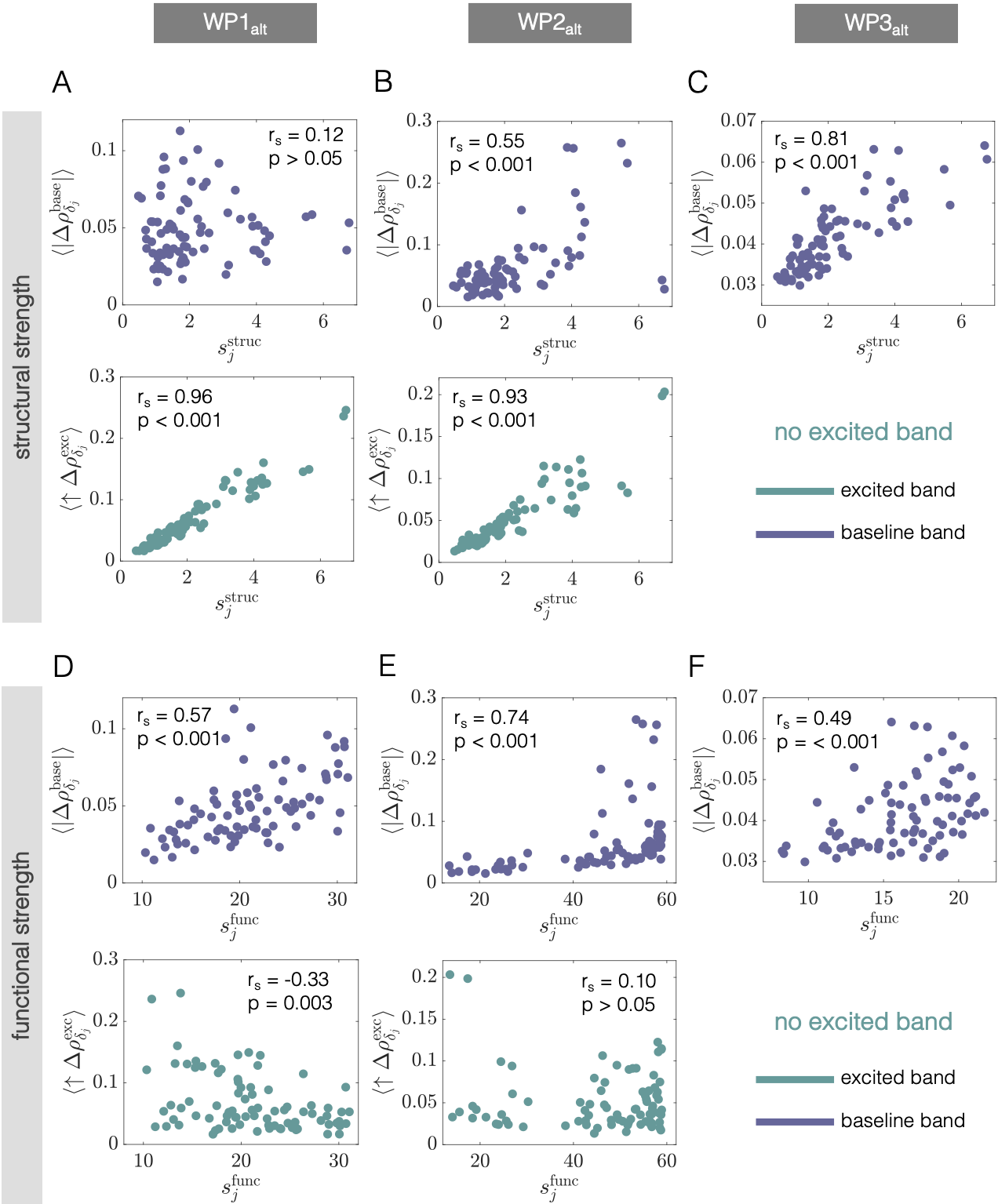}
	\caption{\textbf{Associations between stimulation-induced modulations in phase-locking and structural or functional connectivity hold for alternative working points near WP1, WP2, and WP3.} \textit{(A)} At $\textrm{WP1}_{\textrm{alt}}$, the quantity $\langle | \Delta \rho^{\mathrm{base}}_{\delta_j} | \rangle $ \textit{vs.} structural node strength $s^{\mathrm{struc}}_{j}$ (Top) and the quantity $\langle \uparrow \Delta \rho^{\mathrm{exc}}_{\delta_j} \rangle $ \textit{vs.} structural node strength $s^{\mathrm{struc}}_{j}$ (Bottom).  \textit{(B)} At $\textrm{WP2}_{\textrm{alt}}$, the quantity $\langle | \Delta \rho^{\mathrm{base}}_{\delta_j} | \rangle $ \textit{vs.} structural node strength $s^{\mathrm{struc}}_{j}$ (Top) and the quantity $\langle \uparrow \Delta \rho^{\mathrm{exc}}_{\delta_j} \rangle $ \textit{vs.} structural node strength $s^{\mathrm{struc}}_{j}$ (Bottom). \textit{(C)} At $\textrm{WP3}_{\textrm{alt}}$, the quantity $\langle | \Delta \rho^{\mathrm{base}}_{\delta_j} | \rangle $ \textit{vs.} structural node strength $s^{\mathrm{struc}}_{j}$ (Top). For this working point, there is no excited frequency band. \textit{(D)} At $\textrm{WP1}_{\textrm{alt}}$, the quantity $\langle | \Delta \rho^{\mathrm{base}}_{\delta_j} | \rangle $ \textit{vs.} functional node strength $s^{\mathrm{func}}_{j}$ (Top) and the quantity $\langle \uparrow \Delta \rho^{\mathrm{exc}}_{\delta_j} \rangle $ \textit{vs.} functional node strength $s^{\mathrm{func}}_{j}$ (Bottom). \textit{(E)} At $\textrm{WP2}_{\textrm{alt}}$, the quantity $\langle | \Delta \rho^{\mathrm{base}}_{\delta_j} | \rangle $ \textit{vs.} functional node strength $s^{\mathrm{func}}_{j}$ (Top) and the quantity $\langle \uparrow \Delta \rho^{\mathrm{exc}}_{\delta_j} \rangle $ \textit{vs.} functional node strength $s^{\mathrm{func}}_{j}$ (Bottom). \textit{(F)} At $\textrm{WP3}_{\textrm{alt}}$, the quantity $\langle | \Delta \rho^{\mathrm{base}}_{\delta_j} | \rangle $ \textit{vs.} functional node strength $s^{\mathrm{func}}_{j}$ (Top). For this working point, there is no excited frequency band. In all panels, insets indicate Spearman correlation coefficients between the plotted quantities, and their associated $p$-values.}
	\label{f:PLVmodulations_alternateWPs}
\end{figure*}

We begin by analyzing an alternative working point near WP1, which we term $\textrm{WP1}_{\textrm{alt}}$. WP1 was located at $P_{E}^{\mathrm{base}} = 0.553$ and $C = 2.5$; for $\textrm{WP1}_{\textrm{alt}}$, we consider parameters $P_{E}^{\mathrm{base}} = 0.555$ and $C = 2.5$. Note that because peak global baseline coherence $\rho^{\mathrm{global}}$ is reached rapidly as a function of $P_{E}^{\mathrm{base}}$ (see Fig. 3), in order to consider a second working point located prior to $\rho^{\mathrm{global}}$ but still near WP1, we can only shift $P_{E}^{\mathrm{base}}$ slightly from its value at WP1. We find the same set of relationships between phase-locking modulations and structural or functional strength at $\textrm{WP1}_{\textrm{alt}}$ as we did at WP1 (see Fig.~\ref{f:PLVmodulations_alternateWPs}A,D). Specifically: \textit{(1)} the average phase-locking induced in the excited frequency band $\langle \uparrow \Delta \rho_{\delta_j}^{\mathrm{exc}} \rangle$ is most strongly associated with the structural strength of the perturbed region $s_{j}^{\mathrm{struc}}$ (Fig.~\ref{f:PLVmodulations_alternateWPs}A,Bottom), and \textit{(2)} the average absolute phase-locking modulation in the baseline frequency band $\langle | \Delta \rho_{\delta_j}^{\mathrm{base}} |  \rangle$ is most strongly associated with the functional strength of the perturbed region $s_{j}^{\mathrm{func}}$ (Fig.~\ref{f:PLVmodulations_alternateWPs}D,Top). The relationships between $\langle | \Delta \rho_{\delta_j}^{\mathrm{base}} |  \rangle$ and $s_{j}^{\mathrm{struc}}$, and between $\langle \uparrow \Delta \rho_{\delta_j}^{\mathrm{exc}}  \rangle$ and $s_{j}^{\mathrm{func}}$, are either not statistically significant or weaker, respectively (Fig.~\ref{f:PLVmodulations_alternateWPs}A,Top and Fig.~\ref{f:PLVmodulations_alternateWPs}D, Bottom).

We next analyze an alternative working point near WP2, $\textrm{WP2}_{\textrm{alt}}$. Recall that WP2 was located at $P_{E}^{\mathrm{base}} = 0.57$ and $C = 2.5$; for $\textrm{WP2}_{\textrm{alt}}$, we consider $P_{E}^{\mathrm{base}} = 0.572$ and $C = 2.5$. In order to examine a second working point in close vicinity of the peak in global baseline coherence -- which was the condition used to determine parameters for WP2 -- we must consider only a small change in $P_{E}^{\mathrm{base}}$ away from its value at WP2. This is again because the dynamical state of the system changes quickly as a function of $P_{E}^{\mathrm{base}}$ in this regime (see Fig. 3A). Using the specified parameter choices, we find consistent relationships at WP2 and $\textrm{WP2}_{\textrm{alt}}$ in terms of how perturbation-induced phase-locking modulations are related to structural and functional node strength. First, the average phase-locking induced in the baseline frequency band $\langle | \Delta \rho_{\delta_j}^{\mathrm{base}} |  \rangle$ is significantly correlated with the structural strength of the perturbed region $s_{j}^{\mathrm{struc}}$ (Fig.~\ref{f:PLVmodulations_alternateWPs}B,Top), but remains most strongly related to functional strength $s_{j}^{\mathrm{func}}$ (Fig.~\ref{f:PLVmodulations_alternateWPs}E,Top). Second, the average phase-locking induced in the excited frequency band $\langle \uparrow \Delta \rho_{\delta_j}^{\mathrm{exc}} \rangle$ is strongly associated with the structural strength of the perturbed region (Fig.~\ref{f:PLVmodulations_alternateWPs}B,Bottom), and is not significantly correlated with functional strength (Fig.~\ref{f:PLVmodulations_alternateWPs}E,Bottom). 

Lastly, we analyze an alternative working point near WP3, $\textrm{WP3}_{\textrm{alt}}$. WP3 was located at $P_{E}^{\mathrm{base}} = 0.7$ and $C = 2.5$; for $\textrm{WP3}_{\textrm{alt}}$, we consider $P_{E}^{\mathrm{base}} = 0.68$ and $C = 2.5$. We once more find that the relationships between phase-locking modulations induced by regional stimulation and structural or functional node strength are consistent across WP3 and $\textrm{WP3}_{\textrm{alt}}$ (see Fig.~\ref{f:PLVmodulations_alternateWPs}C,F). In particular, there is a strong positive correlation between the average phase-locking induced in the baseline frequency band $\langle | \Delta \rho_{\delta_j}^{\mathrm{base}} |  \rangle$ and the structural strength of the perturbed region $s_{j}^{\mathrm{struc}}$ (Fig.~\ref{f:PLVmodulations_alternateWPs}C,Top). The relationship between $\langle | \Delta \rho_{\delta_j}^{\mathrm{base}} |  \rangle$ and the functional strength $s_{j}^{\mathrm{func}}$ is weaker (Fig.~\ref{f:PLVmodulations_alternateWPs}F,Top). Finally, note that there is no excited frequency band at WP3 or $\textrm{WP3}_{\textrm{alt}}$.

\section{Effects of perturbation strength}
\label{s:perturbation_strength}

\begin{figure*}
	\centering
	\includegraphics[width=0.6\textwidth]{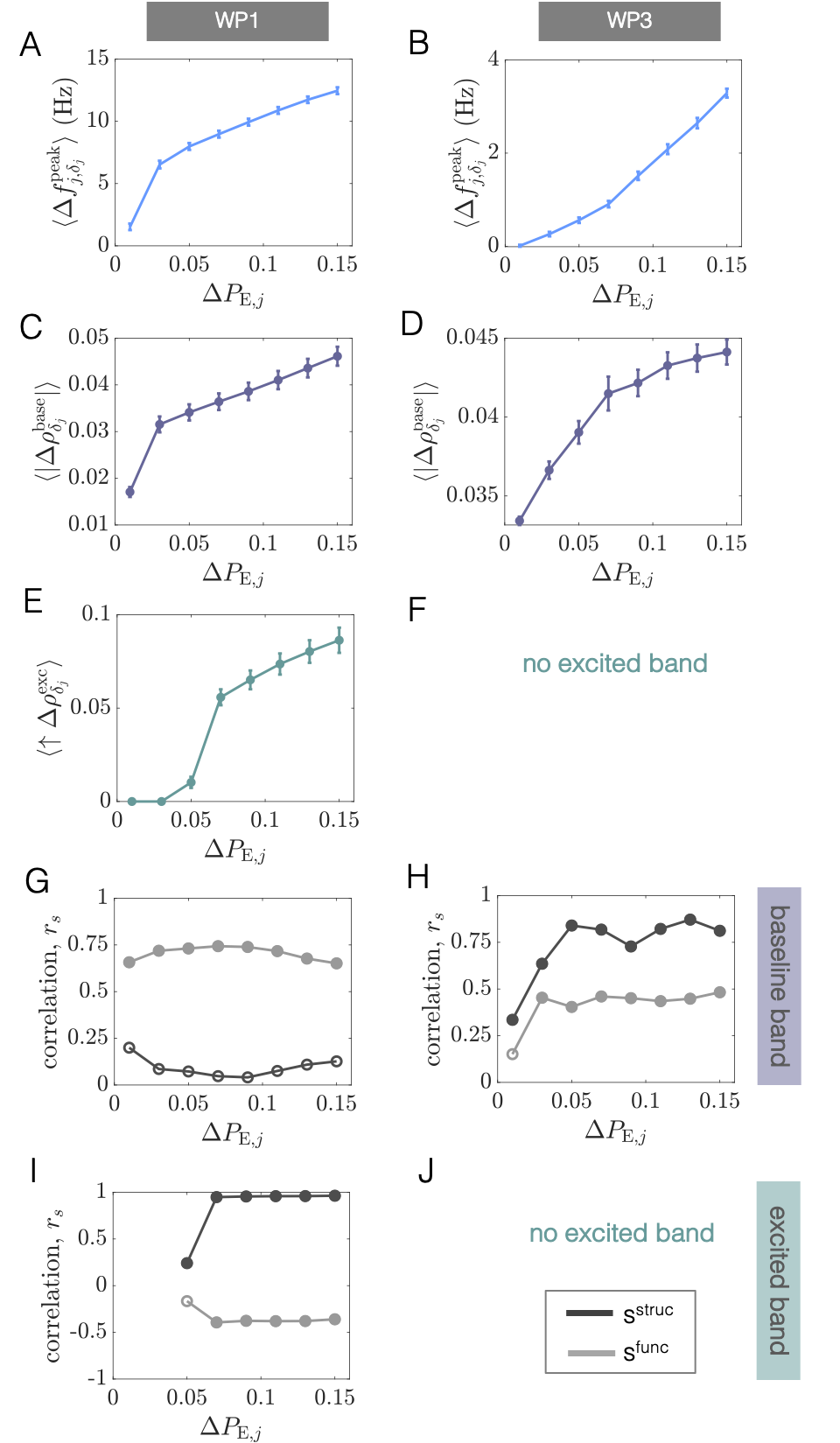}
	\caption{\textbf{Effects of varying the excitation strength at WP1 and WP3.} \textit{(A)} The average shift in peak frequency of the stimulated region $ \langle \Delta f_{j,\delta_j}^{\mathrm{peak}} \rangle$ \textit{vs.} $\Delta P_{\mathrm{E},j}$ at WP1. \textit{(B)} The average shift in peak frequency of the stimulated region $ \langle \Delta f_{j,\delta_j}^{\mathrm{peak}} \rangle$ \textit{vs.} $\Delta P_{\mathrm{E},j}$ at WP3. \textit{(C)} Network-averaged absolute PLV change in the baseline band $\langle | \Delta \rho^{\mathrm{base}}_{\delta_j} | \rangle $ \textit{vs.} $\Delta P_{\mathrm{E},j}$ at WP1. \textit{(D)} Network-averaged absolute PLV change in the baseline band $\langle | \Delta \rho^{\mathrm{base}}_{\delta_j} | \rangle $ \textit{vs.} $\Delta P_{\mathrm{E},j}$ at WP3. \textit{(E)} Network-averaged positive PLV change in the excited band $\langle \uparrow \Delta \rho^{\mathrm{exc}}_{\delta_j}  \rangle $ \textit{vs.} $\Delta P_{\mathrm{E},j}$ at WP1. \textit{(F)} No excited frequency band emerges at WP3 for any of the considered perturbation strengths. \textit{(G)} At WP1, the Spearman correlation $r_{s}$ between \textit{(1)} $\langle | \Delta \rho^{\mathrm{base}}_{\delta_j} | \rangle $ and structural strength $s^{\mathrm{struc}}_{j}$ (dark gray) or \textit{(2)} between $\langle | \Delta \rho^{\mathrm{base}}_{\delta_j} | \rangle $ and functional strength $s^{\mathrm{func}}_{j}$ (light gray), plotted as a function of $\Delta P_{\mathrm{E},j}$. Filled-in circles indicate that the correlation is statistically significant at $p<0.05$. \textit{(H)} At WP3, the Spearman correlation $r_{s}$ between \textit{(1)} $\langle | \Delta \rho^{\mathrm{base}}_{\delta_j} | \rangle $ and structural strength $s^{\mathrm{struc}}_{j}$ (dark gray) or \textit{(2)} between $\langle | \Delta \rho^{\mathrm{base}}_{\delta_j} | \rangle $ and functional strength $s^{\mathrm{func}}_{j}$ (light gray), plotted as a function of $\Delta P_{\mathrm{E},j}$. Filled-in circles indicate that the correlation is statistically significant at $p<0.05$.  \textit{(I)} At WP1, the Spearman correlation $r_{s}$ between \textit{(1)} $\langle \uparrow \Delta \rho^{\mathrm{exc}}_{\delta_j}  \rangle $ and structural strength $s^{\mathrm{struc}}_{j}$ (dark gray) or \textit{(2)} between $\langle \uparrow \Delta \rho^{\mathrm{exc}}_{\delta_j} \rangle $ and functional strength $s^{\mathrm{func}}_{j}$ (light gray), plotted as a function of $\Delta P_{\mathrm{E},j}$. Filled-in circles indicate that the correlation is statistically significant at $p<0.05$.}
	\label{f:vary_stimStrength}
\end{figure*}

In the main text, we studied a single excitation strength of $\Delta P_{\mathrm{E},j} = 0.1$. In this section, we assess the dependence of various results on the level of additional excitatory input $\Delta P_{\mathrm{E},j}$ received by the perturbed unit (Fig.~\ref{f:vary_stimStrength}). In particular, for both WP1 and WP3, we vary $\Delta P_{\mathrm{E},j}$ between 0.01 and 0.15 in steps of 0.02.

We first analyze how the perturbation strength affects the shift in the peak frequency of the stimulated area. As a summary measure, we consider the change in peak frequency averaged across all choices of the stimulated area, $ \langle \Delta f_{j,\delta_j}^{\mathrm{peak}} \rangle$. As expected, this quantity increases with increasing excitation strength for both WP1 (Fig.~\ref{f:vary_stimStrength}A) and WP3 (Fig.~\ref{f:vary_stimStrength}B). We next study the stimulation-induced changes in phase-locking in the baseline frequency band as a function of the excitation strength. In particular, we examine the mean change in coherence $\langle | \Delta \rho^{\mathrm{base}} | \rangle$, where the average is computed first over all pairs of brain areas for a given stimulation site, and then across all choices of the perturbed region. For both working points, this measure also increases monotonically as a function of $\Delta P_{\mathrm{E},j}$ (Fig.~\ref{f:vary_stimStrength}C,D). Hence, as the strength of the stimulation increases, so does the overall amount of functional reconfiguration at the system's baseline frequencies. For WP1, we find that the average of the positive changes in phase-locking in the excited band also grows as a function of the perturbation strength (Fig.~\ref{f:vary_stimStrength}E). Note, however, that no excited frequency band emerges at WP3 for any of the stimulation strengths considered (Fig.~\ref{f:vary_stimStrength}F). 

In the main text, we found that at WP1, the absolute change in baseline band phase-coherence induced by exciting region $j$, $\langle | \Delta \rho^{\mathrm{base}}_{\delta_j} | \rangle $, was strongly correlated with the functional strength of region $j$, $s^{\mathrm{func}}_{j}$ (Fig. 5F,Left). In contrast, there was not a strong association between $\langle | \Delta \rho^{\mathrm{base}}_{\delta_j} | \rangle $ and the structural strength $s^{\mathrm{struc}}_{j}$ at WP1 (Fig. 5E,Left). Here, we observe that the nature of these two relationships remains qualitatively the same across the considered range of stimulation strengths $\Delta P_{\mathrm{E},j}$ (Fig.~\ref{f:vary_stimStrength}G). A second result from the main text was the finding of a strong positive correlation between the phase-locking induced in the excited frequency band, $\langle \uparrow \Delta \rho^{\mathrm{exc}}_{\delta_j} \rangle $, at WP1 and the structural strength $s_{j}^{\mathrm{struc}}$ of the stimulated unit (Fig. 5E,Left). The present analysis reveals that the relationship between $\langle \uparrow \Delta \rho^{\mathrm{exc}}_{\delta_j} \rangle $ and $s_{j}^{\mathrm{struc}}$ holds across a range of perturbation strengths $\Delta P_{\mathrm{E},j} > 0.05$ (Fig.~\ref{f:vary_stimStrength}I), which correspond to perturbation strengths yielding a clear increase in phase-locking in the excited frequency band (see Fig.~\ref{f:vary_stimStrength}E). For values of $\Delta P_{\mathrm{E},j} < 0.05 $, local excitations do not induce an excited frequency band at all, and so the correlation is undefined. Furthermore, at $\Delta P_{\mathrm{E},j} = 0.05 $, only some units in the network induce an excited frequency band; this effect results in a positive but weak correlation between $\langle \uparrow \Delta \rho^{\mathrm{exc}}_{\delta_j} \rangle $ and $s_{j}^{\mathrm{struc}}$. Finally, for the case of $\Delta P_{\mathrm{E},j} = 0.1 $ studied in the main text, there was a strong positive correlation between $\langle | \Delta \rho^{\mathrm{base}}_{\delta_j} | \rangle $ and $s^{\mathrm{struc}}_{j}$ at WP3 (and a weaker positive correlation between $\langle | \Delta \rho^{\mathrm{base}}_{\delta_j} | \rangle $ and $s^{\mathrm{func}}_{j}$). These trends also hold across the range of stimulation strengths examined in this section (Fig.~\ref{f:vary_stimStrength}H).

In conclusion, we note that while an in-depth examination of the effects of the stimulation strength is beyond the scope of the present study, it is important that the main relationships between interareal phase-locking modulations and network properties hold over a range of values for this parameter. 

\section{Results for an alternative value of the global coupling}
\label{s:vary_coupling}

In the main text, we examined the effects of focal excitatory stimulation for a global coupling of $C = 2.5$. Here, we analyze an alternative (but relatively nearby) coupling value of $C = 2.0$, and show that qualitatively similar results are found. As for $C = 2.5$, we consider three different working points by varying the level of background drive $P_{E}^{\mathrm{base}}$, while holding the coupling fixed. Specifically, we consider $P_{E}^{\mathrm{base}} = 0.60$ (WP1), $P_{E}^{\mathrm{base}} = 0.625$ (WP2), and $P_{E}^{\mathrm{base}} = 0.745$ (WP3), which place the system below, at, or above the state of peak global coherence (see Fig. 3A of the main text), respectively. As before, these working points represent three distinct dynamical states of the system.

\begin{figure*}
	\centering
	\includegraphics[width=0.4\textwidth]{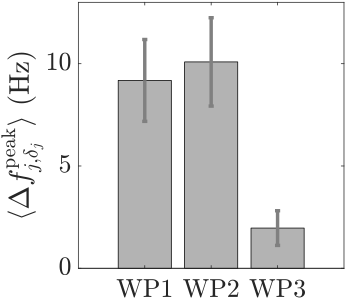}
	\caption{\textbf{Effect of enhanced regional excitation on the stimulated region's power spectra for three different working points at a coupling value of $C = 2$.} The average shift in the peak frequency of the excited region $\langle \Delta f^{\mathrm{peak}}_{j,\delta_j} \rangle$ for WP1, WP2, and WP3 (error bars indicate the standard deviation over all choices of the excited unit).}
	\label{f:varyCoupling_powSpec}
\end{figure*}

We begin by considering the effects of regional stimulation on the power spectra of the excited area. To summarize this, we examine the average shift in the peak frequency of the stimulated area, $\langle \Delta f_{j,\delta_j}^{\mathrm{peak}} \rangle$, for each of the three working points (Fig. \ref{f:varyCoupling_powSpec}). We find that for all three states, additional excitation has the effect of increasing the peak frequency of the stimulated region. However, for WP1 and WP2, the peak frequency shifts by a noticeably larger amount ($\langle \Delta f_{j,\delta_j}^{\mathrm{peak}} \rangle$ = 9.2Hz for WP1 and $\langle \Delta f_{j,\delta_j}^{\mathrm{peak}} \rangle$ = 10.1Hz for WP2) relative to the more modest effect at WP3 ($\langle \Delta f_{j,\delta_j}^{\mathrm{peak}} \rangle$ = 2.0Hz). These general trends are consistent with the results in the main text, and again demonstrate that individual areas are most responsive to additional excitation in states of lower background drive (WP1 and WP2). In contrast, given the same excitation strength, regional dynamics are relatively imperturbable when the system operates in the high background drive state (WP3).

We next examine how focal stimulation affects interareal phase-locking at each of the three working points. For WP1 and WP2 we analyze separate ``baseline" and ``excited" frequency bands, since the peak frequency of the stimulated area becomes well separated from the peak frequencies of the system at baseline. For WP3, we consider a single ``baseline" band, as the peak frequency of the excited area shifts only slightly and tends to overlap with the main frequencies at baseline. For the present analysis, we use the same protocol described in the main text to define baseline and excited frequency bands. In general, we refer the reader to Sec. III for further details and discussion regarding the results presented below.

\begin{figure*}
	\centering
	\includegraphics[width=\textwidth]{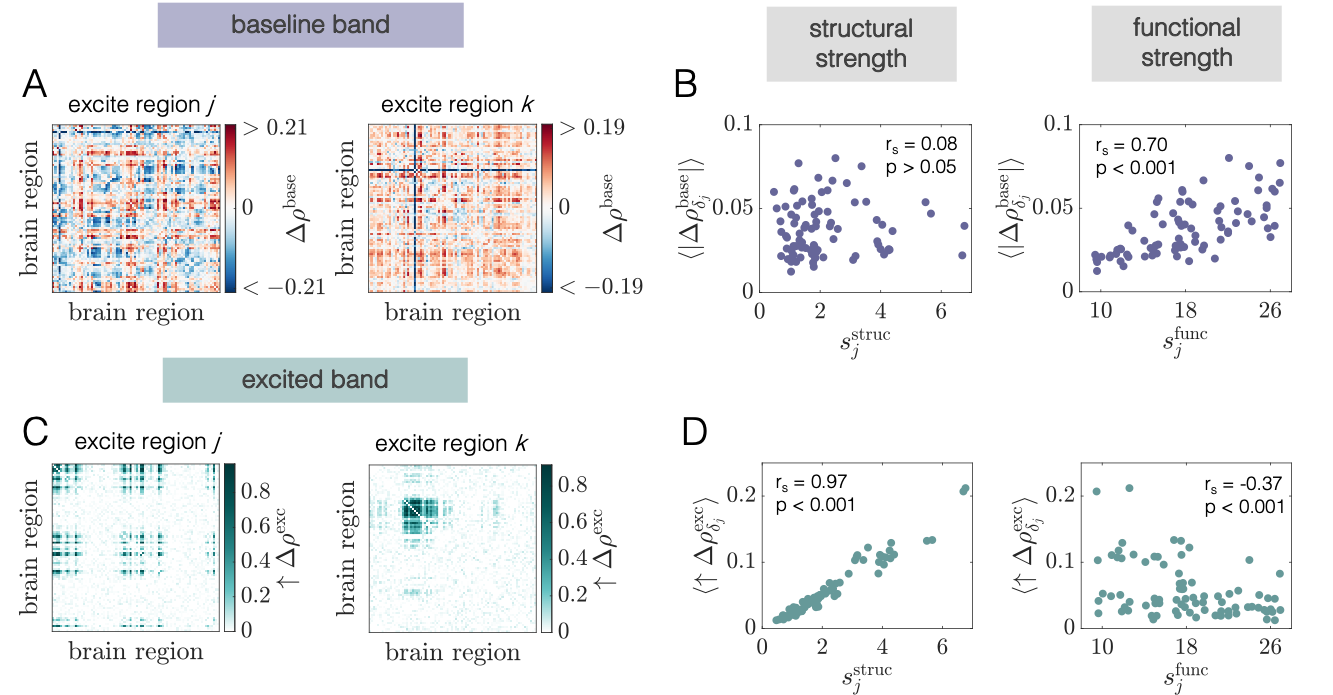}
	\caption{\textbf{Modulations of phase-locking induced by regional stimulation for WP1 at a coupling value of $C = 2$.} \textit{(A)} Pairwise changes in the PLV inside the baseline band $ \Delta \rho^{\mathrm{base}} $ when region $j$ (Left) or region $k \neq j$ (Right) is perturbed. Note that in this figure, regions $j$ and $k$ correspond to regions 4 (R--Medial Orbitofrontal) and 23 (R--Lateral Occipital), respectively. \textit{(B)} The quantity $\langle | \Delta \rho^{\mathrm{base}}_{\delta_j} | \rangle $ \textit{vs.} structural node strength $s^{\mathrm{struc}}_{j}$ (Left), and the quantity $\langle | \Delta \rho^{\mathrm{base}}_{\delta_j} | \rangle $ \textit{vs.} functional node strength $s^{\mathrm{func}}_{j}$ (Right). \textit{(C)} Pairwise increases in the PLV inside the excited band $ \Delta \rho^{\mathrm{exc}} $ when region $j$ (Left) or region $k \neq j$ (Right) is perturbed. \textit{(D)} The quantity $\langle \uparrow \Delta \rho^{\mathrm{exc}}_{\delta_j} \rangle $ \textit{vs.} structural node strength $s^{\mathrm{struc}}_{j}$ (Left) and the quantity $\langle \uparrow \Delta \rho^{\mathrm{exc}}_{\delta_j} \rangle $ \textit{vs.} functional node strength $s^{\mathrm{func}}_{j}$ (Right). In panels \textit{(B)} and \textit{(D)}, insets indicate Spearman correlation coefficients between the plotted quantities, and their associated $p$-values).}
	\label{f:varyCoupling_deltaPLV_WP1}
\end{figure*}

We first show -- for WP1 -- examples of the phase-locking modulations within the baseline and excited frequency bands for two different choices of the stimulated area (Figs.~\ref{f:varyCoupling_deltaPLV_WP1}A,C). As in the main paper (see Figs. 5A,C), we see that the network response to a local perturbation differs between the two frequency bands, and for different choices of the stimulated region. We next study the associations between the network-wide average of the phase-locking modulations induced by regional stimulation and structural or functional strength (Figs.~\ref{f:varyCoupling_deltaPLV_WP1}B,D). In comparing the results presented here for a coupling of $C = 2$ to those in Fig. 5E,F of the primary text for $C = 2.5$, we find similar relations. Specifically, the average absolute change in the PLV for the baseline frequency band $\langle |\Delta \rho_{\delta_j}^{\mathrm{base}}| \rangle$ is most strongly related to the baseline functional strength of the stimulated area $s_{j}^{\mathrm{func}}$ (Fig.~\ref{f:varyCoupling_deltaPLV_WP1}B,Right). In contrast, the network-average of the positive changes in excited band PLV $\langle \uparrow \Delta \rho_{\delta_j}^{\mathrm{exc}} \rangle$ is most strongly related to the structural strength of the stimulated region $s_{j}^{\mathrm{func}}$ (Fig.~\ref{f:varyCoupling_deltaPLV_WP1}D,Left). The other relationships are either weaker or not statistically significant.

\begin{figure*}
	\centering
	\includegraphics[width=\textwidth]{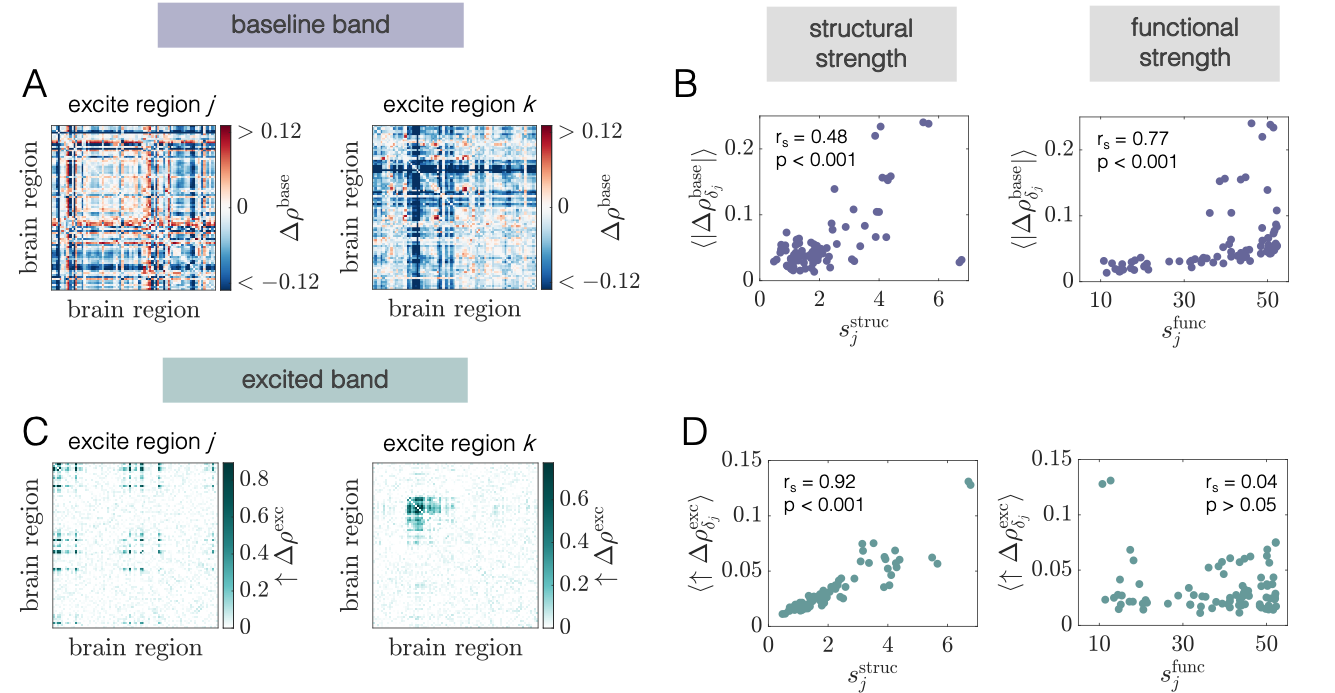}
	\caption{\textbf{Modulations of phase-locking induced by regional stimulation for WP2 at a coupling value of $C = 2$.} \textit{(A)} Pairwise changes in the PLV inside the baseline band $ \Delta \rho^{\mathrm{base}} $ when region $j$ (Left) or region $k \neq j$ (Right) is perturbed. Note that in this figure, regions $j$ and $k$ correspond to regions 4 (R--Medial Orbitofrontal) and 23 (R--Lateral Occipital), respectively. \textit{(B)} The quantity $\langle | \Delta \rho^{\mathrm{base}}_{\delta_j} | \rangle $ \textit{vs.} structural node strength $s^{\mathrm{struc}}_{j}$ (Left), and the quantity $\langle | \Delta \rho^{\mathrm{base}}_{\delta_j} | \rangle $ \textit{vs.} functional node strength $s^{\mathrm{func}}_{j}$ (Right). \textit{(C)} Pairwise increases in the PLV inside the excited band $ \Delta \rho^{\mathrm{exc}} $ when region $j$ (Left) or region $k \neq j$ (Right) is perturbed. \textit{(D)} The quantity $\langle \uparrow \Delta \rho^{\mathrm{exc}}_{\delta_j} \rangle $ \textit{vs.} structural node strength $s^{\mathrm{struc}}_{j}$ (Left), and the quantity $\langle \uparrow \Delta \rho^{\mathrm{exc}}_{\delta_j} \rangle $ \textit{vs.} functional node strength $s^{\mathrm{func}}_{j}$ (Right). In panels \textit{(B)} and \textit{(D)}, insets indicate Spearman correlation coefficients between the plotted quantities, and their associated $p$-values).}
	\label{f:varyCoupling_deltaPLV_WP2}
\end{figure*}

We next conduct the same analyses regarding changes to interareal phase-locking, but for WP2. Here, Figs.~\ref{f:varyCoupling_deltaPLV_WP2}A,C correspond to Figs. 7A,C of the main paper, which show examples of the phase-locking modulations within the baseline and excited frequency bands for two different choices of the stimulated area. Furthermore, Figs.~\ref{f:varyCoupling_deltaPLV_WP2}B,D correspond to Figs. 7E,F of the primary text, and depict the relationships between the different phase-locking modulations induced by regional stimulation and structural or functional network strength. We again find qualitatively similar behavior between the results shown here and those depicted in the main text. Note that for both values of the coupling ($C = 2$ here and $C = 2.5$ in the primary text), the main difference between WP1 and WP2 is that structural strength $s_{j}^{\mathrm{struc}}$ also exhibits a positive correlation with the average absolute change in phase-coherence for the baseline frequency band $\langle |\Delta \rho_{\delta_j}^{\mathrm{base}}| \rangle$ (Fig.~\ref{f:varyCoupling_deltaPLV_WP2}B,Left). However, for both values of $C$, phase-locking modulations at the system's endogenous frequencies continue to be most strongly associated with the stimulated region's baseline functional strength (Fig.~\ref{f:varyCoupling_deltaPLV_WP2}B,Right).

\begin{figure*}
	\centering
	\includegraphics[width=\textwidth]{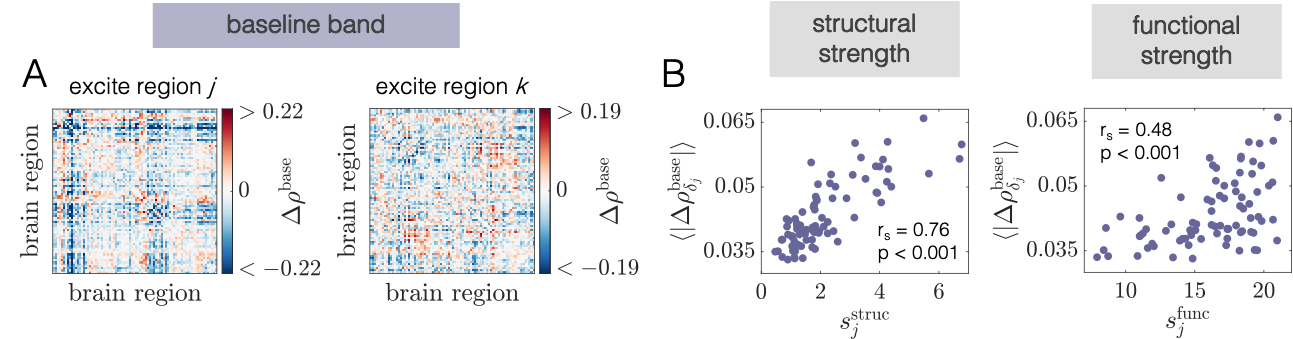}
	\caption{\textbf{Modulations of phase-locking induced by regional stimulation for WP3 at a coupling value of $C = 2$.} \textit{(A)} Pairwise changes in the PLV  inside the baseline band $ \Delta \rho^{\mathrm{base}} $ when region $j$ (Left) or region $k \neq j$ (Right) is perturbed. Note that in this figure, regions $j$ and $k$ correspond to regions 10 (R--Precentral) and 15 (R--Isthmus), respectively. \textit{(B)} The quantity $\langle | \Delta \rho^{\mathrm{base}}_{\delta_j} | \rangle $ \textit{vs.} structural node strength $s^{\mathrm{struc}}_{j}$ (Left), and \textit{vs.} functional node strength $s^{\mathrm{struc}}_{j}$ (Right). Insets indicate Spearman correlation coefficients between these quantities, and their associated $p$-values).}
	\label{f:varyCoupling_deltaPLV_WP3}
\end{figure*}

For completeness, we lastly consider phase-locking reconfigurations induced by regional excitation at WP3. Fig.~\ref{f:varyCoupling_deltaPLV_WP3}A shows the change in phase-coherence between each pair of regions (for the single, baseline frequency band) for two different choices of the stimulated area. As found in the main paper (e.g., Fig. 9A), the response, in general, differs across the choice of the excited region. The relationships between the phase-coherence modulations and structural or functional network strength found here for $C = 2$ (Fig.~\ref{f:varyCoupling_deltaPLV_WP3}B) are also consistent with the analysis performed in the primary text for $C = 2.5$ (Fig. 9C). Namely, there is a strong, positive correlation between $\langle |\Delta \rho_{\delta_j}^{\mathrm{base}}| \rangle$ and $s_{j}^{\mathrm{struc}}$ and a weaker but still significant correlation between $\langle |\Delta \rho_{\delta_j}^{\mathrm{base}}| \rangle$ and $s_{j}^{\mathrm{func}}$.

\section{Details on the Hilbert Transform}
\label{s:HT_appendix}

A common way to extract an instantaneous phase variable from a real-valued oscillatory signal is with the Hilbert transform. To begin, one writes the analytic (complex-valued) signal representation $X_{\mathrm{A}}(t)$ of the real-valued time-series $X(t)$ as 

\begin{equation}
X_{\mathrm{A}}(t) = X(t) + i X_{\mathrm{H}}(t) = A(t)e^{i \theta(t)},
\label{eq:analytic_signal}
\end{equation}

\noindent where $X_{\mathrm{H}}(t)$ is the Hilbert transform of $X(t)$, $A(t)$ is the instantaneous amplitude of $X(t)$, and $\theta(t)$ is the instantaneous phase of $X(t)$. Once one has computed $X_{\mathrm{H}}(t)$ and thus $X_{\mathrm{A}}(t)$, it is easy to see from Eq.~\ref{eq:analytic_signal} that the phase $\theta(t)$ can be computed as

\begin{equation}
\theta(t) = \arg[X_{\mathrm{A}}(t)].
\label{eq:hilbert_phase}
\end{equation}

\noindent The Hilbert transform of a signal $X(t)$ is defined as

\begin{equation}
X_{\mathrm{H}}(t) = \frac{1}{\pi}  \int_{-\infty}^{\infty}\frac{X(t')}{t - t'}dt',
\label{eq:hilbert_transform}
\end{equation}

\noindent where the integral is evaluated as a Cauchy principal value. From Eq.~\ref{eq:hilbert_transform}, one observes that the Hilbert transform is the convolution of $X(t)$ and $1\large{/}\pi t$: $X_{\mathrm{H}}(t) = X(t) *  1\large{/}\pi t$, so the Fourier transform (FT) of $X_{\mathrm{H}}(t)$, $\tilde{X}_{\mathrm{H}}(f)$, is just the product of the FTs of $X(t)$ and $1\large{/}\pi t$. For frequencies $f>0$, we thus have that $\tilde{X}_{\mathrm{H}}(f) = -i\tilde{X}(f)$, from which it becomes clear that the Hilbert transform just induces a phase shift of $\pi/2$ to each frequency component in the signal. 

In this study, we computed Hilbert transforms of the simulated neural activity using the `hilbert' function in MATLAB. As described in Sec. IIF1, the Hilbert Transform was applied after first filtering the raw time-series in a specified frequency band, in order to ensure that the corresponding phase variable is well-defined \cite{Pikovsky:2003a}.

\clearpage
\newpage

\section{Brain region identification numbers and labels}
\label{s:region_names}

\begin{longtable}[c]{c c c c c}
	\caption{Brain region ID numbers with their corresponding hemisphere (L $=$ left hemisphere, R $=$ right hemisphere) and anatomical labels. The ID numbers match the node numbering used in the adjacency matrix (see Fig. 1C), which also applies to all other region-by-region figures in the main text and in the supplementary text.} \\ \\
	\hline
	\hline
	Region ID & Hemisphere--Label & \ & Region ID & Hemisphere--Label \\
	\hline
	\hline
	\\
			1 & R--Lateral Orbitofrontal &  & 42 & L--Lateral Orbitofrontal \\
			2 & R--Pars Orbitalis &  & 43 & L--Pars Orbitalis \\
			3 & R--Frontal Pole & & 44 & L--Frontal Pole  \\
			4 & R--Medial Orbitofrontal & & 45 & L--Medial Orbitofrontal \\
			5 & R--Pars Triangularis & & 46 & L--Pars Triangularis \\
			6 & R--Pars Opercularis & & 47 & L--Pars Opercularis \\
			7 & R--Rostral Middle Frontal & & 48 & L--Rostral Middle Frontal\\
			8 & R--Superior Frontal & & 49 & L--Superior Frontal \\
			9 & R--Caudal Middle Frontal & & 50 & L--Caudal Middle Frontal   \\
			10 & R--Precentral & & 51 & L--Precentral \\
			11 & R--Paracentral & & 52 & L--Paracentral \\
			12 & R--Rostral Anterior Cingulate & & 53 & L--Rostral Anterior Cingulate \\
			13 & R--Caudal Anterior Cingulate & & 54 & L--Caudal Anterior Cingulate \\
			14 & R--Posterior Cingulate & & 55 & L--Posterior Cingulate \\
			15 & R--Isthmus & & 56 & L--Isthmus \\
			16 & R--Postcentral & & 57 & L--Postcentral \\
			17 & R--Supramarginal & & 58 & L--Supramarginal \\
			18 & R--Superior Parietal & & 59 & L--Superior Parietal  \\
			19 & R--Inferior Parietal & & 60 & L--Inferior Parietal \\
			20 & R--Precuneus & & 61 & L--Precuneus  \\
			21 & R--Cuneus & & 62 & L--Cuneus \\
			22 & R--Pericalcarine & & 63 & L--Pericalcarine \\
			23 & R--Lateral Occipital & & 64 & L--Lateral Occipital \\
			24 & R--Lingual & & 65 & L--Lingual \\
			25 & R--Fusiform & & 66 & L--Fusiform \\
			26 & R--Parahippocampal & & 67 & L--Parahippocampal \\
			27 & R--Entorhinal & & 68 & L--Entorhinal \\
			28 & R--Temporal Pole & & 69 & L--Temporal Pole  \\
			29 & R--Inferior Temporal & & 70 & L--Inferior Temporal \\
			30 & R--Middle Temporal & & 71 & L--Middle Temporal \\
			31 & R--Banks of Superior Temporal Sulcus & & 72 & L--Banks of Superior Temporal Sulcus \\
			32 & R--Superior Temporal & & 73 & L--Superior Temporal \\
			33 & R--Transverse Temporal & & 74 & L--Transverse Temporal \\
			34 & R--Insula & & 75 & L--Insula \\
			35 & R--Thalamus Proper & & 76 & L--Thalamus Proper  \\
			36 & R--Caudate & & 77 & L--Caudate  \\
			37 & R--Putamen & & 78 & L--Putamen \\
			38 & R--Pallidum & & 79 & L--Pallidum \\
			39 & R--Accumbens & &  80 & L--Accumbens \\
			40 & R--Hippocampus & & 81 & L--Hippocampus \\
			41 & R--Amygdala & & 82 & L--Amygdala \\
			
			\hline
			\hline
	\label{t:region_names}
\end{longtable}

\clearpage
\newpage

\bibliographystyle{unsrt}
\bibliography{/Users/liapapadopoulos/Dropbox/PhD_Work/My_Projects/bibFiles/master.bib}